\documentclass[usenatbib,preprint]{emulateapj}

\usepackage{graphicx}
\usepackage {natbib,aas_macros}
\usepackage {bm}
\usepackage {amsmath,amssymb}
\usepackage {setspace}

\usepackage{graphicx}
\usepackage {natbib,aas_macros}
\usepackage {bm}
\usepackage {amsmath,amssymb}
\usepackage {setspace}

\newcommand{\cms}{{\rm cm~s^{-1}}} 

\newcommand{\msun}{\thinspace M_\odot} 
\newcommand{\msunyear}{\thinspace M_\odot~{\rm yr}^{-1}} 
\newcommand{\St}{{\rm St}} 
\newcommand{\AU}{{\rm ~AU}} 
\newcommand{\ms}{{\rm ~m~s^{-1}}} 
\newcommand{\gcm}{~{\rm ~g~cm}^{-3} } 
\newcommand{\gscm}{~{\rm ~g~cm}^{-2} } 
\newcommand{\cm}{~{\rm ~cm} }

\begin{document}
\title[Apparent mass reduction of GI disk]{Apparent disk-mass reduction
and planetesimal formation in gravitationally unstable disks in Class 0/I YSOs}

\author{
Y. Tsukamoto\altaffilmark{1,2}, S. Okuzumi\altaffilmark{3}, and A. Kataoka\altaffilmark{4,5}
}
\altaffiltext{1}{
Laboratory of Computational Astrophysics, RIKEN, Saitama, Japan }
\altaffiltext{2}{
Graduate Schools of Science and Engineering, Kagoshima University, Kagoshima, Japan}
\altaffiltext{3}{
Department of Earth and Planetary Sciences, Tokyo Institute of Technology, Meguro, Tokyo, Japan
}
\altaffiltext{4}{
Zentrum f\"ur Astronomie der Universit\"at Heidelberg, Institut f\"ur Theoretische Astrophysik, Albert-Ueberle-Str. 2, 69120 \\ Heidelberg, Germany
 }
\altaffiltext{5}{
National Astronomical Observatory of Japan, Mitaka, Tokyo 181-8588, Japan
}


\begin{abstract}
We investigate the dust structure 
of gravitationally unstable disks undergoing mass accretion
from the envelope, envisioning the application to Class 0/I 
young stellar objects (YSOs)
We find that the dust disk quickly settles into a steady state and that, 
compared to a disk with interstellar medium (ISM) dust-to-gas 
mass ratio and micron-sized dust,
the dust mass in the steady-state decreases by a factor of 1/2 to 1/3,
and the dust thermal emission decreases by a factor of 1/3 to 1/5.
The latter decrease is caused by dust depletion and
opacity decrease owing to dust growth.
Our results suggest that the masses of gravitationally unstable disks 
in the Class 0/I YSOs are underestimated by a factor of 1/3 to 1/5 
when calculated from the dust thermal emission
assuming an ISM dust-to-gas mass ratio and micron-sized dust opacity, 
and that a larger fraction of disks in Class 0/I YSOs
is gravitationally unstable than was previously believed.
We also investigate the orbital radius  $r_{\rm P}$ within
which planetesimals form via coagulation of porous dust aggregates and
show that $r_{\rm P}$ becomes $\sim 20$ AU for a gravitationally 
unstable disk around a solar mass star.
Because $r_{\rm P}$ increases as the gas surface density 
increases and a gravitationally unstable disk has a 
maximum gas surface density,
$r_{\rm P}\sim 20$ AU is the theoretical maximum radius.
We suggest that planetesimal formation in the Class 0/I phase 
is preferable to that in the Class II phase
because large gas surface density is expected and large amount of dust 
is supplied by envelope-to-disk accretion.
\end{abstract}


\section{Introduction}
\label{intro}
Class 0/I young stellar objects (YSOs)
are in the earliest phase of protostar formation.
Recent observations have shown that 
they possess circumstellar disks
\citep{2007A&A...461.1037B,2008A&A...481..141L,
2012ApJ...754...52T,2012Natur.492...83T,
2013A&A...560A.103M,2014ApJ...796...70C,
2014ApJ...796..131O,2014Natur.507...78S,2014ApJ...793....1Y,
2015ApJ...812...27A,2017ApJ...834..178Y}.
They also possess a gaseous envelope
and the envelope-to-disk mass accretion occurs continuously.
The life-times are estimated to be 0.1 Myr 
and 0.5 Myr for the Class 0 YSOs and
 Class I YSOs, respectively \citep{2009ApJS..181..321E,2011ARA&A..49...67W}.

Theoretical studies have suggested that
gravitationally unstable disks frequently form in Class 0/I YSOs.
It is well known that the disk in a Class 0/I YSO
easily becomes gravitationally
unstable when the magnetic field in the cloud core is sufficiently weak
\citep{
1994ApJ...421..640N,
2003ApJ...595..913M,2006ApJ...650..956V,2009ApJ...704..715V,
2010ApJ...719.1896V,2010ApJ...724.1006M,
2011MNRAS.416..591T,2012MNRAS.427.1182S,2012PASJ...64..116K,
2013MNRAS.428.1321T,2013MNRAS.436.1667T,
2013ApJ...770...71T,2014MNRAS.439.3039L,2015MNRAS.446.1175T,
2016MNRAS.461.2257K}.
In particular, Vorobyov and his collaborators
investigated disk evolution 
using long-term simulations (to the end of the Class I phase) 
and showed that gravitationally unstable disks ubiquitously 
form in cloud cores in which the magnetic field is negligible.
Even when the magnetic field in the cloud core is relatively strong 
(e.g., when its mass-to-flux ratio normalized by the critical mass-to-flux
ratio is an order of unity), 
a gravitationally unstable disk  can form during the Class 0/I YSO phase.
Indeed, recent three-dimensional non-ideal magnetohydrodynamics
(MHD) simulations suggest that,
even with a relatively strong magnetic field, a circumstellar 
disk forms immediately after protostar formation
\citep{2011PASJ...63..555M,2015ApJ...810L..26T,2015MNRAS.452..278T,
2015ApJ...801..117T,
2016A&A...587A..32M,2016MNRAS.457.1037W}
and becomes gravitationally unstable 
\citep{2011PASJ...63..555M,2015ApJ...810L..26T,2015MNRAS.452..278T}
(for a review of disk formation in 
magnetized cloud cores, see \citet{2016PASA...33...10T}).
In particular, \citet{2011PASJ...63..555M} investigated 
the long-term evolution of
circumstellar disks (until $10^5$ years after protostar formation)
and showed that gravitationally unstable disks can form even 
in strongly magnetized cloud cores.
Therefore, from a theoretical point of view, 
gravitationally unstable disks may
frequently appear in the Class 0/I phase.

The properties of a gravitationally unstable disk are summarized
as follows.
Disks exhibit gravitational instability (GI) when 
their Toomre's $Q$ value \citep{1964ApJ...139.1217T} fulfills the 
following condition
\begin{equation}
Q \equiv \frac{c_{\rm s}\kappa_{\rm ep}}{\pi G \Sigma_{\rm gas}}\lesssim 1.4
\end{equation}
against non-axisymmetric perturbation \citep{1994ApJ...436..335L},
where $c_{\rm s},~\kappa_{\rm ep},~$, $\Sigma_{\rm gas}$, and $G$ are the 
sound velocity, epicycle frequency, gas surface density, and gravitational constant,
respectively.
To fulfill the $Q$ value criterion, 
the mass of the disk $M_{\rm disk}$
should be $M_{\rm disk}/M_{\rm star} \gtrsim 0.1$,
where $M_{\rm star}$ is the mass of the central star.
Because GI induces spiral arms and promotes 
mass and angular momentum transport.
it is a candidate mechanism for angular momentum
transfer in circumstellar disks (in other words,
GI can be source of viscous $\alpha$ \citep{1973A&A....24..337S}).
Some gravitationally unstable disks can fragment, and
binaries,
brown dwarfs, and wide orbit planets can be formed by such
fragmentation \citep{2008A&A...480..879S,2009MNRAS.392..413S,2010ApJ...714L.133V,
2010MNRAS.408L..36N,2013MNRAS.436.1667T}.
Furthermore, the disk fragmentation and subsequent accretion of the fragments onto
the central star may explain FU Orionis outbursts \citep{2006ApJ...650..956V}.
Another important property of a gravitationally unstable 
disk is that its surface density $\Sigma_{\rm gas}$ 
is at theoretical upper limit
because the GI inevitably develops and 
reduces $\Sigma_{\rm gas}$ at $Q \sim 1.4$.
It has been shown that 
the time and azimuthally averaged disk structures
are well described by a
steady-state viscous disk model with the assumption that $Q={\rm const}$
\citep[e.g.,][]{2015MNRAS.446.1175T}.
This greatly simplifies the gas disk structure because
detailed information about the magnetic 
field and the ionization degree of the disk are not required.

Although theoretical studies 
predict that gravitationally 
unstable disks frequently form in Class 0/I YSOs,
observations suggest that 
most of Class 0/I YSO disks are gravitationally stable.
Observations of disk mass estimated from (sub)millimeter 
dust emissions have shown that disk masses in Class 0/I YSOs
are typically $0.01-0.1 \msun$ 
and the mean disk mass is approximately a few $0.01 \msun$
\citep{2007ApJ...671.1800A,2009A&A...507..861J},
which is factor of 1/2 to 1/10 smaller than that required to develop
GI. 
Therefore, a discrepancy exists between observation and theoretical prediction.

The source of this discrepancy is unclear.
One may imagine that inclusion of 
magneto-rotational instability (MRI), 
which is not incorporated (or resolved) in 
the abovementioned simulations, could remove the discrepancy by promoting
further mass accretion. However, 
we argue that this may not be the case.
To clarify this point, we consider the $\alpha$ 
value at the edge of the disk required to realize 
a typical mass accretion rate from the envelope 
of $\dot{M}_{\rm gas} \sim 10^{-6} \msunyear$.
Using the viscous accretion disk model \citep{1973A&A....24..337S},
the $\alpha$ value is estimated as
\begin{eqnarray}
\alpha = \frac{\dot{M}_{\rm gas}}{3 \pi \Sigma_{\rm gas} c_{\rm s}^2/\Omega} =
\frac{1}{3}\frac{\dot{M}_{\rm gas}}{c_{\rm s}^3/G}Q = 0.74 \left( \frac{\dot{M}_{\rm gas}}{10^{-6} \msunyear} \right) \nonumber \\
 \left( \frac{Q}{10} \right) 
\left( \frac{T}{20 K} \right)^{-3/2},
\end{eqnarray}
where we assume that the typical radius of the disk 
is $r_{\rm disk}=100 {\rm AU}$ \citep{2011ARA&A..49...67W} 
and that disk temperature is typically $20$ K 
at $100$ AU \citep{1997ApJ...490..368C}. We also assume that 
$\kappa_{\rm ep}=\Omega$, where $\Omega$ is the angular velocity
and $c_{\rm s}=\sqrt{k_BT/m_{\rm g}}=1.9\times10^4~(T/10 {~\rm K})^{1/2} \cms$,
where $k_B $ and $m_{\rm g}=3.9\times10^{-24} {\rm g}$ 
are the Boltzmann constant and mean molecular mass, respectively.
This estimate shows that a quite large $\alpha$ of $\sim 1$ is required 
to achieve  $\dot{M}_{\rm gas}=10^{-6} \msunyear$ in a disk 
with $Q=10$ and $r_{\rm disk}=100$ AU. 
On the other hand, the MRI-induced $\alpha$ 
is typically $\alpha\sim 10^{-2}$, even in fully ionized disk,
and may be significantly smaller when the non-ideal MHD effects are
at work \citep{2000ApJ...530..464F,2011ApJ...736..144B, 2013MNRAS.434.2295K}.
Therefore, the angular momentum transfer owing to MRI is 
too weak to attain the typical mass accretion 
rate of $\dot{M}_{\rm gas}=10^{-6} \msunyear$ from the envelope,
and the mass inevitably accumulates in the disk.
Even for a disk with $Q\sim2$, which corresponds
to a marginally gravitationally unstable disk, 
$\alpha$ should be $\sim 0.14$
at $100$ AU. Again, MRI may not play a major role in the outer region.
Because no physical mechanism is known to produce 
$\alpha \sim 1$, and because the above estimate is robust and 
difficult to refute, we conclude that MRI (or other angular 
momentum transfer mechanisms) may not solve the discrepancy.

Thus, we should seek a source of this discrepancy other than MRI.
One possible explanation we pursue in this paper is 
growth and depletion of the dust particles.
Gas disk mass is often estimated from 
dust thermal emission by assuming the typical 
dust-to-gas mass ratio 
of the interstellar medium (ISM), $f_{\rm dg, ISM}=1/100$
and micron-sized dust.
However, whether these assumptions are also justified for the disk is unclear
because dust particles can grow
and can be depleted by radial migration.
The reduction of the dust-to-gas mass ratio causes
an apparent underestimation of the gas mass.
Furthermore, 
dust opacity can decrease by dust growth, which also leads to
an apparent underestimation of the gas mass.
The problem of assuming the ISM dust-to-gas mass ratio
and simply converting the dust mass to a gas mass
have been pointed out by several authors \citep{2005IAUS..231..377K,
2005A&A...434..971D,2007A&A...469.1169B,2014MNRAS.444..887D}.
In fact, the disk observations using HD line
emission, which is a more direct tracer for disk mass
suggest that the disk mass estimated from dust
thermal emission tends to be smaller
\citep{2013Natur.493..644B,2016ApJ...831..167M}.
Note also that \citet{2007ApJ...671.1800A}
\citep[see also][]{1998ApJ...495..385H} 
pointed out that 
the disk mass required to maintain disk-to-star mass accretion during the
Class II phase is much larger than that estimated from
dust thermal emission.
Although these previous studies mostly focus on the Class II phase,
dust growth and depletion also possibly explain
the discrepancy of disk mass in Class 0/I YSOs.

Several observational studies
have suggested the dust growth occurs in YSOs.
\citep{1991ApJ...381..250B,2002ApJ...581..357K,2007ApJ...659..479J,
2009ApJ...696..841K,2010A&A...521A..66R,2010A&A...512A..15R,2012ApJ...760L..17P,
2014A&A...567A..32M,2013ApJ...771...48T,
2015ApJ...813...41P,2016A&A...588A..53T,
2016ApJ...821L..16C}.
It is known that the opacity spectral index $\beta$ decreases
owing to dust growth.
For example, \citet{2010A&A...521A..66R} showed
that the Class II YSOs in Ophiuchus exhibit $\beta\sim 0.5$ 
, which is significantly
smaller than the ISM value, i.e., $\beta~1.7$.
Even in Class 0/I YSOs, the value of $\beta$ can be
smaller than that in the ISM 
\citep{2007ApJ...659..479J,2009ApJ...696..841K,2013ApJ...771...48T}.
These observations suggest that dust growth 
occurs ubiquitously in disks (here, we assume that 
a dust disk is optically thin for millimeter wavelengths) and that the
dust size may be significantly larger than that in the ISM.
In particular, small $\beta$ observed in some Class 0 YSOs suggests
that the dust growth occurs even in the earliest 
phase of star formation. 
Note, however, that we should take care in interpreting these observations
because $\beta$ also decreases when the disk is optically thick.
As shown by the recent observation of a Class I object 
\citep{2016Natur.535..258C},
it is possible that the decreases in the value of $\beta$ comes not from
dust growth but from the large optical depth of the disk.
Thus, verifications of dust growth in Class 0/I objects by
future high-resolution observations are awaited.

From a theoretical point of view, dust growth and subsequent
dust radial drift are also
expected even in the Class 0/I phase because of the small 
timescale of dust growth and radial drift.
Previous studies have pointed out
that the growth and drift timescales of dust particles
are much smaller than the viscous timescale of the
disk \citep{1977MNRAS.180...57W,1986Icar...67..375N,
2005ApJ...627..286T}
and the dust-to-gas mass ratio decreases in a very short duration
\citep{2005A&A...434..971D,
2007A&A...469.1169B,2008A&A...480..859B}.
In these previous studies, however,
isolated disks were investigated
and how dust evolution proceeds in Class 0/I phase (or under the
envelope-to-disk accretion) was not clear.
More recently, \citet{2010A&A...513A..79B}
investigated the evolution of dust particles 
in an evolving circumstellar disk by considering the
envelope-to-disk mass accretion.
They showed that
the dust-to-gas mass ratio become as small as $\sim 1/1000$ 
at $t=1$ Myr after the simulation was initiated, with
the collisional fragmentation being negligible.
Although \citet{2010A&A...513A..79B} clearly showed the possibility of 
significant reduction of the dust-to-gas mass ratio at the end of
the Class I phase,
the quantitative reduction rate 
of the dust-to-gas mass ratio during the Class 0/I phase was not shown.
Furthermore, the dependence of the reduction rate 
on parameters, such as mass accretion rate onto the disk, mass 
of the central star, dust porosity, 
and the gas disk structure, remains unclear.

Another important issue regarding dust evolution
in a gravitationally unstable disk is the 
maximum orbital radius of planetesimal formation.
\citet{2012ApJ...752..106O} and \citet{2013A&A...557L...4K} 
proposed a planetesimal formation scenario
in which icy planetesimals form from highly porous dust aggregates.
In their scenario, as noted by
\citet{2012ApJ...752..106O}, the orbital radius within which 
planetesimals form
is an increasing function of the gas surface density.
On the other hand, the gas surface density of a gravitationally unstable disk
is the theoretical upper limit.
Therefore,
we can determine the maximum orbital radius within
which planetesimals form
by considering planetesimal formation in a gravitationally unstable disk.

In this paper, we investigate the dust 
structure and planetesimal formation in gravitationally unstable disks
undergoing mass accretion from envelopes.
This paper is organized as follows:
In \S 2, we describe the models and 
governing equations for dust evolution.
The results are given in \S 3. 
We summarize and discuss our results in \S 4.

\section{Models}
In this section, we describe the gas disk model and the
governing equations for dust evolution adopted in this paper.
To clarify the dust evolution in a gas disk 
and simplify the system behavior,
we assume that the gas disk is in a steady state and does not 
evolve. As discussed in this section, this assumption is valid.

\subsection{Steady-state structure of gravitationally unstable disks}
\label{method}
We construct the steady-state profile of a gravitationally unstable disk 
as a function of the radius $r$, mass of the central star $M_{\rm star}$, 
and mass accretion rate of gas $\dot{M}_{\rm gas}$. 
The fundamental assumptions of our disk model are as follows:
\begin{enumerate}
\item The disk can be described by the viscous 
$\alpha$ accretion disk model \citep{1973A&A....24..337S};
\item The disk is steady, meaning that $\dot{M}_{\rm gas}={\rm const}$;
\item The Toomre's $Q$ value of the disk is on the order of unity and constant.
\end{enumerate}

With these assumptions, the physical quantities of 
the disk should satisfy the following equation:

\begin{equation}
\label{sec2_dynamics}
\begin{split}
\left| \frac{d \ln \Omega}{d \ln R} \right| \alpha \frac{c_{\rm s}^2}{\Omega} \Sigma_{\rm gas} = \frac{1}{2 \pi}\dot{M}_{\rm gas} = {\rm const.}~ (\propto r^0),
\end{split}
\end{equation}
where, 
$\dot{M}_{\rm gas}$ is the mass accretion rate of the gas,
$\alpha= \nu \frac{\Omega}{c_{\rm s}^2}$, and $\nu$ is the kinematic viscosity.
The $Q$ value for a marginally gravitationally unstable disk takes constant value
$Q_{\rm crit}$,
\begin{equation}
\label{sec2_qvalue}
Q= Q_{\rm crit} ~( \propto r^0).
\end{equation}
We set 
$Q_{\rm crit}=2$ because spiral arms develop at $Q\sim 1.4$
\citep{1994ApJ...436..335L} and
a marginally unstable disk may have a slightly larger $Q$ value than $1.4$.
Here and in the following, we approximate the epicycle frequency as
$\kappa_{\rm ep}=\Omega$.

Equations (\ref{sec2_dynamics}) and (\ref{sec2_qvalue}) yield
\begin{equation}
\begin{split}
\Sigma_{\rm gas} \propto T^{1/2} \Omega, \\
\alpha \propto \dot{M}_{\rm gas} T^{-3/2},
\end{split}
\end{equation}
where we have used $c_{\rm s}\propto T^{1/2}$.
Thus, from equations (\ref{sec2_dynamics}) and
(\ref{sec2_qvalue}), we can determine the 
profile of the gravitationally unstable steady disk 
by specifying a rotation profile, temperature profile 
(or energy balance equation), 
and mass accretion rate.

In this paper, we assume the Keplerian rotation as
\begin{equation}
\label{sec2_omega_kep}
\Omega=\sqrt{\frac{G M_{\rm star}}{r^3}},
\end{equation}
for simplicity.
Note, however, that the rotational profile 
of a gravitationally unstable disk may
differ from simple Keplerian rotation, because the  gravitational potential
is modified from that of the point gravity source by the disk self-gravity
\citep[see,][]{2015MNRAS.446.1175T}.

\subsubsection{Temperature profile}
We assume that the disk temperature $T$
is determined by the stellar irradiation and obeys 
the profile given in \citet{1970PThPh..44.1580K} 
and \citet{1997ApJ...490..368C} as
\begin{equation}
\label{sec2_irr}
T=T_{\rm irr} \equiv 150 \left( \frac{r}{1 {\rm AU}} \right)^{-3/7}.
\end{equation}
The assumption that the disk temperature is determined by 
irradiation is valid because the viscous heating is
negligible for $r\gtrsim10$ AU which is our primary focus.
In Appendix B, 
we estimate the temperature profile determined by the viscous heating and 
confirm that the viscous heating is negligible.

\subsubsection{Gas disk structure}
By solving equations
(\ref{sec2_dynamics}), (\ref{sec2_qvalue}), (\ref{sec2_omega_kep}),
and (\ref{sec2_irr}),
we obtain the steady-state solution 
for the gravitationally unstable gas disk as
\begin{eqnarray}
\label{GI_disk_sigma}
\Sigma_{\rm gas}&=&1.3 \times 10^1 
\left( \frac{M}{M_\odot}\right)^{1/2} \left( \frac{r}{100 {\rm AU}}\right)^{-12/7} {\rm g~ cm^{-2}},\\
%
\label{GI_disk_alpha}
\alpha&=&1.4 \times 10^{-1} \left( \frac{\dot{M}_{\rm gas}}{10^{-6} M_\odot {\rm yr}^{-1}} \right) 
\left( \frac{r}{100 {\rm AU}}\right)^{9/14}.
\end{eqnarray}
Note that $\alpha$ has a radial dependence of $\alpha\propto r^{9/14}$,
which is a general feature of realistic disks.


The diffusion timescale of our disk is estimated as 
\begin{eqnarray}
t_{\rm diff} \equiv \frac{r^2}{\nu} = 
 1.4\times10^5 \left( \frac{r}{100 \AU} \right)^{2/7} \nonumber  \\
\left( \frac{\dot{M}_{\rm gas}}{10^{-6} \msunyear} \right)^{-1}  
\left( \frac{M_{\rm star}}{\msun} \right)^{1/2} {\rm years}
\end{eqnarray}
where $\nu=\alpha c_{\rm s}^2/\Omega$ is the kinematic viscosity.
This value is smaller than or comparable to
the lifetime of the Class 0/I phase $\sim 0.5$ Myr,
and our assumption of the steady-state is valid.

\subsubsection{Assumptions for the viscous $\alpha$}
\label{visc_alpha_discussion}
We assume that $\alpha$ values
that lead to the gas advection and internal turbulence are identical
for simplicity.
While the former includes both turbulent viscosity and gravitational
torque, i.e., $\alpha_{\rm turb}+\alpha_{\rm grav}$.
The latter incorporates $\alpha_{\rm turb}$ only.
Here $\alpha_{\rm turb}=\langle \Sigma_{\rm gas} \delta v_R \delta v_\phi \rangle$ 
is the $\alpha$ value induced by the Reynolds stress.
and $\alpha_{\rm grav}=\langle \int d z g_R g_\phi/(4 \pi G) \rangle$
is that induced by the gravitational torque,
where $\langle \rangle$ indicates
the time and azimuthal average, and $\delta v_R,~\delta v_\phi$
and $g_R,~g_\phi$ indicate the radial and azimuthal 
components of the velocity fluctuation and
of the gravitational force induced by the spiral arms, respectively.
This simplification is valid if 
$\alpha_{\rm turb}/\alpha_{\rm grav}\gg 1$.
Note that, with this simplification, we overestimate the
collision velocity induced by the disk turbulence,
and the realistic collision velocity is
smaller than the value obtained in this paper.
We neglect the effect of MRI, because
it is expected to be weak ($\alpha\lesssim 10^{-2}$) compared 
to the GI in the outer region ($\alpha \sim 10^{-1}$).

\subsection{Dust model}
\label{sec_dust_model}
An important quantity that determines the dust dynamics is the 
stopping time of dust particles $t_s$, which
is the timescale on which the dust particle
momentum is relaxed by gas drag.
In this paper, we consider two regimes of 
the stopping time depending on 
the dust size:
\begin{equation}
t_s = 
     \begin{cases}
       \frac{\rho_{\rm int} a_{\rm dust}}{\rho_{\rm gas} v_{\rm th}} &\quad a_{\rm dust}<\frac{9}{4}\lambda_{\rm mfp} ~~(\text{Epstein drag law}) \\
       \frac{4 \rho_{\rm int} a_{\rm dust}^2}{9\rho_{\rm gas} v_{\rm th}\lambda_{\rm mfp}}  &\quad a_{\rm dust}>\frac{9}{4}\lambda_{\rm mfp} ~~(\text{Stokes drag law}) \\
     \end{cases}
\end{equation}
where $\rho_{\rm int},~a_{\rm dust}$, and $\rho_{\rm gas}$ are 
the internal density, radius of the dust particles, 
and gas density, respectively.
We adopt  $\rho_{\rm int}=1.4f \gcm$ where $f$ is volume filling factor.
The thermal velocity $v_{\rm th}$ is given by 
$v_{\rm th}=\sqrt{8/\pi}c_{\rm s}$.
The mean free path is given by $\lambda_{\rm mfp}=m_{\rm gas}/(\sigma_{\rm mol} \rho_{\rm gas})$,
where $\sigma_{\rm mol}=2\times10^{-15} {\rm cm^2}$ is the 
collisional cross section of the gas molecules 
and $m_{\rm gas}=3.9\times10^{-24} {\rm g}$ is the mean molecular mass.
We do not consider Newton's drag regime, which
applies to very large particles
around which the gas flow has high Reynolds numbers.
The motion of the dust particles is characterized by the 
Stokes number $\St$, which is defined as 
\begin{equation}
\label{Stokes_num}
\St \equiv \Omega t_s=
     \begin{cases}
       \frac{\pi \rho_{\rm int} a_{\rm dust}}{2\Sigma_{\rm gas}}, &\quad a_{\rm dust}<\frac{9}{4}\lambda_{\rm mfp} ~~(\text{Epstein drag law}), \\
       \frac{2 \rho_{\rm int} a_{\rm dust}^2}{9\Sigma_{\rm gas}\lambda_{\rm mfp}},  &\quad a_{\rm dust}>\frac{9}{4}\lambda_{\rm mfp} ~~(\text{Stokes drag law}). \\
     \end{cases}
\end{equation}

In this study, we employ simplified dust coagulation equations
in which the dust size distribution 
is characterized by the single representative mass $m_{\rm dust}(r)$.
This single-size approximation have been employed in
many previous studies on dust evolution in 
protoplanetary disks 
\citep{2001A&A...378..180K,2012A&A...539A.148B,
2016A&A...589A..15S,2016ApJ...821...82O,2016A&A...586A..20K}.
The governing equations for dust evolution are
\begin{eqnarray}
\label{governing_eq1}
\frac{\partial \Sigma_{\rm dust}}{\partial t}+\frac{1}{r} 
\frac{\partial}{\partial r}(r v_{\rm r, dust}\Sigma_{\rm dust})=0,\\
\label{governing_eq2}
\frac{\partial m_{\rm dust}}{\partial t}+v_{\rm r, dust}
\frac{\partial m_{\rm dust}}{\partial r}=\frac{m_{\rm dust}}{t_{\rm coll}},
\end{eqnarray}
where $\Sigma_{\rm dust}$ and $v_{\rm r, dust}$ are the dust surface density
and the dust radial velocity, respectively.
The first equation represents the mass conservation of dust particles, where
we neglect the turbulent diffusion term for simplicity.
In this paper, we consider two forms for the dust radial velocity.
One is the standard form, which is given as
\begin{equation}
\label{vr_dust}
v_{\rm r, dust}=-(\frac{v_{\rm r, gas}}{1+\St^2}+\frac{2 \St}{1+\St^2}\eta v_{\rm K}),
\end{equation}
where $v_{\rm K}=r \Omega $, $v_{\rm r, gas}$ is the gas radial velocity
given as $v_{\rm r, gas}=\dot{M}_{\rm gas}/(2 \pi r \Sigma_{\rm gas})$, 
and $\eta$ is a parameter that determines the sub-Kepler motion of the gas
and is expressed as \citep{1977MNRAS.180...57W},
\begin{equation}
\eta=-\frac{1}{2}\left(\frac{c_{\rm s}}{v_{\rm K}}\right)^2 \frac{d \ln P}{d \ln r}.
\end{equation}

The first and second terms in the right hand side describe the dust radial
motion caused by gas advection \citep{2001A&A...378..180K} and
the radial drift of the dust particles \citep{1977MNRAS.180...57W},
respectively.
We also consider the other form for the dust radial velocity, which is given by
\begin{equation}
\label{vr_dust2}
v_{\rm r, dust}=-(v_{\rm r, gas}+\frac{2 \St}{1+\St^2}\eta v_{\rm K}).
\end{equation}
In this form, 
the radial drift caused by gas advection is artificially enhanced 
for $\St\gtrsim1$.
The reason why we consider this form is to investigate
the orbital radius of planetesimal formation in the steady-state solution.
When $v_{\rm r, dust}$ is calculated using equation (\ref{vr_dust}),
the radial migration of planetesimals essentially stops because
their Stokes number is $\St\gg 1$, and
the orbital radius  within
which planetesimals form  $r_{\rm P}$ is inevitably
influenced by the initial condition.
The radius of the planetesimal formation 
calculated using equation (\ref{vr_dust}) 
indicates the maximum radius of planetesimal formation $r_{\rm P, max}$ 
during the time evolution of the dust disk for a given parameter set
because the initial disk has a larger dust 
surface density and larger dust mass accretion rate than tare presented 
in the steady disk.
On the other hand, 
when $v_{\rm r, dust}$ is calculated using equation (\ref{vr_dust2}),
the planetesimals migrate with gas advection velocity and
are swept away from the disk.
Therefore, $r_{\rm P}$ obtained with equation (\ref{vr_dust2}) is
the planetesimal formation radius expected 
from the steady-state solution. We denote 
this radius as $r_{\rm P, steady}$, which
corresponds to the minimum value of $r_{\rm P}$ for
a given parameter set.

In a realistic situation, whether the planetesimals form
at $r_{\rm P, max}$ or $r_{\rm P, steady}$ is unclear, in fact,
it largely depends on the formation process of the gas disk.
If the disk formation process is sufficiently rapid and 
the disk maintains the ISM dust-to-gas mass ratio, the planetesimals
form at $r_{\rm P, max}$. On the other hand,
they form at $r_{\rm P, steady}$ 
if the disk formation process is slow and the dust particles
are already depleted in the inner region.
We can expect, however, that planetesimals form 
between $r_{\rm P, max}$ and $r_{\rm P, steady}$.

Note that the results other than $r_{\rm P}$ discussed in this paper 
are independent of the choice of the dust radial velocity.
We therefore use equation (\ref{vr_dust}) unless otherwise noted.

Equation (\ref{governing_eq2}) represents the dust growth that
can be derived by taking the first moment
of the dust coagulation equation 
\citep[see the Appendix of][]{2016A&A...589A..15S}.
The collision time $t_{\rm coll}$ is given as 
\begin{eqnarray}
t_{\rm coll}=\frac{1}{4\pi a_{\rm dust}^2 n_{\rm dust} \Delta v}.
\end{eqnarray}
where $n_{\rm dust}$ is the dust number density and $\Delta v$ is 
the collision velocity between dust particles.
$n_{\rm dust}$ can be rewritten using 
$\Sigma_{\rm dust}$, the dust scale height $H_{\rm dust}$, 
and the mass of the dust aggregate $m_{\rm dust}$ as
\begin{eqnarray}
n_{\rm dust}=\frac{\Sigma_{\rm dust}}{\sqrt{2\pi} H_{\rm dust} m_{\rm dust}}.
\end{eqnarray}
By assuming a balance between vertical settling and turbulent diffusion, 
the dust scale height is given as \citep{1995Icar..114..237D,2007Icar..192..588Y},
\begin{eqnarray}
\label{eq_H_dust}
H_{\rm dust}=(1+\frac{\St}{\alpha}\frac{1+2 \St}{1+\St})^{-1/2} H_{\rm gas},
\end{eqnarray}
where $H_{\rm gas}=c_{\rm s}/\Omega$ is the gas scale height.
We assume that the collision velocity of the dust particles is given as
\begin{eqnarray}
\Delta v = \sqrt{\Delta v_{\rm B}^2 +\Delta v_{\rm r}^2 +\Delta v_{\phi}^2 +\Delta v_{\rm z}^2 +\Delta v_{\rm turb}^2},
\end{eqnarray}
where $\Delta v_{\rm B} ,~\Delta v_{\rm r} ,
~\Delta v_{\phi} ,~\Delta v_{\rm z} $and $\Delta v_{\rm turb}$ are 
the collision velocity induced by Brownian motion, radial drift,
azimuthal drift, vertical settling, and disk turbulence, respectively.
We evaluate these components using the prescription described 
in \citet{2012ApJ...752..106O}.

\subsection{Initial and outer boundary condition}
We assume that the　initial dust-to-gas mass ratio in the disk is 
$f_{\rm dg, ISM}=1/100$, and that
the dust surface density profile is initially given as
$\Sigma_{\rm dust}=f_{\rm dg, ISM}\Sigma_{\rm gas}$. 
We also assume that the 
initial dust size is constant in the disk and given as 
$a_{\rm dust,init}=1 f^{-1/3} {\rm \mu m}$, where $f$ is the filling factor.
For consistency, we introduced the factor $f^{-1/3}$ 
to the internal density $\rho_{\rm int} \propto a^3 f$.

To mimic mass accretion from the envelope, 
the mass flux at the outer boundary is kept constant
during the simulation.
The dust-to-gas mass ratio and dust size at the outer boundary
are set as $f_{\rm dg, ISM}$ and $a_{\rm dust, init}$, respectively.
Thus, $\dot{M}_{\rm dust}=f_{\rm dg,ISM} \dot{M}_{\rm gas}$ at the
boundary.
With this treatment, we implicitly assume that the mass loading from
the envelope primarily occurs at the disk edge.

Mass loading from the disk edge well describes
realistic envelope-to-disk mass accretion.
Previous studies employing MHD simulations have reported the formation of
pseudo-disks and outflow 
\citep[e.g.,][]{2003ApJ...599..363A,2011PASJ...63..555M,2015MNRAS.452..278T}.
A pseudo-disk is a flattened disk-like structure that forms around a disk
and connects to the disk edge.
Because mass accretion primarily occurs through the pseudo-disk, 
almost all of the gas accretes onto the disk edge.
Furthermore, as the outflow has a large opening angle and 
sweeps up  gas residing above the disk 
\citep{2008ApJ...676.1088M,2012MNRAS.423L..45P},
the gas cannot accrete from the vertical
direction. Based on these considerations, we assume 
that the gas and dust mass are primarily loaded from the disk edge.

\subsection{Opacity of dust aggregate}
To estimate the radiative flux of dust thermal emission from 
the simulated dust disk,
we calculate the absorption opacity of the dust aggregates,
$\kappa_{\rm d,\lambda}$ using the analytic formula
given by \citet{2014A&A...568A..42K}.
The dust monomers are assumed to be 
composed of silicate, carbonaceous materials, and water ice
The mass fraction abundances are
identical to those adopted by \citet{1994ApJ...421..615P},
$\zeta_{\rm silicate}:\zeta_{\rm carbon}:\zeta_{\rm ice}=2.64:3.53:5.55$.
We employ the values for the refractive indices 
of astronomical silicate, amorphous carbon, and water ice given by
\citet{2001ApJ...548..296W}, \citet{1996MNRAS.282.1321Z}, and  
water ice  given by \citet{1984ApOpt..23.1206W}, respectively.
The effective monomer refractive index
is calculated using the Bruggeman mixing rule.

As we consider porous dust aggregates in this paper,
it is necessary to know their opacity.
We can regard a porous aggregate as a mixture of monomers and 
vacuum, and effective medium theory can be applied in order to obtain
the effective refractive index. This is calculated 
using the Maxwell-Garnett rule \citep[for details, see][]{2014A&A...568A..42K}.
We assume that the dust size distribution obeys a power law 
$dn/da \propto a^{-2.5}$ with cut-off radii 
of $a_{\rm min}=a_{\rm dust, init}$ and 
$a_{\rm max}=a_{\rm dust}$.
This power law is slightly shallower than that estimated for the ISM 
$dn/da \propto a^{-3.5}$, or $q=3.5$ \citep{1977ApJ...217..425M},
because, as discussed in \citet[][]{1993Icar..106...20M}, 
a smaller $q$ is expected 
when the coagulation process dominates the fragmentation
process that is true in the situation we consider in this paper.
The shallower size distribution is  also expected 
to explain the observed small value of the  opacity spectral index 
$\beta$ \citet{2010A&A...512A..15R}.

\begin{figure}
\includegraphics[width=50mm,angle=-90]{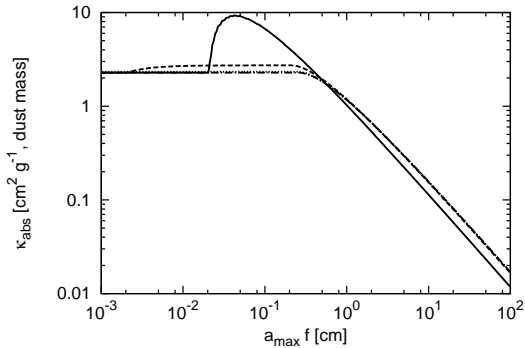}
\caption{
Absorption opacity  at $\lambda=1.3~{\rm mm}$  as a function of 
the product of the maximum dust radius and the filling factor, $a_{\rm max} f$
The solid, dashed, dotted, and dashed-dotted lines show
the opacity for $f=1,~10^{-1},~10^{-2},$ and $10^{-3}~$,
respectively.
}
\label{opacity_fig}
\end{figure}

Figure \ref{opacity_fig}
show the dust absorption opacity at $\lambda=1.3~{\rm mm}$ 
(which corresponds to ALMA Band 6)
as a function of the product of maximum 
dust size and  filling factor, $a_{\rm max} f$.
To obtain the opacity for this figure, 
we assume that $a_{\rm min}=0.1 {\rm \mu m}$.
As noted in \citet{2014A&A...568A..42K}, 
the filling factor and the dust size degenerate,
thus, the dust opacity is identical 
for $af \ll 1 {\rm cm}$ and $af \gg 1 {\rm cm}$.
However,  the opacity is enhanced 
at $10^{-2} {\rm cm} \lesssim a_{\rm max} \lesssim 1 {\rm cm}$ 
in the compact case ($f=1$).
This enhancement causes overestimation of the dust mass based on
the dust thermal emission (see figure \ref{disk_mass_porus}).
We expect, however, that this enhancement is not important 
in a realistic situation because 
the realistic dust aggregates may have 
$f \lesssim 10^{-1}$, as suggested by  observation 
of comets \citep[][]{2005Sci...310..258A,2016Natur.530...63P},
as well as by recent theoretical studies on dust coagulation
incorporating porosity evolution
\citep{2007A&A...461..215O,2012ApJ...752..106O}.
The value of the opacity is consistent with
previous works \citep{2010A&A...512A..15R,2016ApJ...821...82O}.

\subsection{Parameters and Models}
In this paper, as parameters, 
we choose the mass accretion rate ${\dot M_{\rm gas}}$, 
radius of the disk $r_{\rm disk}$, 
the mass of the central star $M_{\rm star}$, and
the filling factor $f$.
Table 1 lists the model names and parameter choices that are
investigated in \S 3.
Furthermore, to derive 
the empirical formula shown in equation (\ref{empirical_form_muf}),
we executed a total of  144 simulations.

\begin{table*}
\label{initial_conditions}

\begin{center}
\caption{
Models investigated in \S 3 and their parameters.
Note that we execute 144 simulations in total in order to derive 
our empirical formula, shown in equation (\ref{empirical_form_muf}) and
most of them are not shown in this table.
}		
\begin{tabular}{cccccccc}
\hline\hline
\shortstack{ Model \\ name}  & \shortstack{Stellar \\ mass \\ ($\msun$)}&  \shortstack{Gas accretion \\ rate \\($\msunyear$)}& \shortstack{Disk \\ radius \\($\AU$)}& \shortstack{Filling \\ factor} & Comment \\
\hline
 M1Mdot37r100f1  & 1 & $3 \times 10^{-7}$ & $100$ & $10^{-1}$ & fiducial model \\
\hline
 M05Mdot37r100f1  & 0.5 & $3 \times 10^{-7}$ & $100$ & $10^{-1}$ & \\
 M2Mdot37r100f1  & 2 & $3 \times 10^{-7}$ & $100$ & $10^{-1}$ & \\
\hline
 M1Mdot36r100f1  & 1 & $3 \times 10^{-6}$ & $100$ & $10^{-1}$ & \\
 M1Mdot16r100f1  & 1 & $1 \times 10^{-6}$ & $100$ & $10^{-1}$ & \\
 M1Mdot17r100f1  & 1 & $1 \times 10^{-7}$ & $100$ & $10^{-1}$ & \\
\hline
 M1Mdot37r50f1  & 1 & $3 \times 10^{-7}$ & $50$ & $10^{-1}$ & \\
 M1Mdot37r200f1  & 1 & $3 \times 10^{-7}$ & $200$ & $10^{-1}$ & \\
\hline
 M1Mdot37r100f0  & 1 & $3 \times 10^{-7}$ & $100$ & $10^{0}$ & \\
 M1Mdot37r100f2  & 1 & $3 \times 10^{-7}$ & $100$ & $10^{-2}$ & \\ 
 M1Mdot37r100f4  & 1 & $3 \times 10^{-7}$ & $100$ & $10^{-4}$ & \\
\hline
 M2Mdot37r100f4  & 2 & $3 \times 10^{-7}$ & $100$ & $10^{-4}$ & \\
 M2Mdot37r100f5  & 2 & $3 \times 10^{-7}$ & $100$ & $10^{-5}$ & \\
 M1Mdot37r100f5  & 1 & $3 \times 10^{-7}$ & $100$ & $10^{-5}$ & \\
\hline
\end{tabular}
\end{center}
\footnotesize
\end{table*}

\section{Results}

\begin{figure*}
\includegraphics[width=60mm,angle=-90]{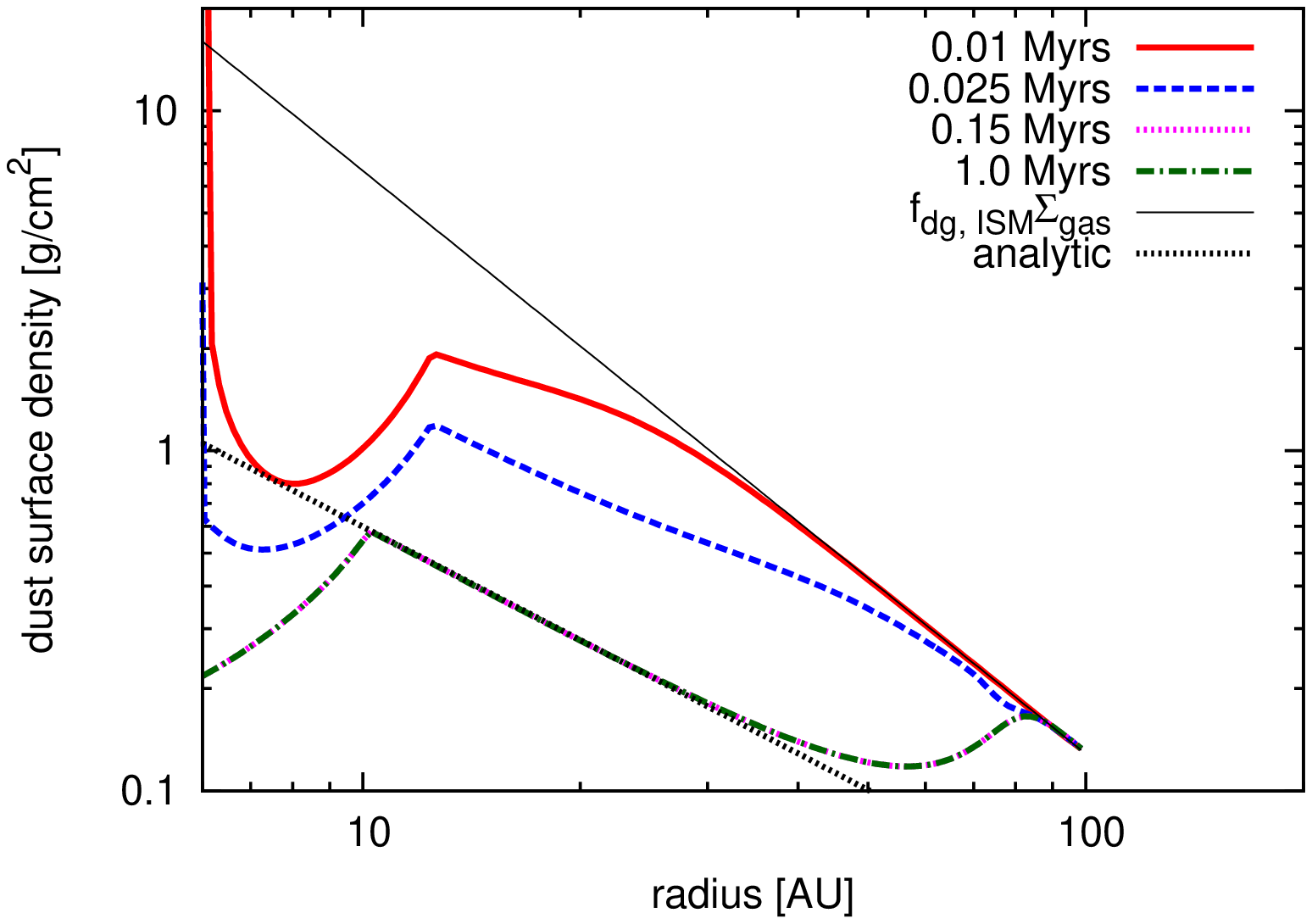}
\includegraphics[width=60mm,angle=-90]{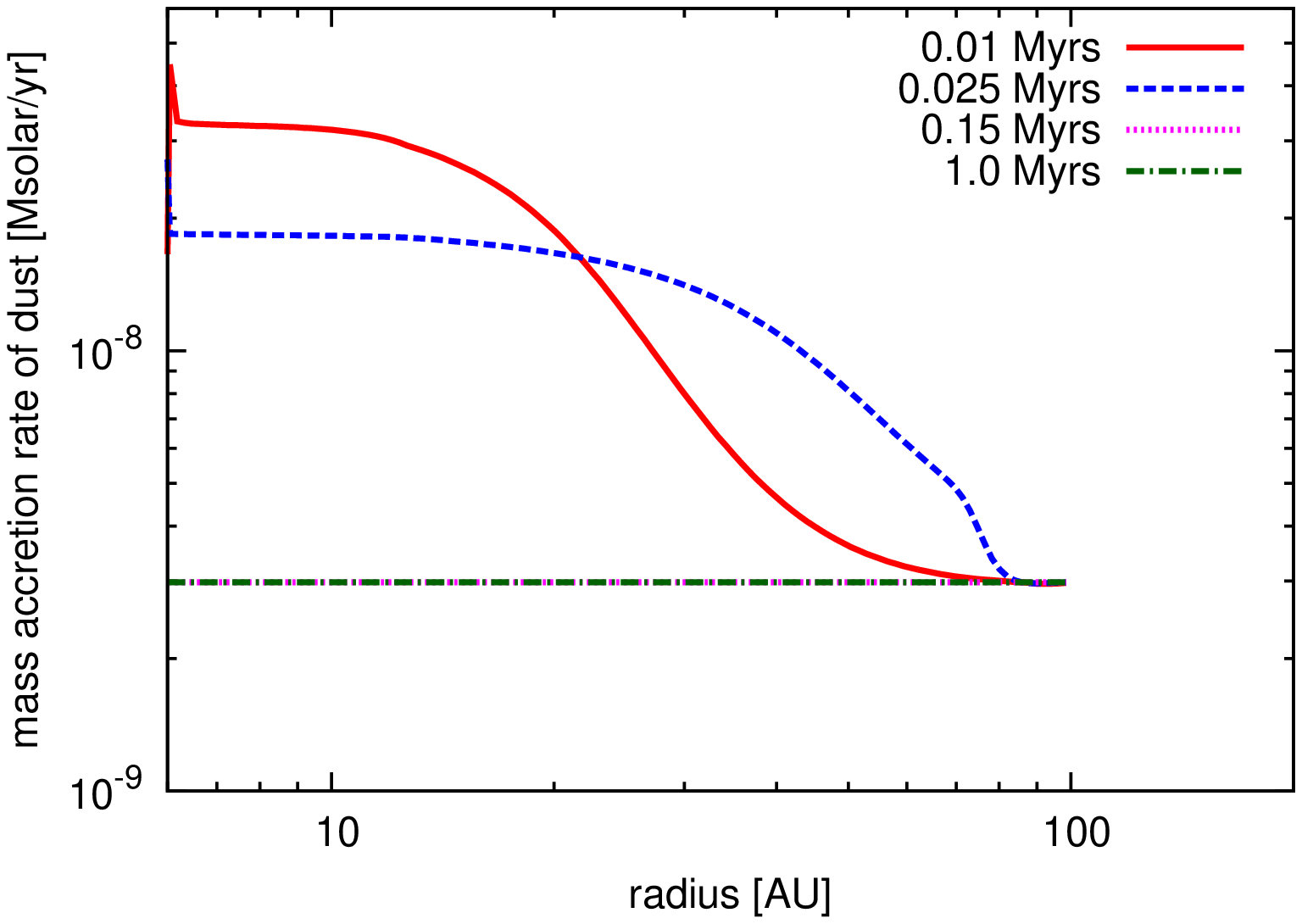}
\includegraphics[width=60mm,angle=-90]{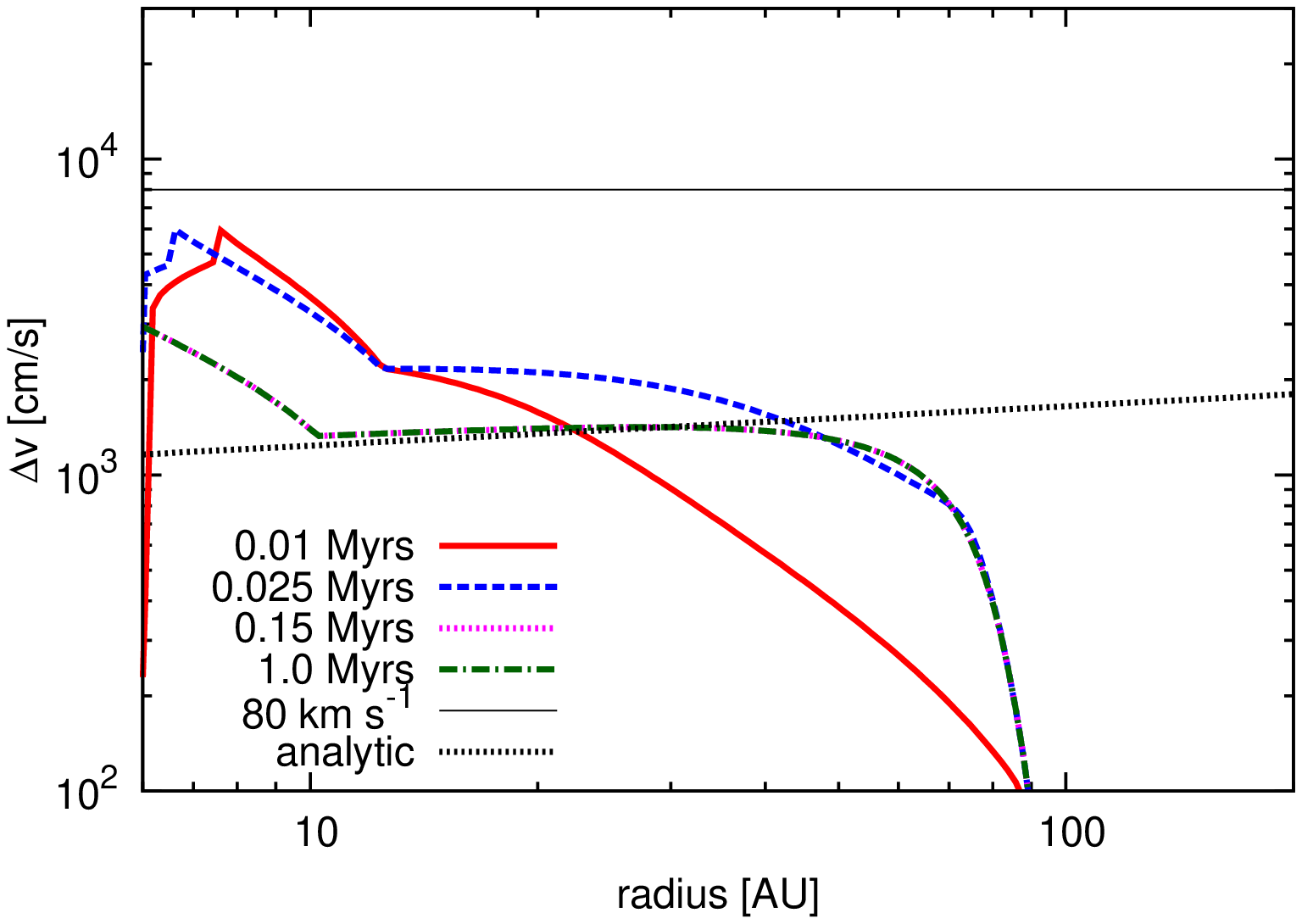}
\includegraphics[width=60mm,angle=-90]{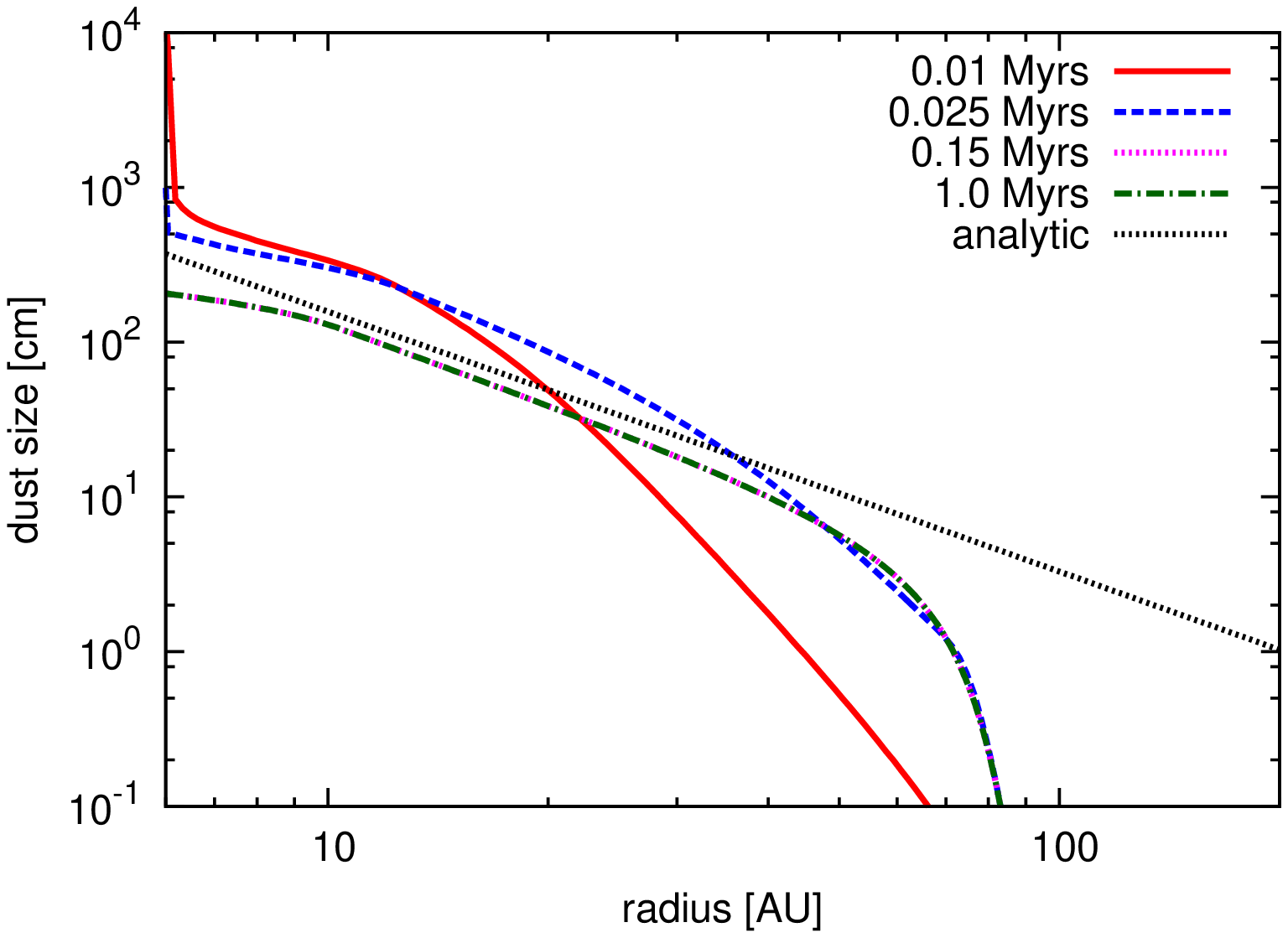}
\includegraphics[width=60mm,angle=-90]{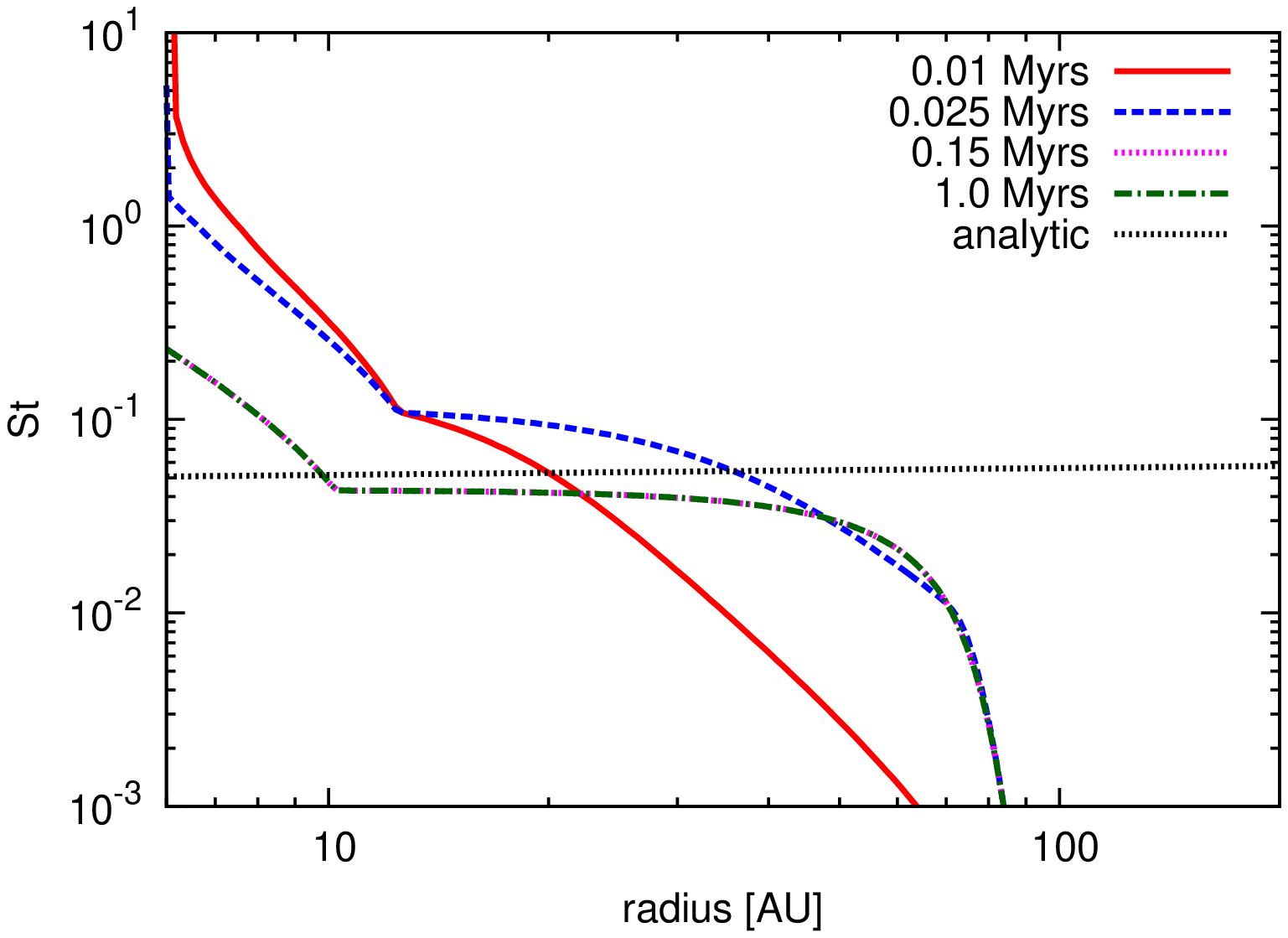}
\caption{
Time evolution of the radial profile of dust density,
mass accretion rate, collision velocity, dust size, and Stokes number
of our fiducial model,  M1Mdot37r100f1.
The red solid, blue dashed, magenta dotted, and green dashed-dotted lines
show the profiles at $t=0.01,~0.025,~0.15,$ and $1.0$ Myr, respectively.
The black solid lines in the surface density and 
in the collision velocity plots show $f_{\rm dg, ISM} \Sigma_{\rm gas}$ and
the threshold velocity $v_{\rm th}=80 \ms$, respectively.
The black dashed lines show 
the analytic steady-state solutions under the condition where
$\Delta v=\sqrt{2 \alpha c_{\rm s}^2 \St},~H_{\rm dust}=(\St/\alpha)^{-1/2}H_{\rm gas}$, and
the Epstein law is followed (equations (\ref{asymptotic_steady_solution_sigma})--
(\ref{asymptotic_steady_solution_dv})).
}
\label{profile3e-7}
\end{figure*}

\subsection{Results from the fiducial model}
In this subsection, we describe the results obtained from our chosen 
fiducial model, M1Mdot37r100f1.
The dependence on the 
model parameters is discussed in subsequent subsections.

\subsubsection{Timescale for settling into the steady state}
In figure \ref{profile3e-7}, we show the time evolution of the dust structure.
As shown in the top left panel,
the dust surface density quickly decreases owing to rapid dust growth
and subsequent radial migration.
The dust disk settles into the steady state at $t \sim 0.15$ Myr.
As a result, the structures at $t=0.15$ Myr and 
$t=1.0$ Myr (at the end of the simulation) are identical.
The steady-state settling is 
particularly clear in terms of $\dot{M}_{\rm dust}$ (top-right panel),
which becomes radially constant at $t\sim0.15$ Myr.
The timescale for steady-state settling can be estimated 
from the timescale when the dust grows
to the size at which radial drift begins at the disk edge,
which is estimated as \citep{2012ApJ...752..106O}
\begin{eqnarray}
\label{time_grow}
t_{\rm grow} \equiv \left(\frac{d \ln m_{\rm dust}}{dt} \right)^{-1}=\frac{4\sqrt{2 \pi}}{3}\frac{H_{\rm dust} \rho_{\rm int} a}{\Delta v \Sigma_{\rm dust}} \nonumber \\ 
\sim 3.4\times10^4 
\left( \frac{M_{\rm star}}{\msun} \right)^{-1/2}\left( \frac{r}{100 {\rm AU}} \right)^{3/2} {\rm years},
\end{eqnarray}
where we assume the gas disk structure
of equations (\ref{GI_disk_sigma}) and (\ref{GI_disk_alpha}) 
, $H_{\rm dust}=(\St/\alpha)^{-1/2} H_{\rm gas}$,  $\Delta v=\sqrt{\alpha c_{\rm s}^2 \St}$, and the Epstein drag law.
We also assume that the dust-to-gas mass ratio at the edge of the disk is 1/100.
This timescale corresponds to the mass doubling time and 
our simulation results show that the timescale for dust growth 
to cm-sized dust is slightly longer ($\sim 10^5 $ yr) than this estimate.
The dust growth timescale is shorter 
than the lifetimes of Class 0/I YSOs, which are
typically 0.5 Myr \citep{2009ApJS..181..321E,2011ARA&A..49...67W}.
Therefore, it is expected that the dust structure in the Class 0/I phase
can be described by the steady-state solution for the dust disk.
The analytic solutions for the steady-state structure are
discussed in detail in Appendix A.

\subsubsection{Steady-state structure of dust disk}
Hereafter, we focus on the steady-state structure of the dust disk
(shown as dotted and dashed-dotted lines).
The top-left panel of figure \ref{profile3e-7}
shows the dust surface density. 
The dust surface density at the outer region, $r\gtrsim 80 \AU$,
is identical to its initial value
because the dust particles are small and 
move with the gas advection.
As a result, the initial dust-to-gas mass ratio is maintained
in this region.
Once the dust particles grow and the dust drift velocity overtakes 
the gas advection velocity, inward drifting of the dust particles begins.
In our fiducial model, the radius at which the radial drift begins is 
$r_{\rm drift}\sim 80 \AU$.
For $r<r_{\rm drift}$, the dust-to-gas mass ratio decreases 
from its initial values owing to the faster radial motion of the 
dust particles.
The dust surface density for $20\AU \lesssim r\lesssim 80 \AU$
asymptotically obeys the power law $\Sigma_{\rm dust}\propto r^{-31/28}$,
which is the asymptotic steady-state 
solution (equation (\ref{asymptotic_steady_solution_sigma}), dotted black line).

$r_{\rm drift}$
can be estimated by considering the radius at which the
gas advection velocity is equal to the dust drift velocity
\begin{equation}
v_{\rm r, gas}= v_{\rm r, dust} \sim 2 \St \eta v_{\rm K},
\end{equation}
where we assume $\St \ll1 $.
By assuming the gas disk structure of equations 
(\ref{GI_disk_sigma}) -- (\ref{GI_disk_alpha}), we 
obtain 
\begin{equation}
r_{\rm drift}=2.2 \times 10^2 \left(\frac{\St}{10^{-2}} \right)^{14/9} \left(\frac{\dot M_{\rm gas}}{10^{-7} \msunyear} \right)^{-14/9} \AU.
\end{equation}
If we assume $\St=0.01-0.02$, as is suggested by the bottom panel,
and $M_{\rm gas}=3 \times 10^{-7} \msunyear$, then $r_{\rm drift}=40-120 \AU$.
This estimate crudely explains our numerical results.

As the dust migrates inwards, $a_{\rm dust}$ exceeds 
the mean free path and the drag law changes to the Stokes' law
at a certain radius $r_{\rm Stokes}$.
The change in the drag law is identified by the 
change in the profile at $r\sim 10$ AU.
In the region of the Stoke regime, 
$\Sigma_{\rm dust}$ is an increasing function of $r$.
Through the analytic discussion in Appendix A, we can show that
the dust surface density asymptotically obeys the power law 
$\Sigma_{\rm dust}\propto r^{19/42}$ when $\Delta v=\sqrt{\alpha c_{\rm s}^2 \St}$
and $H_{\rm dust}=(\St/\alpha)^{-1/2} H_{\rm gas}$, and
we can confirm the positive power law of $\Sigma_{\rm dust}$.
However, because of the narrow Stokes drag region in the disk, 
the structure does not converge into the asymptotic solution. Rather, it is 
steeper than the solution.
$r_{\rm Stokes}$ can be estimated from the condition 
\begin{equation}
a_{\rm dust}=\frac{9}{4}\lambda_{\rm mfp}.
\end{equation}
From the gas 
disk structure of equations (\ref{GI_disk_sigma}) and (\ref{GI_disk_alpha}), 
the radius is given as,
\begin{equation}
r_{\rm Stokes}=9.6 \left(\frac{M_{\rm star}}{\msun} \right)^{7/22}
\left(\frac{\St}{0.01} \right)^{7/33} \left(\frac{\rho_{\rm int}}{0.1 \gcm} \right)^{-7/33} \AU.
\end{equation}
This agrees well with our results.

The middle-left panel of figure \ref{profile3e-7}
shows the collision velocity of the dust particles.
In the steady state 
$\Delta v\lesssim 15 \ms$ at $r>10 \AU$, 
and asymptotically obeys the power law $\Delta v \propto r^{1/8}$ 
in $10\AU \lesssim r\lesssim 80 \AU$
(equation (\ref{asymptotic_steady_solution_dv}), dotted black line).
The collision velocity is significantly smaller than the 
threshold velocity $v_{\rm th}$ for collisional fragmentation.
Simulations of aggregate collisions by \citet[][]{2013A&A...559A..62W} 
showed that $v_{\rm th} \sim 80 \ms$ for aggregates composed of $0.1~{\rm \mu m}$ 
sized icy monomer grains, and we chose $v_{\rm th}=80 \ms$ (black solid line).
Furthermore, as we noted in section \S\ref{visc_alpha_discussion},
the collision velocity in our simulations is slightly overestimated.
Thus, we conclude that collisional fragmentation does not play 
a major role in our model.

As shown in the middle-right panel of figure \ref{profile3e-7},
$a_{\rm dust}$ quickly increases and  becomes greater than 1 cm 
in $r\lesssim 70 \AU$,
asymptotically obeying the power law $a_{\rm dust} \propto r^{-47/28}$.
Figure \ref{profile3e-7} shows that the dust particles migrate in the 
form of centimeter-scale particles 
or "pebbles" in the disk during the Class 0/I phase.
Because the amount of dust that passes through the disk is quite large
in the Class 0/I phase (approximately 1 \% of the central-star mass), 
the pebble accretion scenario for planet formation 
\citep{2010A&A...520A..43O,2012A&A...544A..32L,2014A&A...572A.107L} seems to be
preferred in the Class 0/I phase compared to the Class II phase, 
in which dust depletion at the disk outer edge limits 
the embryo growth by pebble accretion \citep{2016A&A...591A..72I}.

\subsubsection{Reduction of the dust mass and thermal emission} 
As implied from the top-left panel of figure \ref{profile3e-7},
the total dust mass in the steady disk
$M_{\rm dust}$  systematically decreases from 
the disk with the ISM dust-to-gas mass ratio $f_{\rm dg, ISM}=1/100$.
We define the reduction rate of dust mass in the disk
owing to the dust depletion $\mu_{\rm M}$ as
\begin{eqnarray}
\label{mu_M}
\mu_{\rm M}\equiv \frac{M_{\rm dust, steady}}{f_{\rm dg, ISM} M_{\rm gas}}.
\end{eqnarray}
Here, $M_{\rm dust, steady}\equiv \int_{r_{\rm min}}^{r_{\rm max}} \Sigma_{\rm dust, steady}(r) 2 \pi r dr$ and $M_{\rm gas} \equiv \int_{r_{\rm min}}^{r_{\rm max}} \Sigma_{\rm gas}(r) 2 \pi r dr$
where $\Sigma_{\rm dust, steady}$ is the steady-state dust surface density.
$\mu_{\rm M}$ is the ratio of the dust mass 
in the steady-state disk to that in the disk with $f_{\rm dg, ISM}$.
(or the initial dust disk in our simulation).
By numerically integrating equation (\ref{mu_M})
taking the inner and outer cut-off radius as 
$r_{\rm min}=10$ AU and $r_{\rm max}=100$ AU, 
we obtain 
\begin{eqnarray}
\mu_{\rm M}= 0.34,
\end{eqnarray}
for our fiducial model.
Therefore, the dust-to-gas mass ratio of the disk for our fiducial
model becomes approximately
\begin{eqnarray}
f_{\rm dg, steady} \equiv \mu_{\rm M} f_{\rm dg, ISM} \sim 1/300,
\end{eqnarray}
and is smaller than the value for the ISM.

Furthermore, as dust growth reduces its opacity,
the apparent mass of the dust disk that is estimated from 
the dust thermal emission assuming
the opacity of micron-sized dust becomes smaller than $M_{\rm dust}$.
In this paper, we consider the dust thermal emission at $\lambda=1.3 {\rm mm}$
which corresponds to ALMA Band 6.
The radiative flux from the dust disk in the steady 
state can be calculated as
\begin{eqnarray}
\label{Fdust}
F_{\rm steady, 1.3mm}= 
\frac{4 \pi}{D^2}\int_{r_{\rm min}}^{r_{\rm max}} 
\{ [1-\exp(\tau_{\rm steady, 1.3mm}(r))] \nonumber \\
 B_{\rm 1.3mm}(T(r))\} 2 \pi r dr.
\end{eqnarray}
where 
\begin{eqnarray}
\label{tau_steady}
\tau_{\rm steady, 1.3mm}(r)= \kappa_{\rm 1.3mm}(a_{\rm dust, steady}(r))\Sigma_{\rm dust, steady}(r),
\end{eqnarray}
and $\kappa_{\rm 1.3mm},~a_{\rm dust, steady},~B_{\rm 1.3mm}$ and $D$ are
the dust opacity at $\lambda= 1.3 {\rm mm}$,
the dust radius in the steady state,
the Plank function at $\lambda= 1.3 {\rm mm}$, and
the distance of the source, respectively.
To calculate the dust opacity, we set $a_{\rm min}=a_{\rm init}$,
$a_{\rm max}=a_{\rm dust}$.
Because the disk mass is often estimated from the dust thermal emission by
assuming micron-sized dust opacity, ISM dust-to-gas mass
ratio, and that dust disk is optically thin,
we can define the "effective" reduction rate of the dust mass
owing to dust depletion and opacity reduction  $\mu_{\rm F}$ 
as satisfying
\begin{eqnarray}
\label{Ffixed}
\mu_{\rm F}=\frac{F_{\rm steady, 1.3mm}}{F_{\rm ISM, 1.3mm}}.
\end{eqnarray}
where $F_{\rm ISM, 1.3mm}$ is the radiative flux defined as
\begin{eqnarray}
\label{Ffixed}
F_{\rm ISM,1.3 mm} \equiv \frac{4 \pi}{D^2}\int_{r_{\rm min}}^{r_{\rm max}} \left\{\kappa_{\rm 1.3mm}(a_{\rm ISM}) ( f_{\rm dg, ISM} \Sigma_{\rm gas}(r))  \right. \nonumber \\
\left.B_{\rm 1.3mm}(T(r))\right\} 2 \pi r dr ,
\end{eqnarray}
where we assume the typical dust size of the ISM $a_{\rm ISM}$ to be
$a_{\rm ISM}=0.1 {\rm \mu m}$.

In the steady state of our fiducial model, 
\begin{equation}
\mu_{\rm F}=0.17.
\end{equation}
Thus, the "effective" dust-to-gas mass ratio for our fiducial
model is calculated as 
\begin{equation}
f_{\rm dg, eff} \equiv \mu_{\rm F} f_{\rm dg, ISM} \sim 1/500.
\end{equation}
The reduction of dust thermal emission causes underestimation
of $M_{\rm gas}$ because the gas mass is often estimated by
assuming $f_{\rm dg, ISM}$ and the opacity of micron-sized dust.
The apparent gas disk mass $M_{\rm app}$ is calculated as 
\begin{equation}
M_{\rm app}\equiv \mu_{\rm F} M_{\rm gas}.
\end{equation}
In our fiducial model, the gas disk mass is
$M_{\rm gas}=\int \Sigma_{\rm gas} 2 \pi r dr=0.16 \msun$ where
the inner and outer cut-off radii are chosen to be
$r_{\rm in}=10 \AU$ and $r_{\rm out}=r_{\rm disk}=100 \AU$, respectively.
The apparent gas disk mass
estimated from the dust thermal emission 
is $M_{\rm app} =0.027 \msun$ and 
is apparently gravitationally stable.
Thus, even when a gravitationally unstable disk exists 
in a Class I YSO, it appears to be gravitationally stable.
Note that $M_{\rm app}$ is 
consistent with the observed disk mass  $M_{\rm gas,obs}$ of Class I YSOs
in the range $0.01 \msun< M_{\rm gas, obs}<0.1 \msun$ 
\citep{2007ApJ...671.1800A,2009A&A...507..861J}.

\subsection{Parameter study}
In this subsection, we investigate how
the steady-state structure and apparent disk mass
depend on the model parameters.
The parameters we consider in this subsection
are the mass accretion rate onto the disk $\dot{M}_{\rm gas}$, 
the filling factor of the dust aggregate $f$, 
the radius of the disk $r_{\rm disk}$, and
the central-star mass $M_{\rm star}$.
In  the models considered in this section, the steady
state is reached within $0.2 $ Myr and it is expected 
that the dust disk  is in its steady state 
in the Class I phase.
Therefore, we focus on the dependence of the steady-state structure
on the parameters.

\subsubsection{Dependence on mass accretion rate}

\begin{figure*}
\includegraphics[width=60mm,angle=-90]{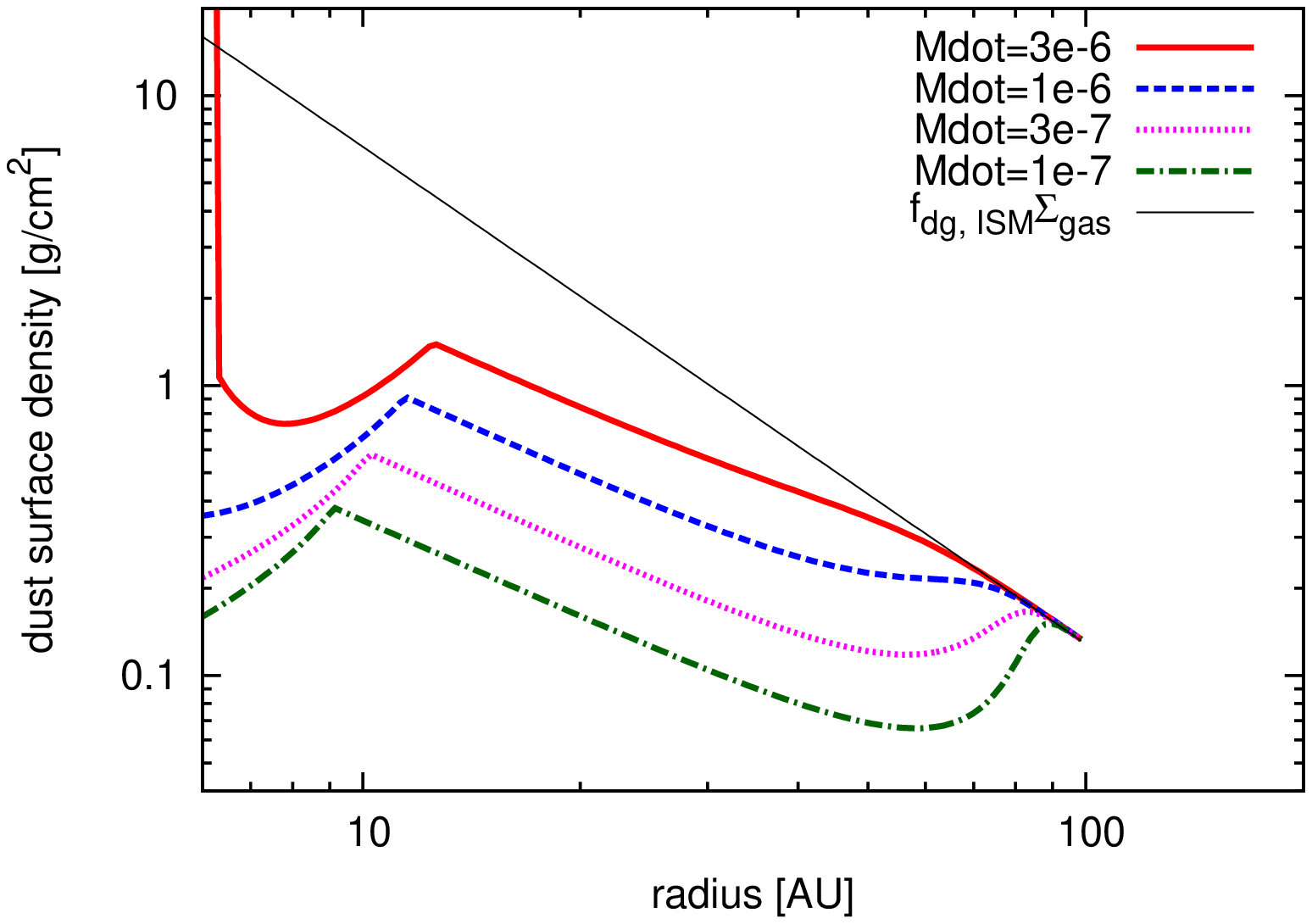}
\includegraphics[width=60mm,angle=-90]{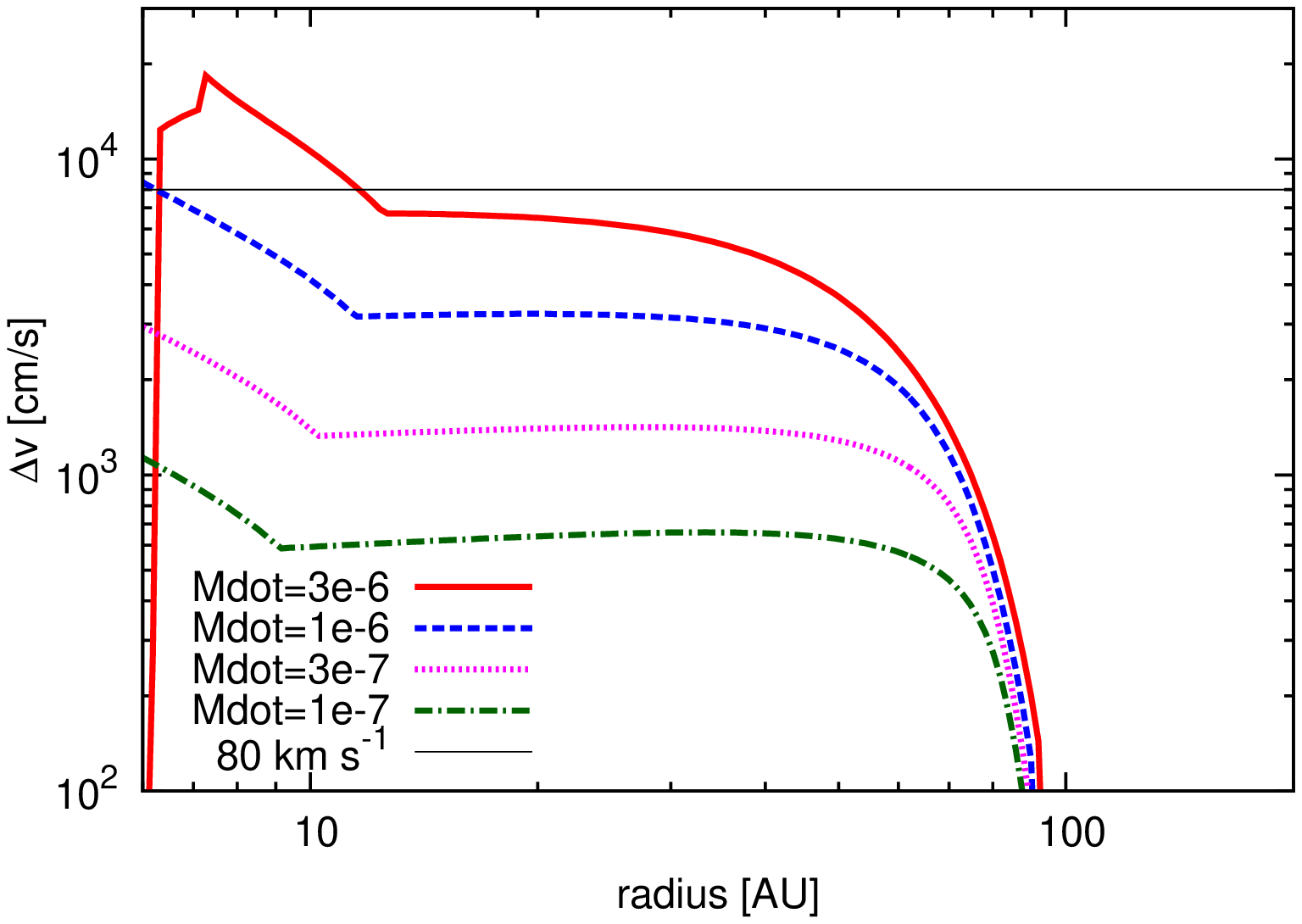}
\includegraphics[width=60mm,angle=-90]{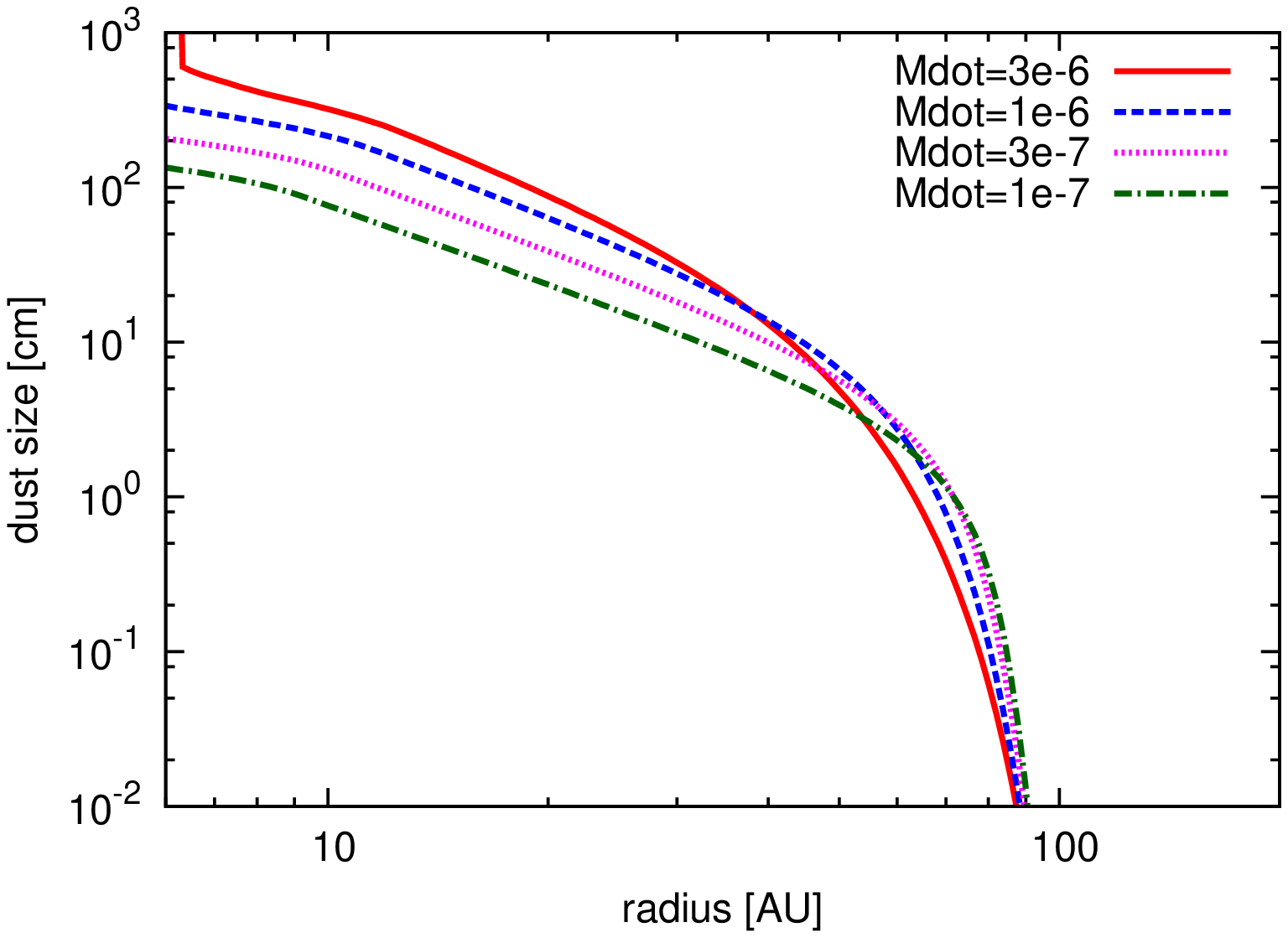}
\includegraphics[width=60mm,angle=-90]{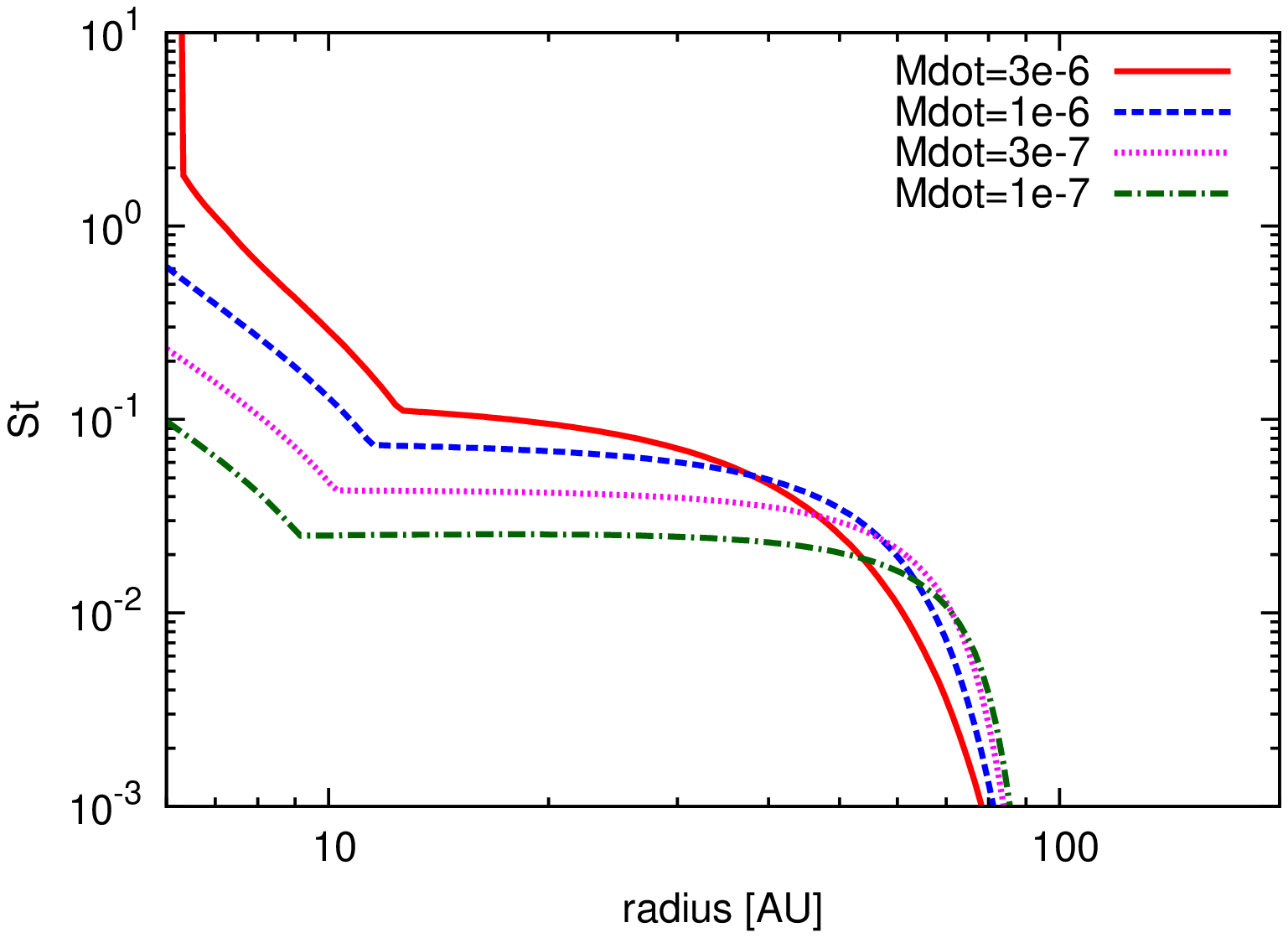}
\caption{
Radial profiles of dust surface density (top left), 
collision velocity (top right)
dust size (bottom left), 
and Stokes number (bottom right)
in the steady state (at $t= 0.2$ Myr) 
for various mass accretion rates.
The red solid, blue dashed, magenta dotted, and green dashed-dotted lines
show the profile 
of M1Mdot36r100f1, M1Mdot16r100f1, M1Mdot37r100f1, and M1Mdot17r100f1, respectively.
The black solid lines in the surface density and in the collision velocity plots
show $f_{\rm dg, ISM} \Sigma_{\rm gas}$ and the threshold velocity, respectively.
}
\label{profile_mdot}
\end{figure*}

In the top-left panel of figure \ref{profile_mdot}, 
we show the surface density profiles of the 
the steady-state dust disk for various mass accretion rate
($\dot{M}_{\rm gas}=1\times10^{-7}, 3\times10^{-7},~1\times10^{-6},~3\times10^{-6}~\msunyear$).
The surface density of the dust for $r_{\rm Stokes}<r<r_{\rm drift}$ 
is an increasing function of $\dot{M}_{\rm gas}$.
It depends on the mass accretion rate as 
$\Sigma_{\rm dust}\propto \dot{M}_{\rm gas}^{1/2}$ 
(see equation (\ref{asymptotic_steady_solution_sigma})).
Thus, as the mass accretion decreases, $M_{\rm app}$ decreases
although the actual gas mass $M_{\rm gas}$ is independent of $\dot{M}_{\rm gas}$
(equation (\ref{GI_disk_sigma})).

The top-right panel of figure \ref{profile_mdot} shows the collision velocity.
Even with the relatively large mass accretion rate,
($\dot{M}_{\rm gas}=3 \times 10^{-6} \msunyear$), 
the collision velocity is smaller than the threshold 
velocity ($v_{\rm th}=80 \ms$) for $r\gtrsim 10 \AU$,
and our assumption of perfect sticking is still justified.
However, if we consider a slightly larger mass accretion
rate, e.g., $\dot{M}_{\rm gas}=10^{-5} \msunyear$, which may occur
in some Class 0/I YSOs, collisional fragmentation
plays a dominant role in determining the dust structure.
Note also that the threshold velocity adopted in this paper is derived
with $0.1~{\rm \mu m}$  sized monomer 
and it may decrease if the monomer size is large. 
If this is the case, the collisional fragmentation
becomes dominant with smaller mass accretion rate.
The collision velocity exhibits the following dependence on the mass accretion
rate: $\Delta v \propto (\alpha \St)^{1/2} \propto \dot{M}_{\rm gas}^{3/4} $
(see, (\ref{GI_disk_alpha}) and (\ref{asymptotic_steady_solution_St})).

The bottom-left panel of figure \ref{profile_mdot} shows the radial 
profile of dust size.
In the outer part of the disk ($r \gtrsim80 \AU \sim r_{\rm drift}$), 
the dust size increases as the accretion rate decreases
owing to the small gas advection velocity
in the small mass accretion models. 
Because of the small advection velocity, the dust particles
can remain in the outer region for a long period of time
and have sufficient time to grow to a larger size.
On the other hand, in the inner part of the disk
($r_{\rm Stokes}< r < r_{r_{\rm drift}}$),
the dust size increases as $a_{\rm dust} \propto \dot{M}_{\rm gas}^{1/2}$
(equation (\ref{asymptotic_steady_solution_size})).

\begin{figure*} 
\includegraphics[width=60mm,angle=-90]{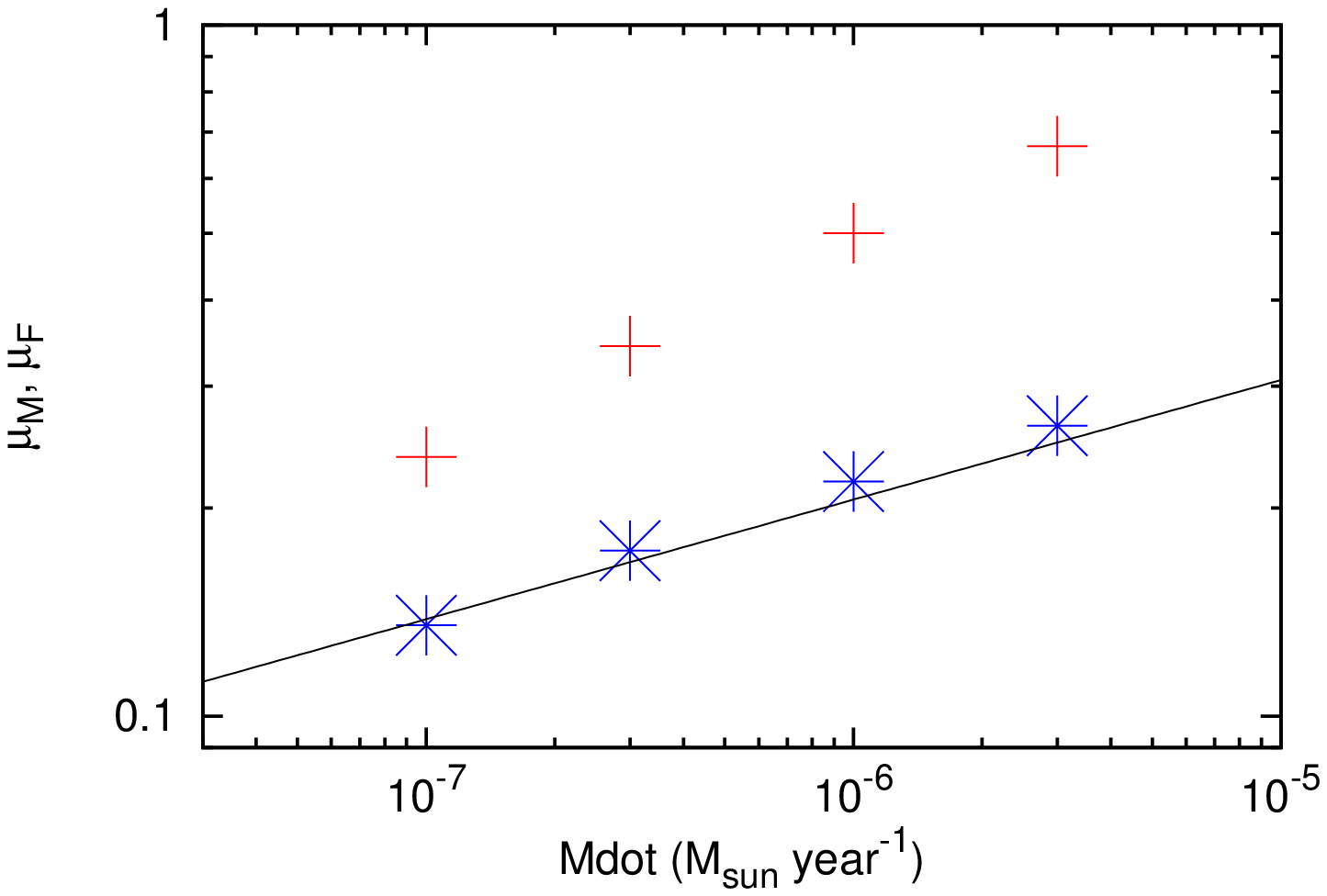}
\includegraphics[width=60mm,angle=-90]{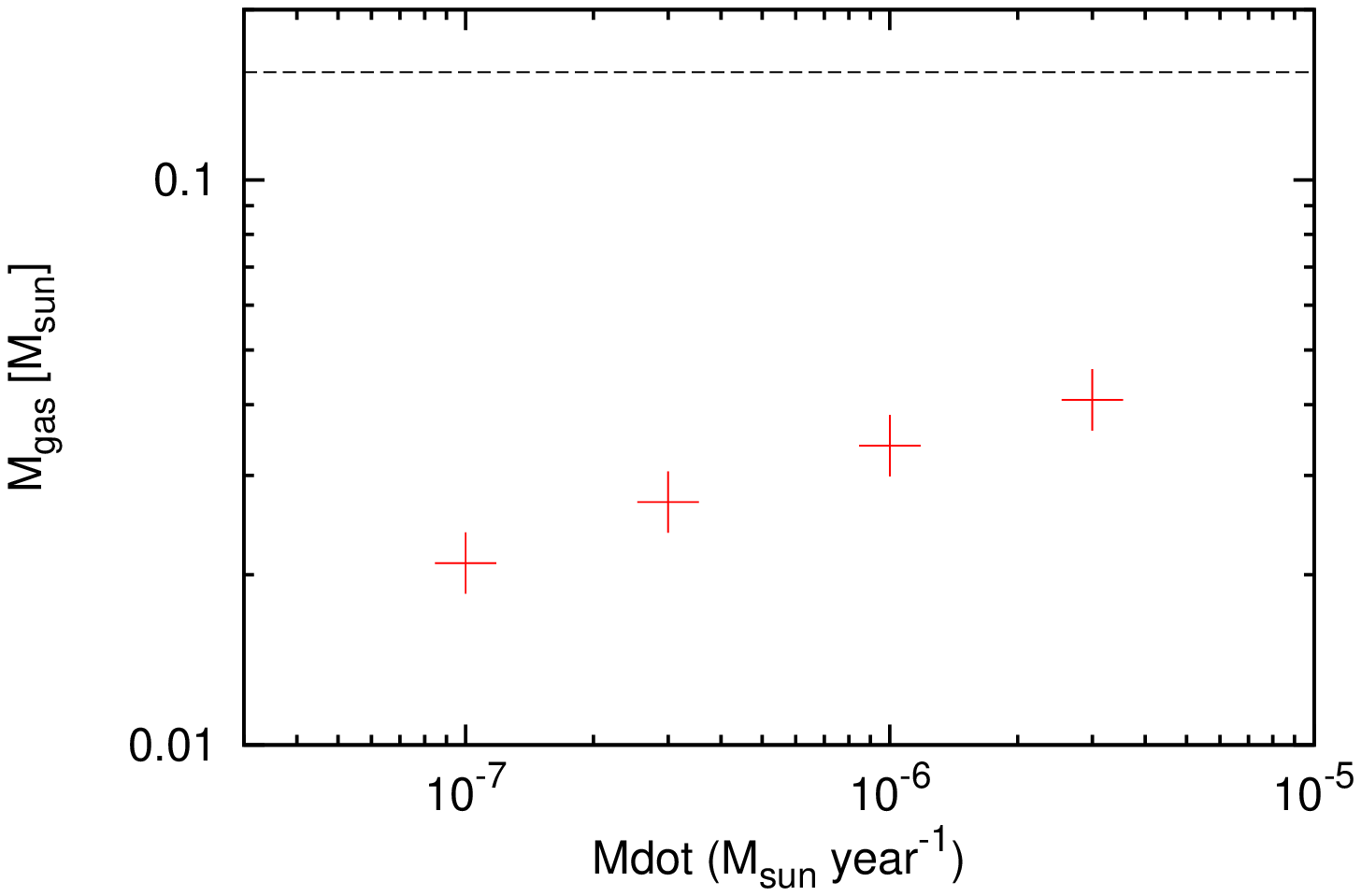}
\caption{
Left panel 
shows the ratio of the dust mass in the steady-state disk (at $t= 0.2$ Myr) to
that in the disk with ISM dust-to-gas mass ratio  $\mu_{\rm M} $,
and ratio of the radiative flux from the steady-state disk to
that from the disk with ISM dust-to-gas mass ratio and micron-sized dust
$\mu_{\rm F}$ for various mass accretion rates.
The red crosses and blue asterisks show $\mu_{\rm M}$ and $\mu_{\rm F}$, respectively,
of M1Mdot36r100f1, M1Mdot16r100f1, M1Mdot37r100f1, and M1Mdot17r100f1.
The black solid line shows the empirical formula for $\mu_{\rm F}$,
equation (\ref{empirical_form_muf}).
The right panel shows the apparent mass $M_{\rm app}$ of the gas disk.
The red crosses show  $M_{\rm app}$.
Here, $M_{\rm app}=\mu_{\rm F} M_{\rm gas}$,
where $M_{\rm gas}=\int \Sigma_{\rm gas} 2 \pi r dr$.
The black dashed line shows the actual mass of the gas disk
in the simulations, $M_{\rm gas}=0.16 \msun$.
}
\label{disk_mass_mdot}
\end{figure*}

In the left panel of figure \ref{disk_mass_mdot}, 
we show $\mu_{\rm M} $ and $\mu_{\rm F}$
for various mass accretion rates.
As pointed out above, $\mu_{\rm F}$ 
indicates the effective reduction rate of the gas mass.
Both $\mu_{\rm M} $ and $\mu_{\rm F}$
are increasing functions of the mass accretion rate. 
The dashed line shows our empirical formula for  $\mu_{\rm F}$, 
equation (\ref{empirical_form_muf}), which indicates 
that $\mu_{F}\propto \dot{M}_{\rm gas}^{0.17}$.

The right panel of figure \ref{disk_mass_mdot} shows the apparent mass
of the gas disk as a function of the mass accretion rate
calculated according to
$M_{\rm app}=\mu_{\rm F} M_{\rm gas}$,
The black line shows the actual gas mass in the disk, 
$M_{\rm gas}=0.16 \msun$.
Because the mass of a gravitationally unstable 
disk does not depend on the
mass accretion rate, the actual gas disk mass is constant. 
In all cases shown in the figure, the apparent mass is within the 
mass range suggested by observations of Class I YSOs,
$0.01 \msun\lesssim M_{\rm gas,obs} \lesssim 0.1 \msun$.

\subsubsection{Dependence on central-star mass}
\begin{figure*}
\includegraphics[width=60mm,angle=-90]{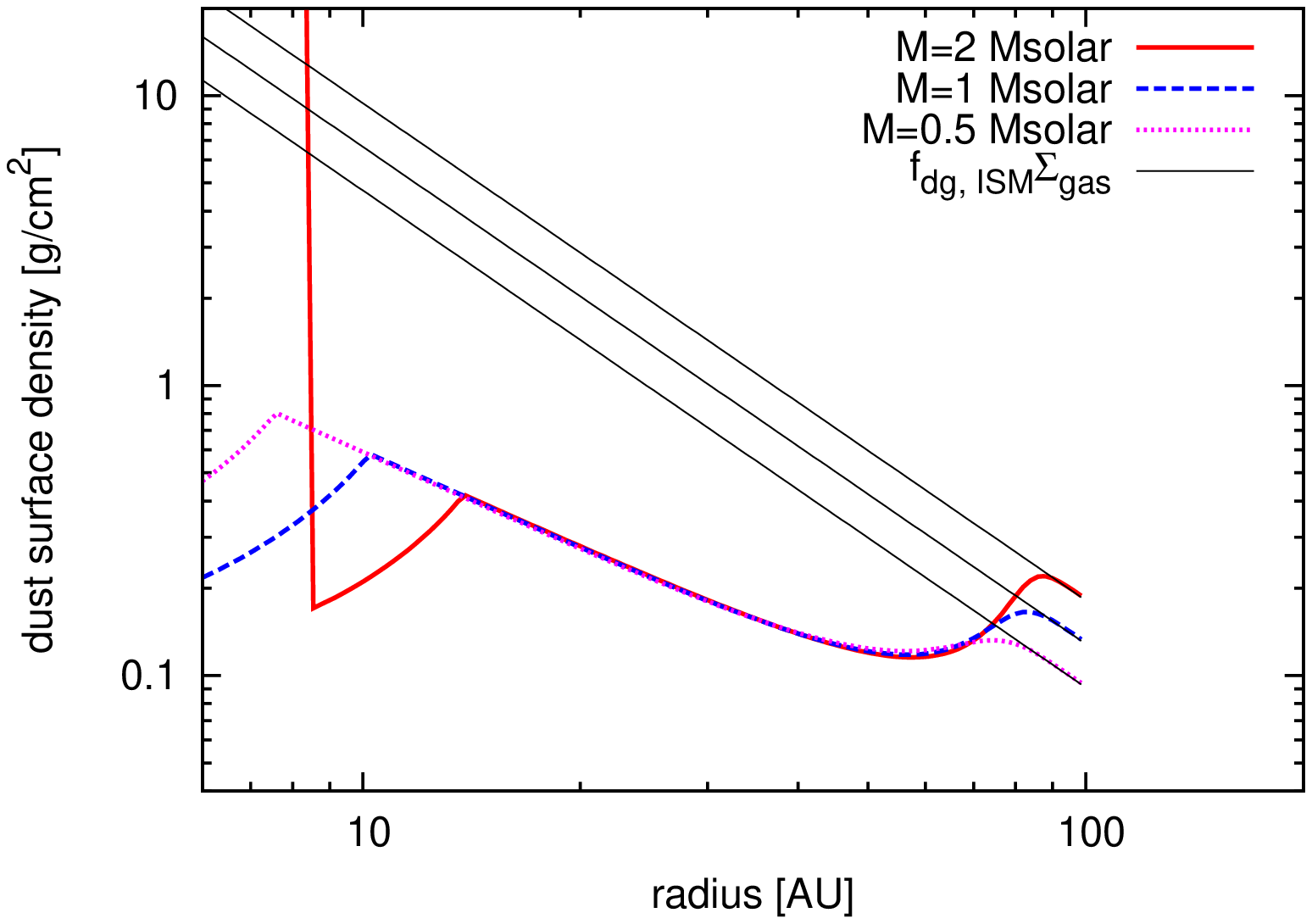}
\includegraphics[width=60mm,angle=-90]{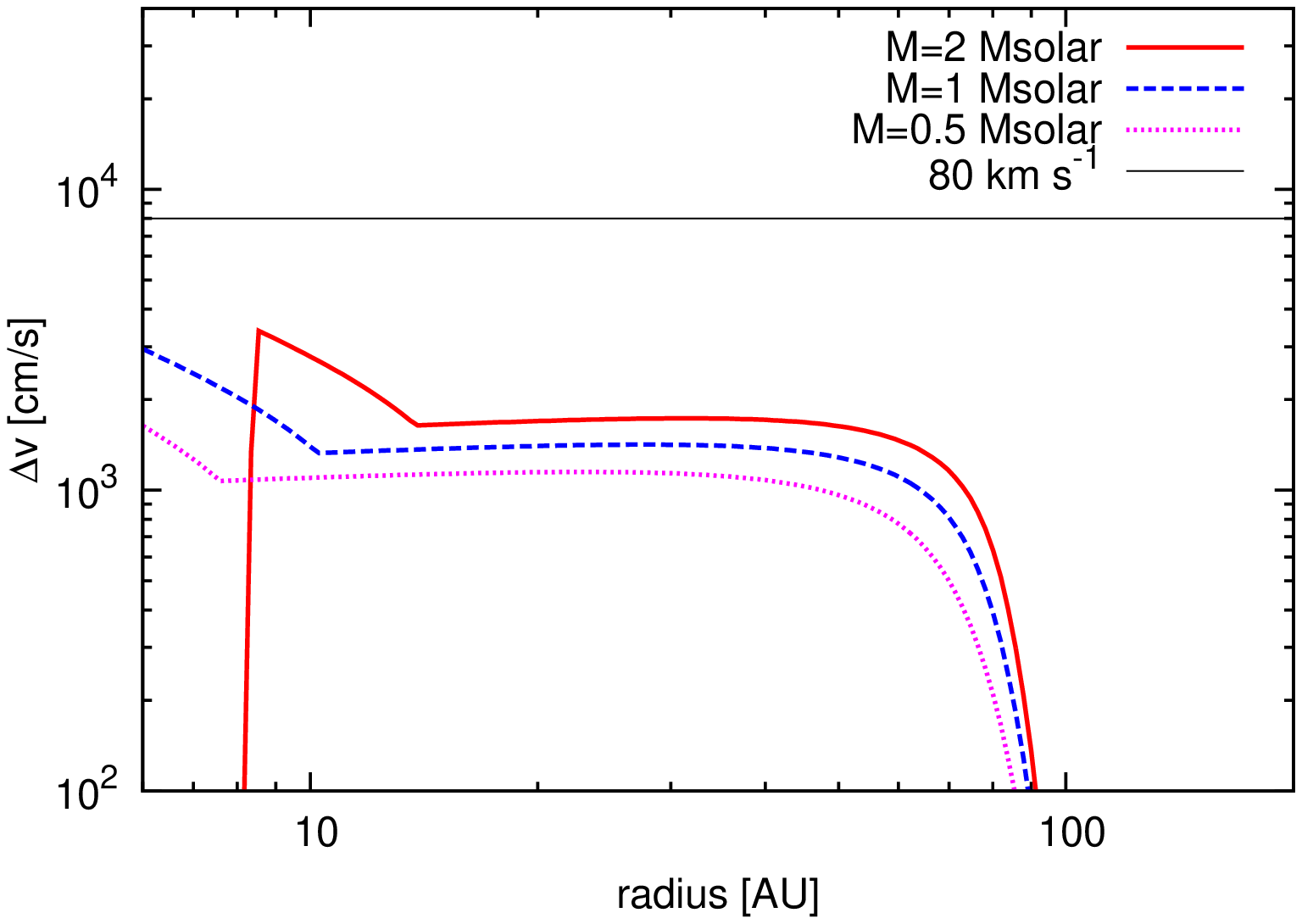}
\includegraphics[width=60mm,angle=-90]{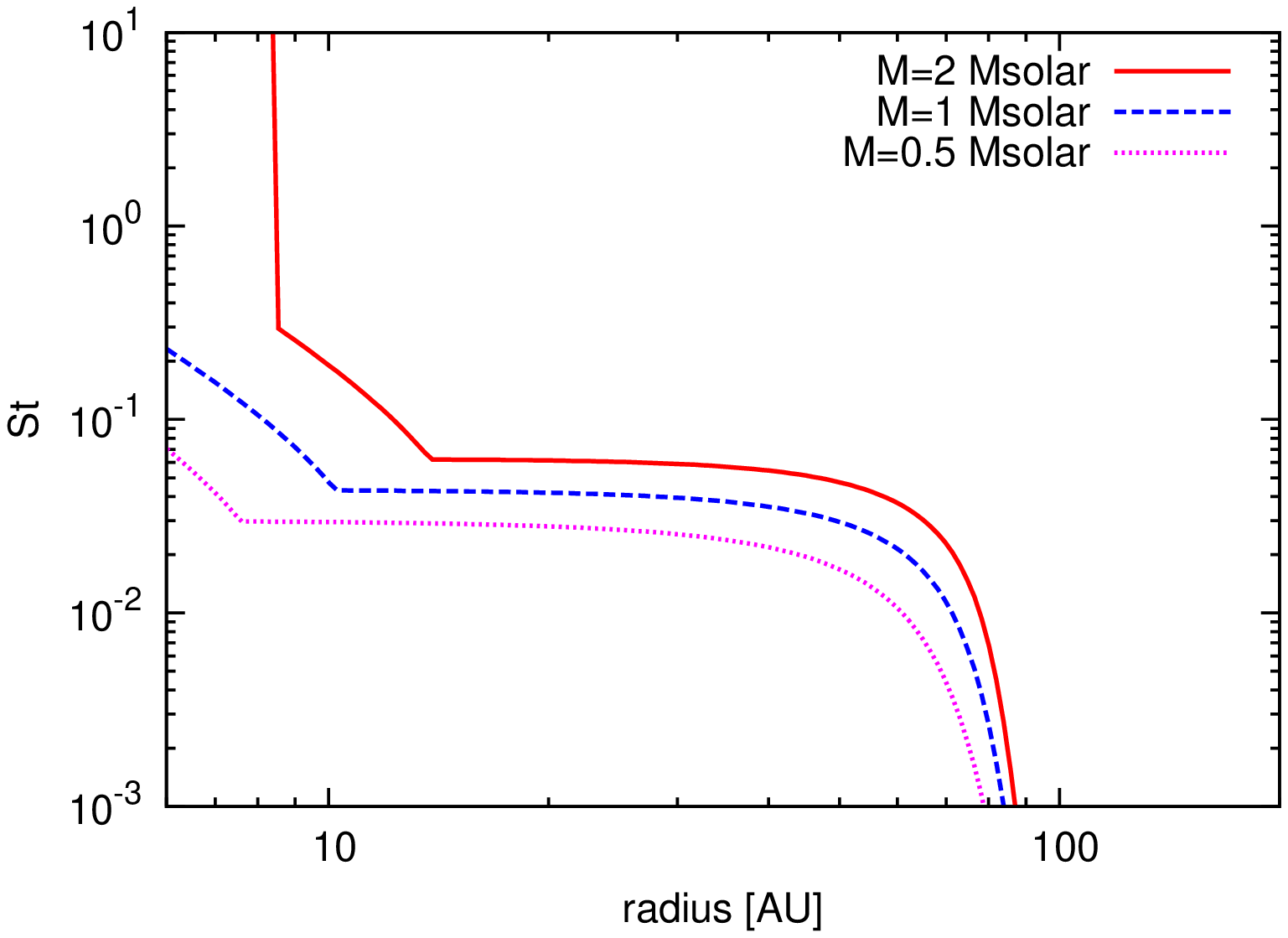}
\caption{
Radial profiles of dust surface density (top left), 
collision velocity (top right),
and Stokes number (bottom)
in the steady state (at $t= 0.2$ Myr) for various central-star masses.
The red solid, blue dashed, and magenta dotted lines
show the profile 
of M2Mdot37r100f1, M1Mdot37r100f1, and M05Mdot37r100f1, respectively.
The black solid lines in the surface density and 
in the collision velocity plots show $f_{\rm dg, ISM} \Sigma_{\rm gas}$ and
the threshold velocity, respectively.
}
\label{profile_mass}
\end{figure*}

In the top-left panel of figure \ref{profile_mass}, 
we show the surface density profile of dust for
various central-star masses ($M_*=0.5,~1.0,~2.0 \msun$).
Interestingly, once radial drift begins,
$\Sigma_{\rm dust}$ converges to the same steady-state 
solution for $r_{\rm Stokes} < r < r_{\rm drift}$,
independent of the central-star mass,
which can be understood as follows.
Because we consider a gravitationally unstable disk,
$\Sigma_{\rm gas}\propto \Omega\propto M_{\rm star}^{1/2}$.
From equation (\ref{asymptotic_steady_solution_St}),
by assuming $\Delta v=\sqrt{\alpha c_{\rm s}^2\St}$,
 $H_{\rm dust}=(\St/\alpha)^{-1/2} H_{\rm gas}$, and 
Epstein drag, 
we can find $t_{\rm coll}/t_{\rm drift}\propto \Omega  \Sigma_{\rm gas} \St^{-2}\propto
M_{\rm star}^{0}$ and hence,
$\St \propto (\Omega \Sigma_{\rm gas})^{1/2}\propto M_{\rm star}^{1/2}$.
Then, from (\ref{sigma_d1}), 
$\Sigma_{\rm dust} \propto (\eta v_{\rm K} \St) \propto 
(\St/v_{\rm K})^{-1}\propto M_{\rm star}^0$.
Thus, the dust surface density in the Epstein regime
is independent of the central-star mass.
Because we use the fact that $\Sigma_{\rm gas}\propto \Omega$,
this is a unique feature of dust disks 
in gravitationally unstable gas disks.

$r_{\rm Stokes}$ increases with $ M_{\rm star}$
because the mean free path depends on $ M_{\rm star}$ as
$\lambda_{\rm mfp}\propto (\Sigma_{\rm gas} \Omega)^{-1} \propto M_{\rm star}^{-1}$.
In addition,
$a_{\rm dust}$ depends on $ M_{\rm star}$, as 
$a_{\rm dust}\propto \St \Sigma_{\rm gas} \propto M_{\rm star}$ at a certain $r$.
As a result, the condition 
$a_{\rm dust}=(9/4) \lambda_{\rm mfp}$ is realized at a larger $r$,
and $r_\St$ increases with increasing $M_{\rm star}$.

As shown in the top right panel of figure \ref{profile_mass}, 
the collision velocity is also an increasing 
function of central-star mass,
it scales as $\Delta v\propto \St^{1/2} \propto M_{\rm star}^{1/4}$ 
(equation (\ref{asymptotic_steady_solution_dv})).
All models considered in this section satisfy the
condition of $\Delta v<v_{\rm th}$,
and the collisional fragmentation does not 
change our results significantly.

\begin{figure*} 
\includegraphics[width=60mm,angle=-90]{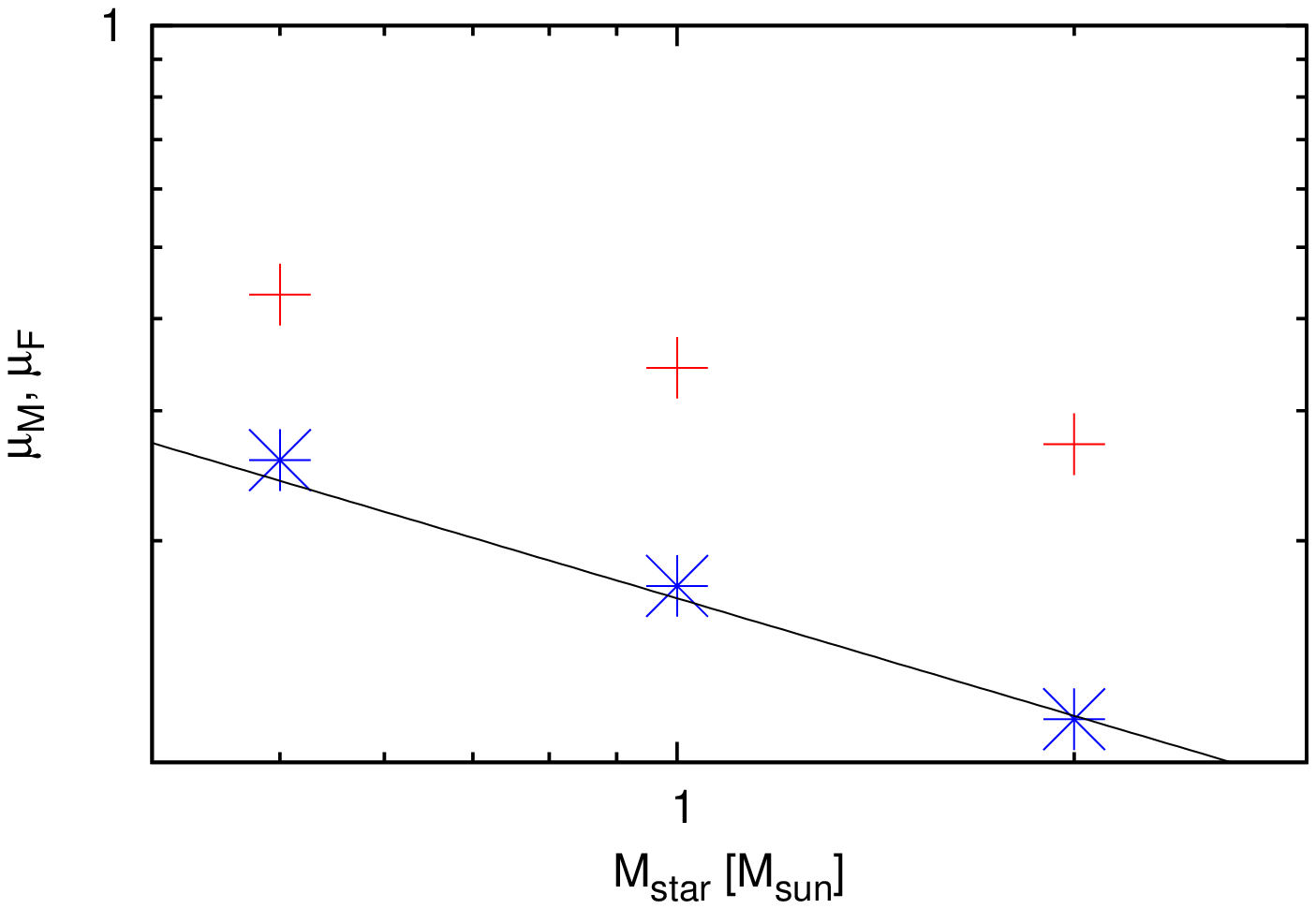}
\includegraphics[width=60mm,angle=-90]{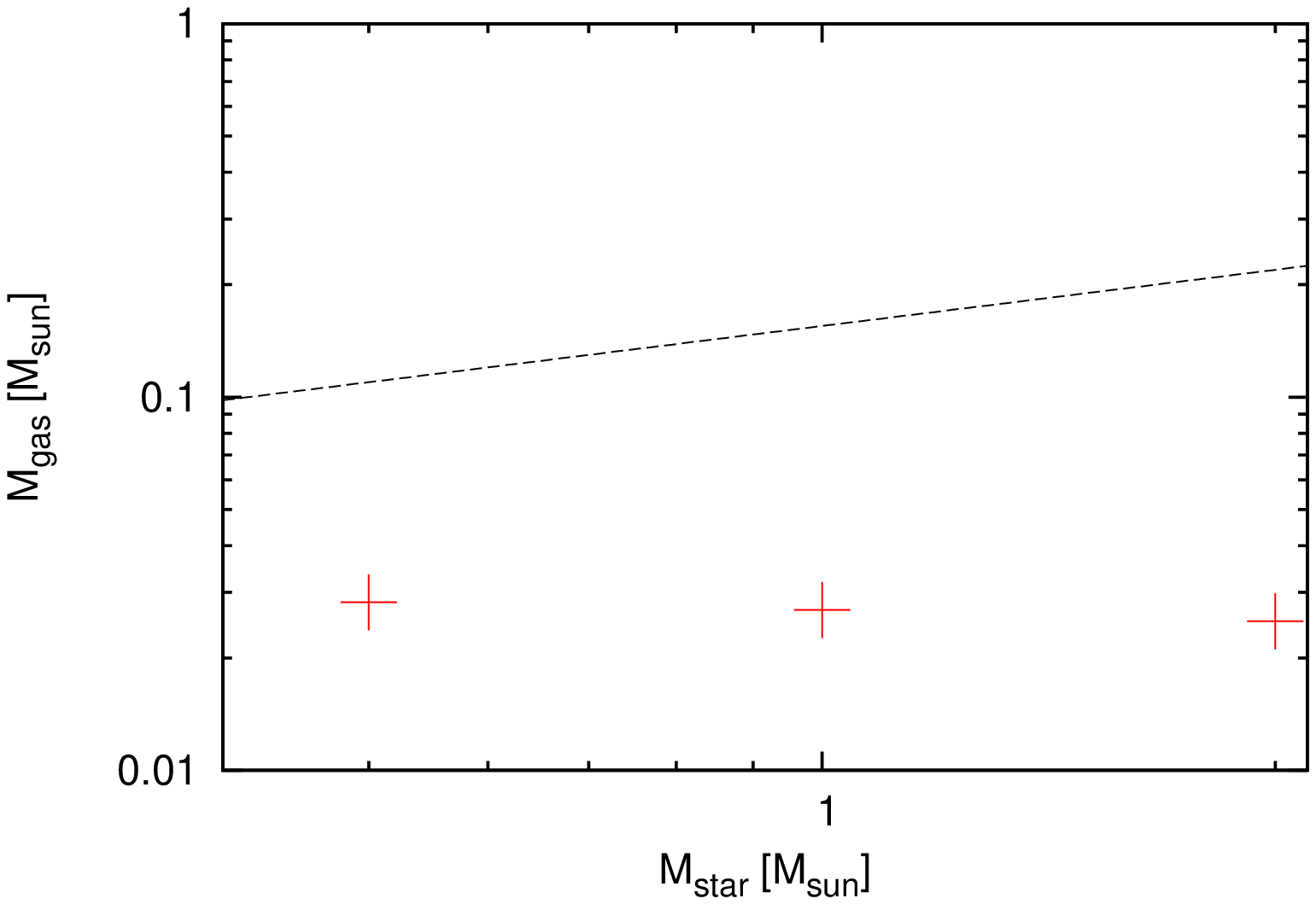}
\caption{
Left panel 
shows the ratio of the dust mass in the steady-state disk (at $t= 0.2$ Myr) to
that in the disk with ISM dust-to-gas mass ratio  $\mu_{\rm M} $,
and ratio of the radiative flux from the steady-state disk to
that from the disk with ISM dust-to-gas mass ratio and micron-sized dust
$\mu_{\rm F}$ for the various central-star masses.
The red crosses and blue asterisks show $\mu_{\rm M}$ and $\mu_{\rm F}$, respectively,
of M2Mdot37r100f1, M1Mdot37r100f1, and M05Mdot37r100f1. 
The black solid line shows the empirical formula for $\mu_{\rm F}$,
equation (\ref{empirical_form_muf}).
The right panel shows the apparent mass $M_{\rm app}$ of the gas disk.
The red crosses show  $M_{\rm app}$, which is 
calculated from $M_{\rm app}=\mu_{\rm F} M_{\rm gas}$,
where $M_{\rm gas}=\int \Sigma_{\rm gas} 2 \pi r dr$.
The black dashed line shows the actual mass of the gas disk
in the simulations having $M_{\rm star}^{1/2}$ dependency 
on the central-star mass.
}
\label{disk_mass_mass}
\end{figure*}

In the left panel of figure \ref{disk_mass_mass}, 
we show  $\mu_{\rm M} $ and $\mu_{\rm F}$ 
as functions of the central-star mass.
Both $\mu_{\rm M} $ and $\mu_{\rm F}$ are decreasing 
functions of the central-star mass.
This dependence exists because,
although $\Sigma_{\rm dust}$ converges to the same 
steady-state solution in the Epstein regime and 
the dust disk has roughly the same mass,
the mass of the gas disk 
is an increasing function of the central-star mass 
as $M_{\rm gas}\propto \Omega \propto M_{\rm star}^{1/2}$.
As a result, $\mu_{\rm M} $ and $\mu_{\rm F}$  decrease as 
the mass of the gas disk increases.
Our empirical formula, equation (\ref{empirical_form_muf}),
shows that $\mu_{F}\propto M_{\rm star}^{-0.53}$, 
having a relatively strong dependence on the central-star mass.

The right panel of figure \ref{disk_mass_mass} 
shows the apparent mass, $\mu_{\rm F}M_{\rm gas}$,
and the actual mass of the gas disk, $M_{\rm gas}$,
as functions of the central-star mass.
The apparent mass has very weak dependence on
the central-star mass.
Although this result appears to contradict the results 
shown in the left panel of figure \ref{disk_mass_mass},
there is no contradiction. As shown by the
dashed line, the mass of a gas disk with constant
$Q$ value is an increasing function of $M_{\rm star}$
as $M_{\rm gas}\propto M_{\rm star}^{1/2}$.
This positive dependence almost cancels the negative dependence
of $\mu_{\rm F}$.
The range of the apparent mass is again within the observed
mass range from the dust thermal emission.
The lack of correlation between the apparent mass and the
central star mass stems primarily from the fact 
that the dust disk converges to the same steady-state solution 
in $r_{\rm Stokes} \lesssim  r \lesssim r_{\rm drift}$ and has roughly 
the same total dust mass.

\subsubsection{Dependence on the dust porosity}
As shown in \citet{2012ApJ...752..106O}, 
dust aggregates may grow to
highly porous aggregates with $\rho_{\rm int} \ll 1 \gcm$.
Therefore, it is expected that the filling factor 
of dust aggregates in the protoplanetary disk 
has a small value ($f \ll 1$).
In this section, we investigate how 
the dust profiles depend on the porosity.
Note that we do not consider
porosity evolution  directly, rather, we treat it as a parameter.

\begin{figure*}
\includegraphics[width=60mm,angle=-90]{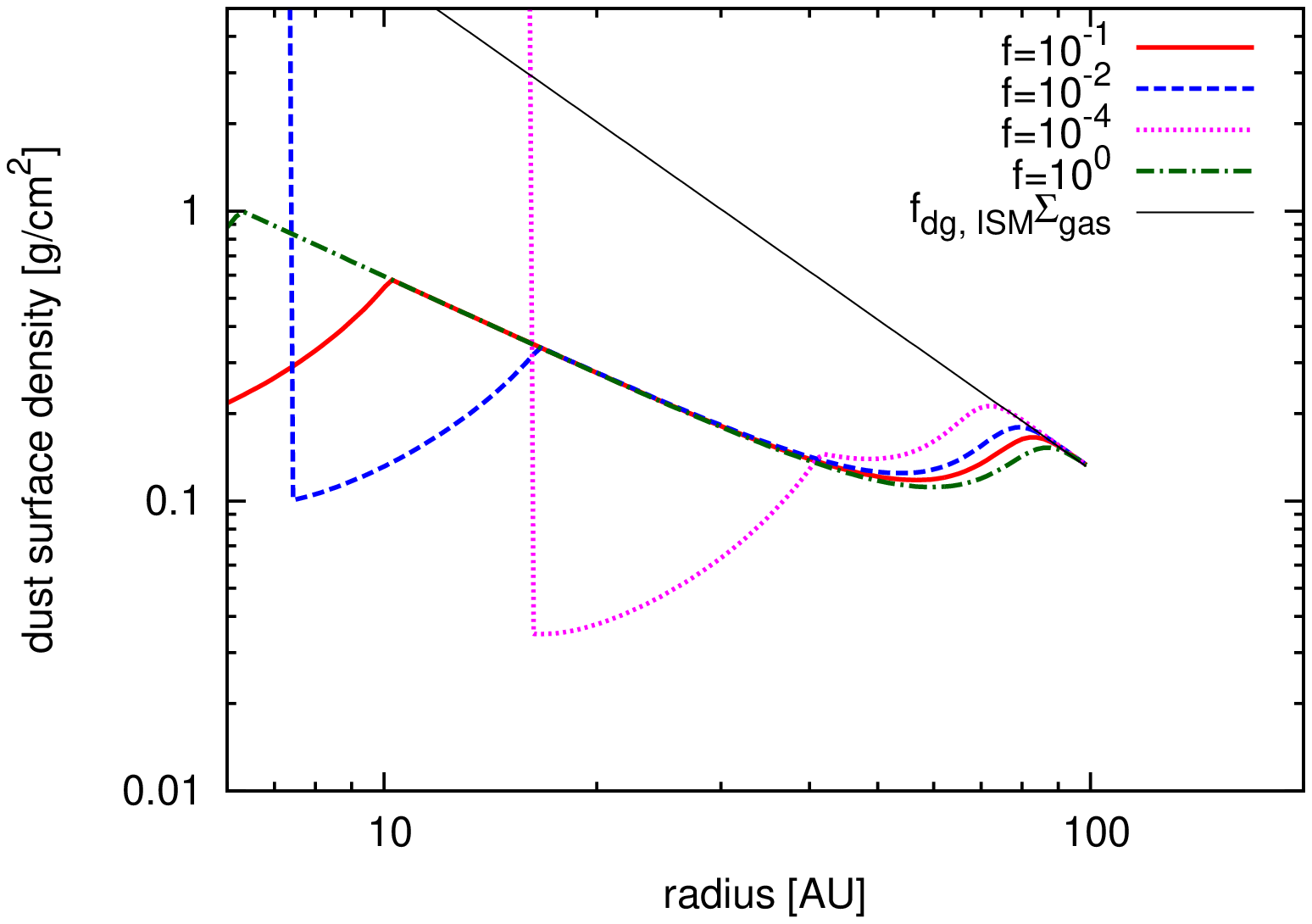}
\includegraphics[width=60mm,angle=-90]{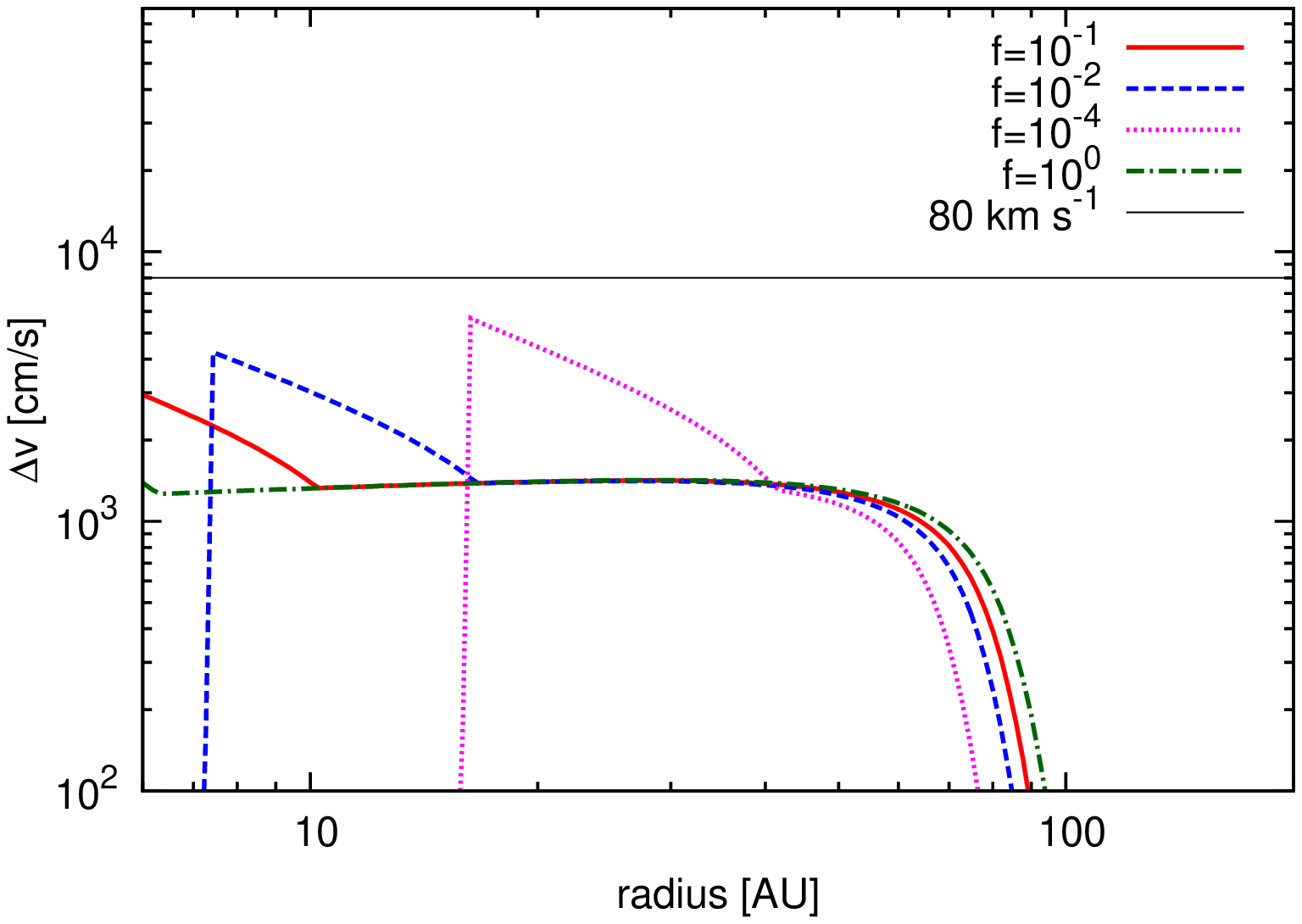}
\includegraphics[width=60mm,angle=-90]{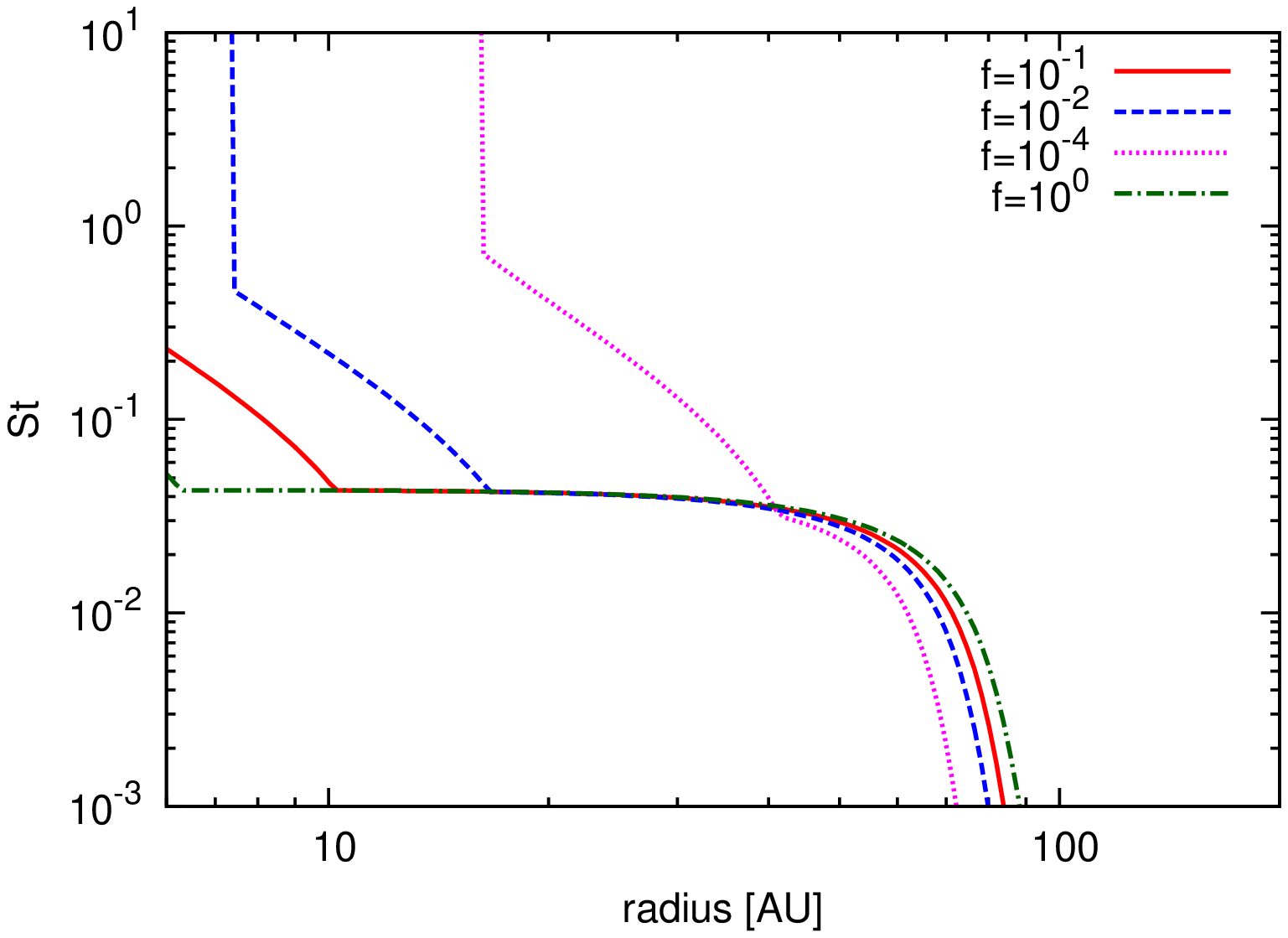}
\caption{
Radial profile of dust surface density (top left), 
collision velocity (top right), and Stokes number (bottom) 
in steady state (at $t= 0.2$ Myr)  for various filling factors.
The red solid, blue dashed, magenta dotted, and green dashed-dotted lines
show the profile 
of M1Mdot37r100f1, M1Mdot37r100f2, M1Mdot37r100f4, 
and  M1Mdot37r100f0, respectively.
The black solid lines in the top-left and top-right panels
show $f_{\rm dg, ISM} \Sigma_{\rm gas}$ and the threshold velocity, respectively.
}
\label{profile_porus}
\end{figure*}

Figure \ref{profile_porus} shows the profiles of dust surface density,
collision velocity, and Stokes number
in the steady state for various filling factors.
As shown in this figure, the dust surface density 
converges to the same steady-state solution for $r_{\rm Stokes}<r<r_{\rm drift}$.
This is because, in the Epstein regime,
$\St\propto \rho_{\rm int} a_{\rm dust}$ and
$\rho_{\rm int}$ and $a_{\rm dust}$ can be simultaneously eliminated 
from equation (\ref{St_mu})
and the Stokes number does not depend on the filling factor. In
addition, the dust surface density is solely determined by the Stokes number.

$r_{\rm Stokes}$ increases as porosity decreases because 
$a_{\rm dust}$ is calculated from $a_{\rm dust}=(2 \Sigma_{\rm gas} \St)/(\pi \rho_{\rm int}) \propto f^{-1}$
and the mean free path does not depend on the internal density.
The steady state in the Stokes regime, which 
is also determined by (\ref{St_mu}), depends on
$\rho_{\rm int}$ because we cannot eliminate the dust radius
and internal density simultaneously from (\ref{St_mu}) in the Stokes regime,
and $\Sigma_{\rm dust}$ converges to the different steady solutions.
In the model with $f=10^{-4}$ (magenta dotted line),
dusts grow quickly and enter the
Stokes regime before the dust disk structure converges to the 
steady-state solution of the Epstein regime.

\begin{figure*}
\includegraphics[width=60mm,angle=-90]{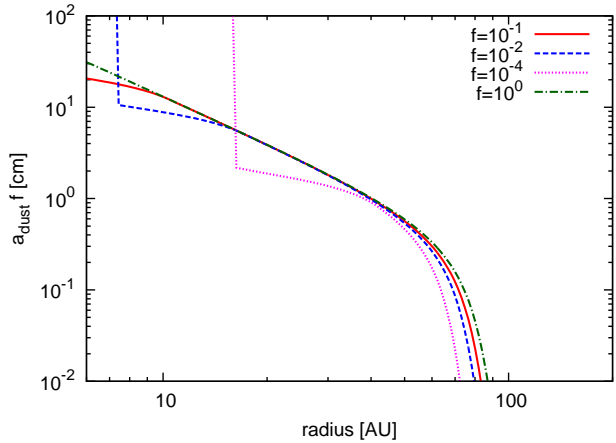}
\caption{
Radial profile of $a_{\rm dust} f$
in steady state (at $t= 0.2$ Myr)  for various filling factors.
The red solid, blue dashed, magenta dotted, and green dashed-dotted lines
show the profiles 
of M1Mdot37r100f1, M1Mdot37r100f2, M1Mdot37r100f4, 
and  M1Mdot37r100f0, respectively.
}
\label{size_porus}
\end{figure*}

In the model with $f=10^{-4}$,
the surface density and Stokes number rapidly increase
at $r \sim 20$ AU,
indicating  planetesimal formation.
Therefore, planetesimal formation is possible at 
$r\gtrsim 10 \AU$, if the filling factor decreases
to $f\lesssim 10^{-4}$ in the gravitationally unstable disk.
Note that the collision velocity is smaller than the threshold 
velocity ($v_{\rm th}=80 \ms$)
and our assumption of perfect sticking is valid in the models
discussed in this subsection.
Although porous dust aggregates can grow to large radii
(in $f=10^{-4}$ case, the radius becomes $a_{\rm dust} \sim 10^{4} \cm$ 
at $r\sim 20$ AU),
this growth does not significantly change the absorption opacity, because
the opacity does not depend on the size $a_{\rm dust}$, but on the product of 
the filling factor and the size, $a_{\rm dust}f$ \citep{2013A&A...557L...4K}.
In figure \ref{size_porus}, we show $a_{\rm dust}f$ in the steady state
for various filling factors.
$a_{\rm dust}f$ converges to the same steady state,
independent of $f$ in the Epstein regime, because 
$a f\propto a\rho_{\rm int}\propto\St$, and $\St$ does not depend on the
filling factor.

\begin{figure*} 
\includegraphics[width=60mm,angle=-90]{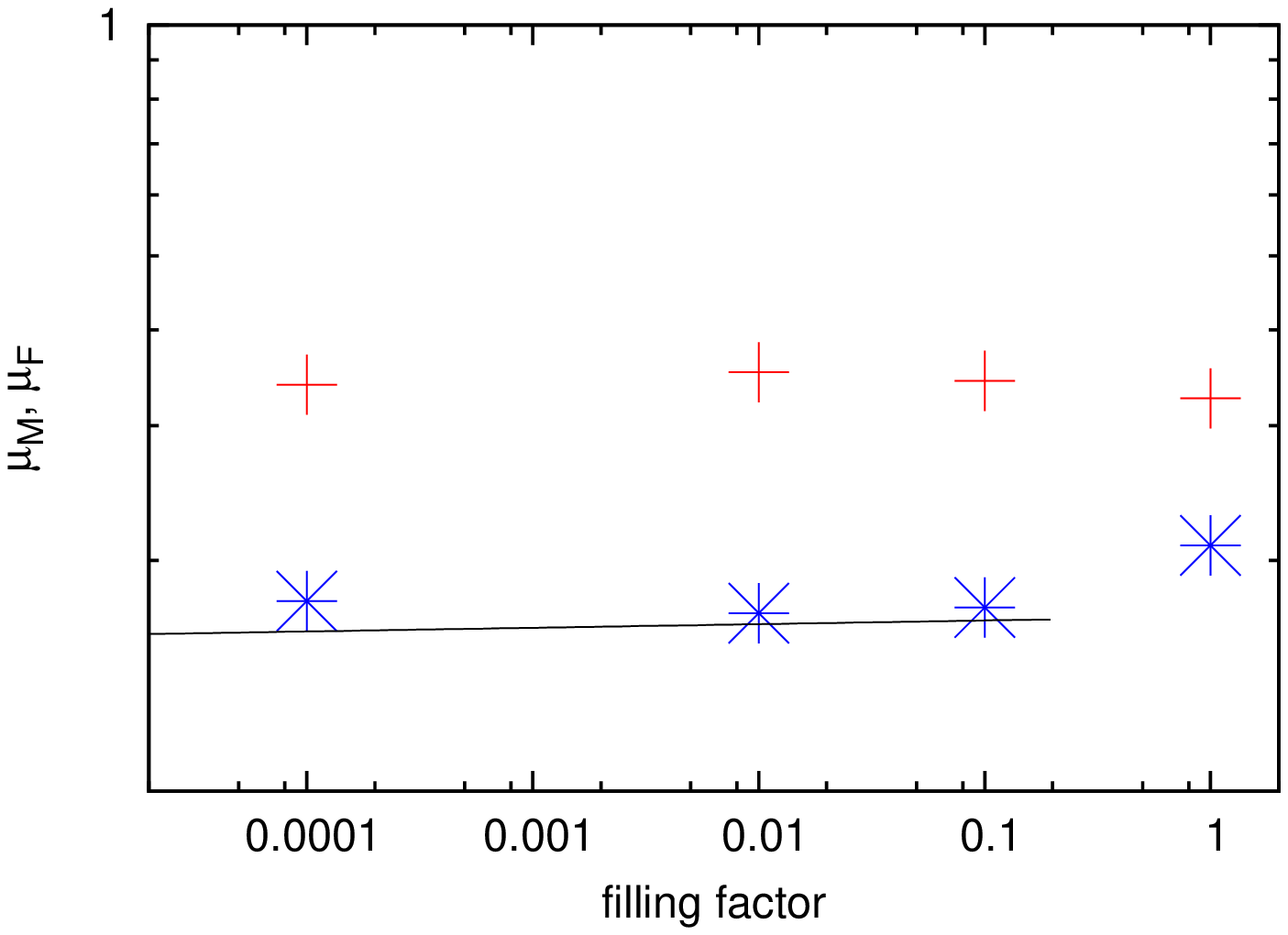}
\includegraphics[width=60mm,angle=-90]{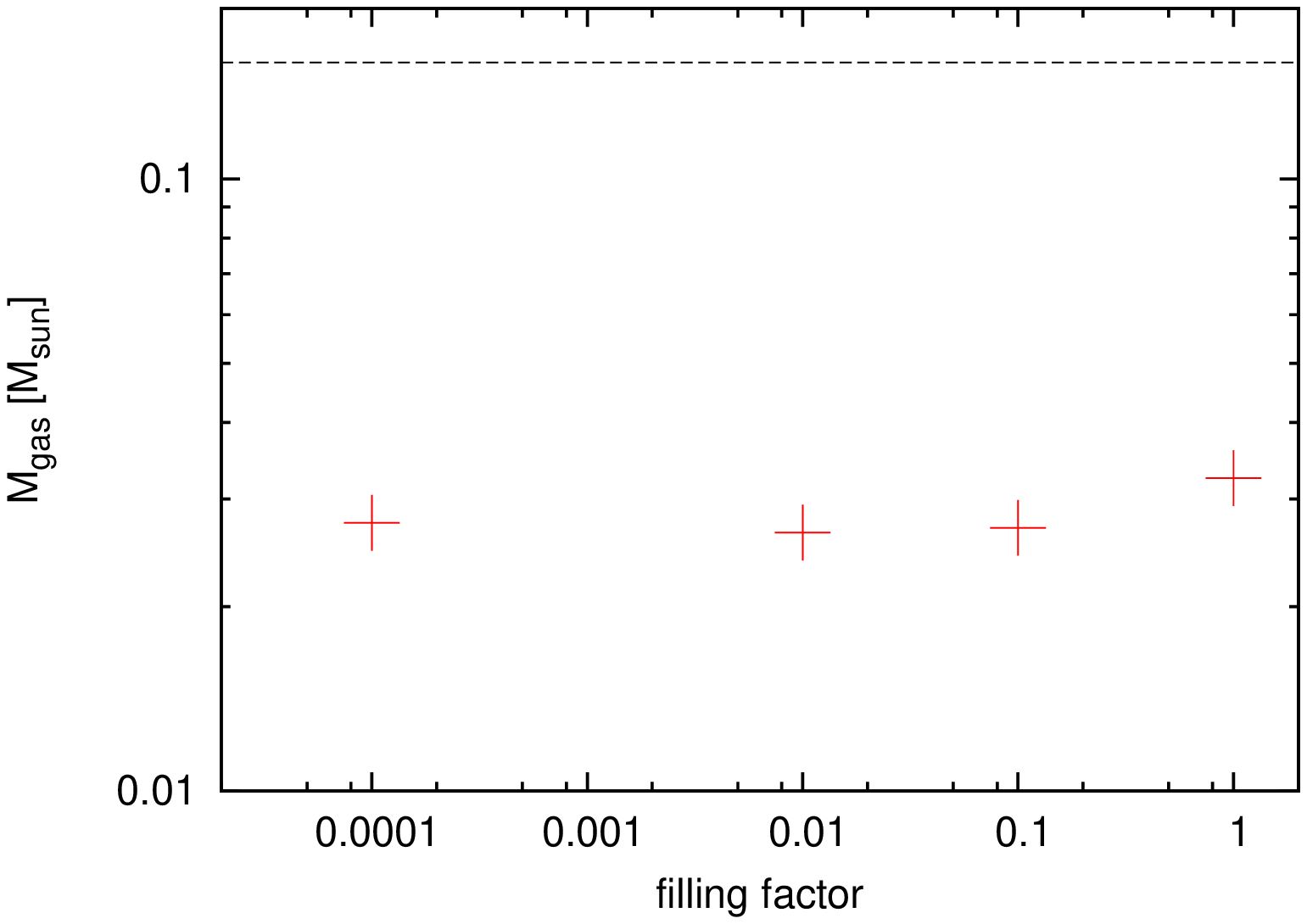}
\caption{
Left panel 
shows the ratio of dust mass in the steady-state disk (at $t= 0.2$ Myr) to
that in a disk with the ISM dust-to-gas mass ratio  $\mu_{\rm M} $,
as well as the ratio of the radiative flux from the steady-state disk to
that from a disk with the ISM dust-to-gas mass ratio and micron-sized dust
$\mu_{\rm F}$ for various filling factors.
The red crosses and blue asterisks show $\mu_{\rm M}$ and $\mu_{\rm F}$, respectively,
of M1Mdot37r100f1, M1Mdot37r100f2, M1Mdot37r100f4, and  M1Mdot37r100f0.
The black solid line shows the empirical formula for $\mu_{\rm F}$,
equation (\ref{empirical_form_muf}), which is applicable for $f \le 10^{-1}$.
The right panel shows the apparent mass $M_{\rm app}$ of the gas disk.
The red crosses show $M_{\rm app}$.
The black dashed line shows the actual mass of the gas disk
in the simulations, $M_{\rm gas}=0.16 \msun$.
}
\label{disk_mass_porus}
\end{figure*}

In the left panel of figure \ref{disk_mass_porus},
we show $\mu_{\rm M}$ and $\mu_{\rm F}$ for various filling factors 
($f=1,~10^{-1},~10^{-2},~10^{-3},$ and $10^{-4}$).
For the calculation of the  $f=10^{-4}$ model, we set $r_{\rm min}=r_P$ 
to limit complexity.
The figure reveals a very weak dependence of $\mu_{\rm M}$
on the filling factor.
Although the dust is depleted in the inner region in the models
with a small filling factor, 
its contribution to the total mass is not significant owing to
the metric $2\pi r$ in the integral. 
Furthermore, in the $50 \AU\lesssim r \lesssim 80 \AU$ region,
the surface densities for the models with small filling 
factor are slightly larger
than those for the models with a large filling factor. 
As a result,  $\mu_{\rm M}$
is almost constant against changes in porosity.

On the other hand, $\mu_{\rm F}$ exhibits a dependence
on filling factor for  $f\ge 10^{-1}$.
From $f=10^{-1}$ to $f=1$, $\mu_{\rm F}$ suddenly 
increases even though the dust mass is almost constant.
This change is caused by the increase in absorption opacity. 
As shown in figure \ref{opacity_fig}, the opacity for
compact dust ($f=1$) increases in the range o
 $10^{-2} {\rm cm}\lesssim a_{\rm dust} \lesssim 1 {\rm cm}$.
Because the dust size in the $f=1$ model enters this range
in the outer region of the disk,
$40 \AU \lesssim r \lesssim 80 \AU$, 
the thermal emission at $\lambda=1.3 {\rm mm}$ from this region 
becomes larger than that with micron-sized dust.
This causes an increase of $\mu_{\rm F}$ in the $f=1$ case.
Note, however, that the compact case ($f=1$)
seems to be unlikely, both theoretically and observationally 
\citep[][]{2005Sci...310..258A,2016Natur.530...63P,2007A&A...461..215O,2012ApJ...752..106O}.
For $f<10^{-1}$, $\mu_{\rm F}$ is almost independent of $f$
because the dust porosity mainly influences
the inner structure of the disk and the contribution of the
inner region to the radiative flux is small owing to the opacity
decrease by dust growth and the metric $2\pi r$ in the integral.
As indicated by our empirical formula,  equation (\ref{empirical_form_muf}),
$\mu_{\rm F}$ scales as $f^{0.0048}$ for $f \le 10^{-1}$ and is
almost independent of the filling factor.

In the left panel of figure \ref{disk_mass_porus}, 
we show the apparent disk mass.
The dependence of the disk mass on the porosity is so weak that 
it does not influence the apparent disk mass.
Again, the apparent disk mass is within 
the range $0.01 \msun < M_{\rm app} < 0.1 \msun $ as suggested by observations.

\subsubsection{Dependence on the disk radius} 
\begin{figure*}
\includegraphics[width=60mm,angle=-90]{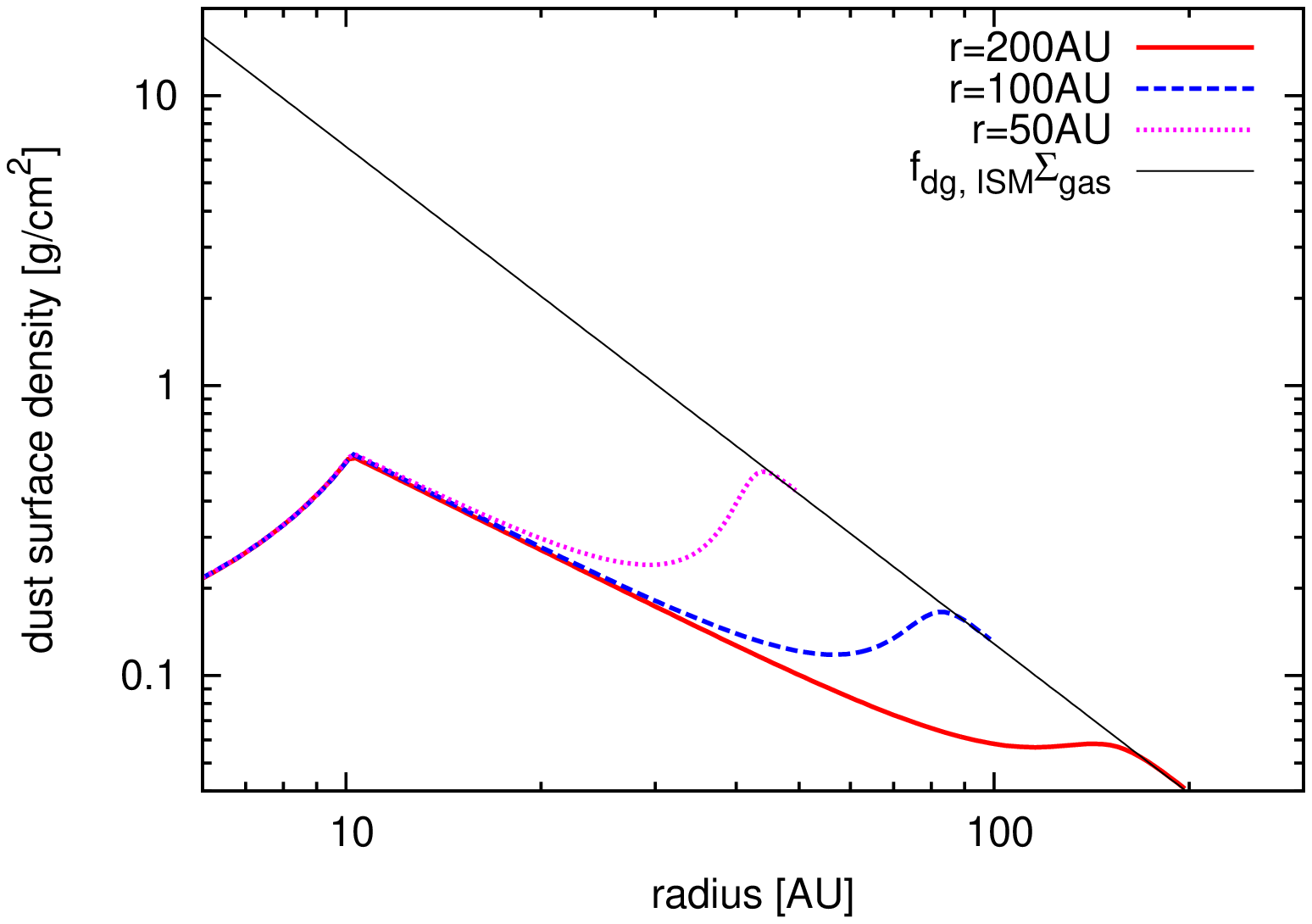}
\includegraphics[width=60mm,angle=-90]{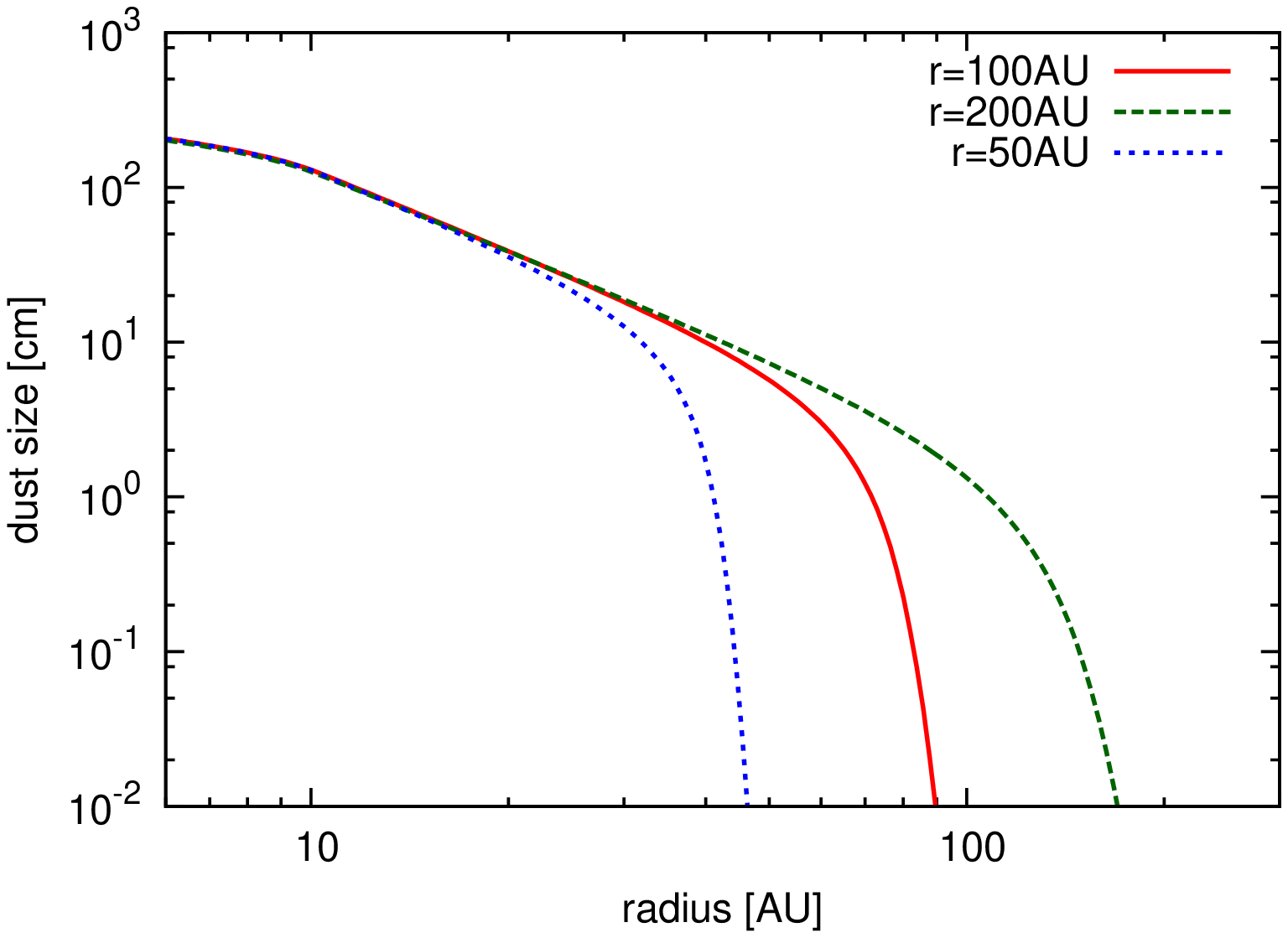}
\caption{
Radial profile of dust surface density and dust size
in steady state (at $t= 0.2$ Myr) for various disk radii.
The red solid, blue dashed, and magenta dotted lines
show the profiles
of M1Mdot37r200f1, M1Mdot37r100f1, and M1Mdot37r50f1, respectively.
The black solid line in the surface density profile shows $f_{\rm dg, ISM} \Sigma_{\rm gas}$. 
}
\label{profile_radius}

\end{figure*}

In figure \ref{profile_radius}, 
we show the profiles of the dust surface density and dust size
in the steady state for various disk radii ($r=50,~100,$ and $200 {\rm AU}$).
Similar to previous cases,
until the dust particles grow sufficiently for radial drift to begin,
they move with the gas advection velocity.
Once the dust radial drift begins, 
the dust surface density decreases from the initial value.
Because the steady-state solution does not depend on 
the disk radius, all of the models converge
to the same solution, both 
in the Epstein and the Stokes regimes, 
once the radial drift begins.

\begin{figure*} 
\includegraphics[width=60mm,angle=-90]{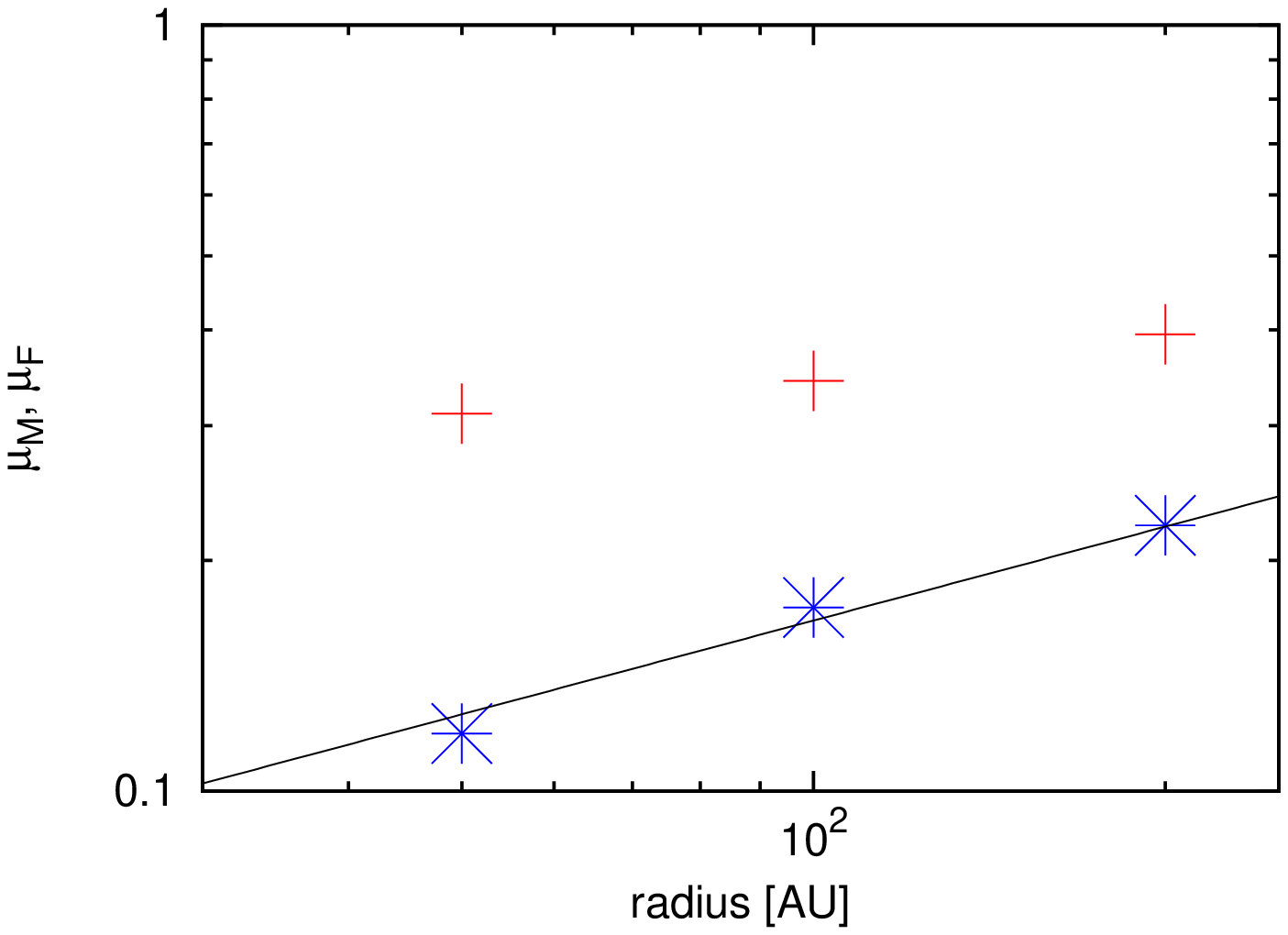}
\includegraphics[width=60mm,angle=-90]{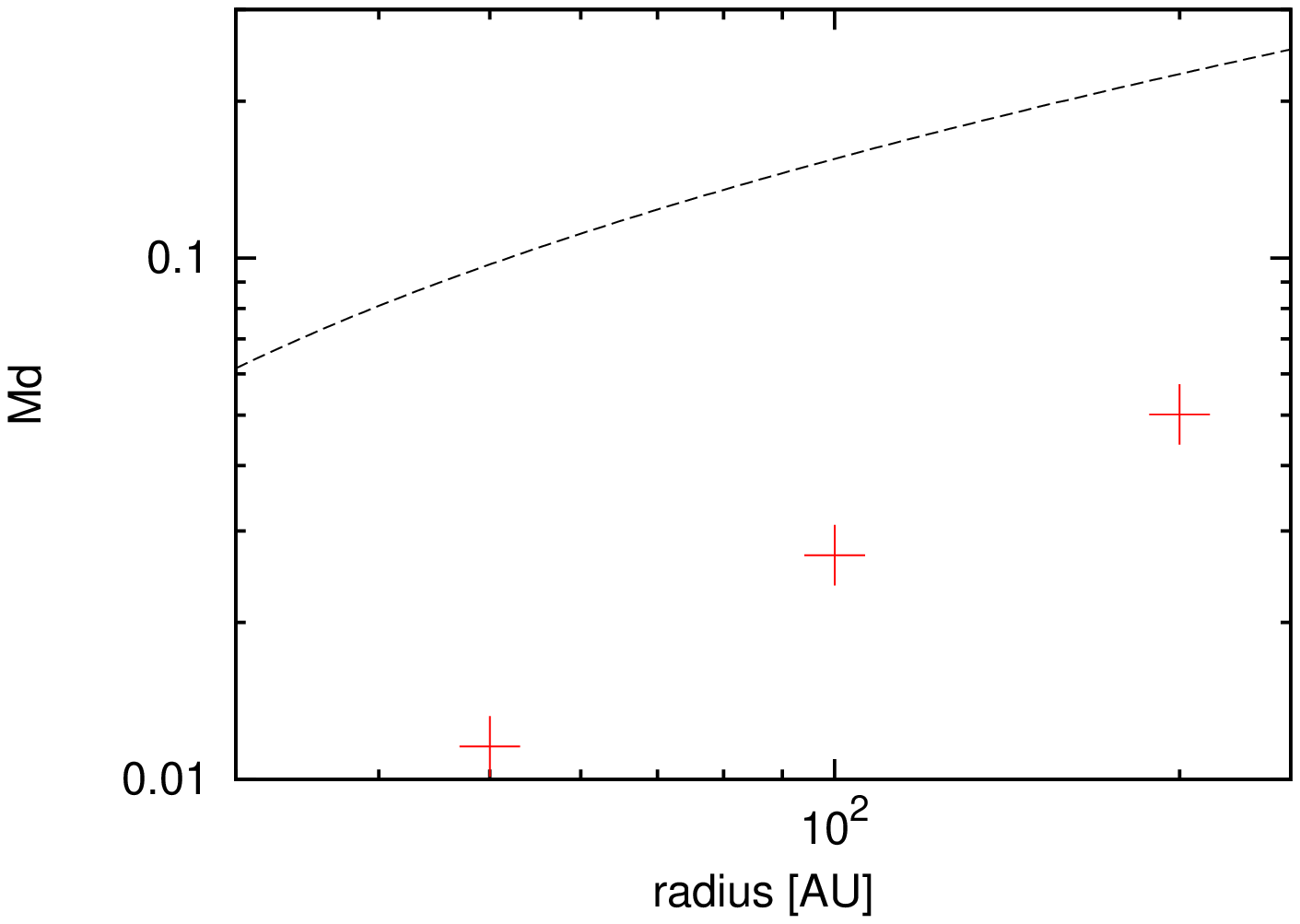}
\caption{
Left panel 
shows the ratio of the dust mass in the steady-state disk (at $t= 0.2$ Myr) to
that in the disk with the ISM dust-to-gas mass ratio $\mu_{\rm M} $,
and ratio of the radiative flux from the steady-state disk to
that from the disk with the ISM dust-to-gas mass ratio and micron-sized dust
$\mu_{\rm F}$ for various disk radii $r_{\rm disk}$.
The red crosses and blue asterisks show $\mu_{\rm M}$ and $\mu_{\rm F}$, respectively,
of M1Mdot37r200f1, M1Mdot37r100f1, and M1Mdot37r50f1.
The black solid line shows the empirical formula for $\mu_{\rm F}$,
equation (\ref{empirical_form_muf}).
The right panel shows the apparent mass $M_{\rm app}$ of the gas disk.
The red crosses show  $M_{\rm app}$.
The black dashed line shows the actual mass of the gas disk
in our simulation, which asymptotically obeys 
$M_{\rm gas}\propto r_{\rm disk}^{2/7}$.
}
\label{disk_mass_radius}
\end{figure*}

As shown in figure \ref{disk_mass_radius}, 
$\mu_{M}$ is $0.3<\mu_{\rm M}<0.4$, and
the dust-to-gas mass ratio is less sensitive to the disk radius.
On the other hand, the dependence of $\mu_{\rm F}$ on the disk radius is relatively strong.
This is because, in the disk with $r_{\rm disk}=50$ AU, 
the dust size becomes $a_{\rm dust} f>1 \cm$  (see right panel of 
figure \ref{profile_radius}),
and the dust opacity decreases over nearly the entire region of the disk.
As shown in our empirical formula, equation (\ref{empirical_form_muf}),
$\mu_{\rm F}$ scales as $\mu_{\rm F}\propto r_{\rm disk}^{0.41}$.
The apparent mass shown in the right panel of figure \ref{disk_mass_radius}, 
increases rapidly as the disk radius increases because both 
$M_{\rm gas}$ and $\mu_{\rm F}$ are increasing functions of the disk radius.
Note, however, that even for a relatively large disk with $r=200 \AU$, 
apparent mass is $0.05 \msun$ and within the range 
suggested by the observations.

\subsection{Maximum radius for planetesimal formation}
\label{sec_planetesimal}
As pointed out by \citet{2012ApJ...752..106O} and \citet{2013A&A...557L...4K},
highly porous aggregates grow faster than radial drift and finally 
form planetesimals in the inner region of a disk
for which Stokes drag law determines the dust stopping time.
According to this scenario, the orbital radius within
which planetesimals form $r_{\rm P}$ increases
with the gas surface density \citep{2012ApJ...752..106O}.
On the other hand, a disk with a larger surface 
density than that of a gravitationally
unstable disk cannot exist.
Therefore, by investigating the parameter dependence of 
$r_{\rm P}$ in a gravitationally unstable disk, we can determine
the maximum value for $r_{\rm P}$.

As noted in \S \ref{sec_dust_model}, unlike the other results 
discussed in this paper,
$r_{\rm P}$ inevitably depends on the initial conditions
when we employ the radial drift velocity of equation (\ref{vr_dust}).
The $r_{\rm P}$ of a model using equation (\ref{vr_dust}) indicates
the maximum value for the parameter set of the model.
On the other hand,
$r_{\rm P}$ becomes independent of the initial condition
when we employ the radial drift velocity of equation 
(\ref{vr_dust2}), and 
the $r_{\rm P}$ of a model using equation (\ref{vr_dust2}) indicates
the orbital radius within
which planetesimals form under a steady-state solution and
is the minimum value for the parameter set.
We first show the results obtained using equation (\ref{vr_dust2}) and
then discuss the difference between 
the results with equation (\ref{vr_dust}) and equation (\ref{vr_dust2}).

\begin{figure*} 
\includegraphics[width=60mm,angle=-90]{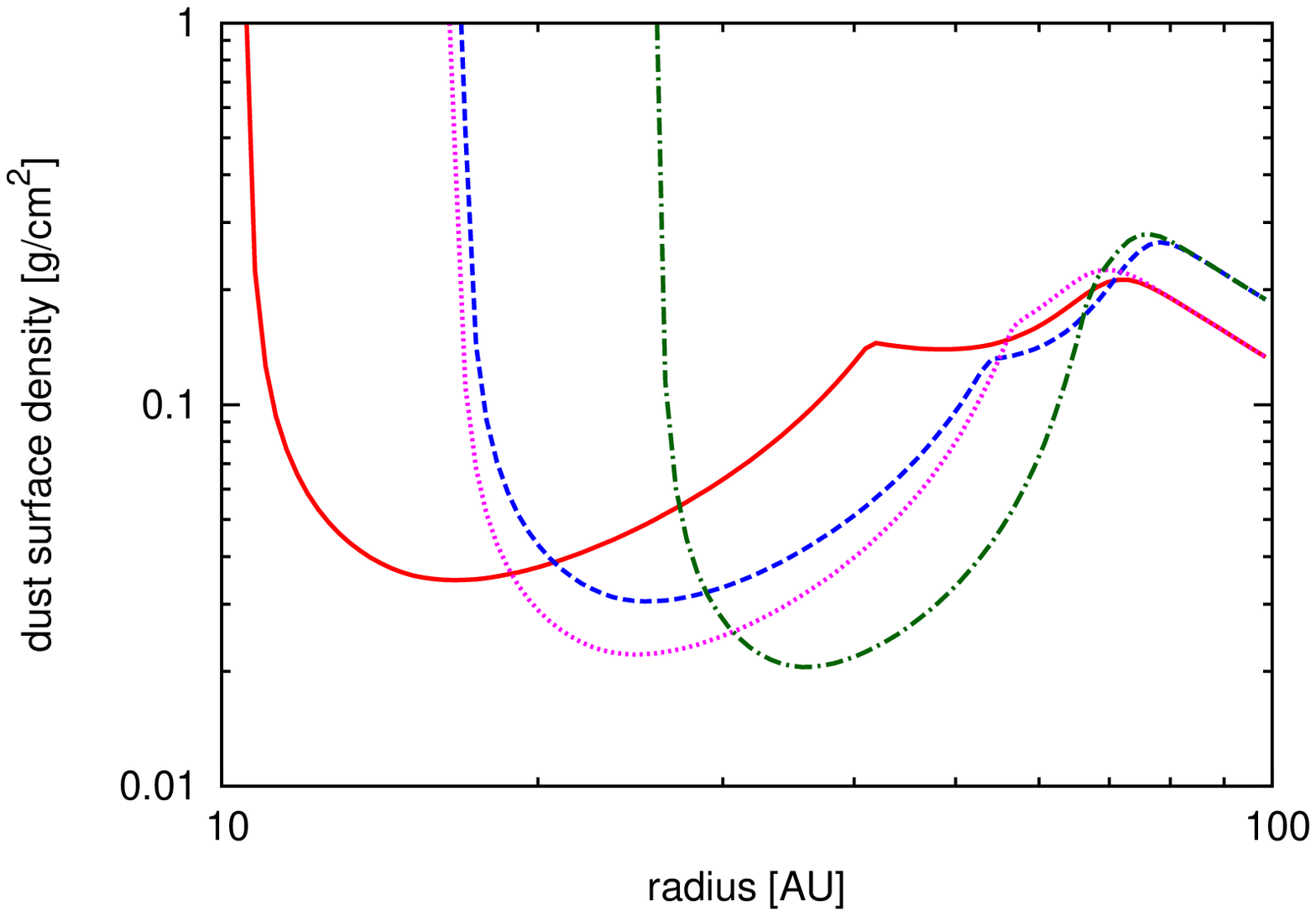}
\includegraphics[width=60mm,angle=-90]{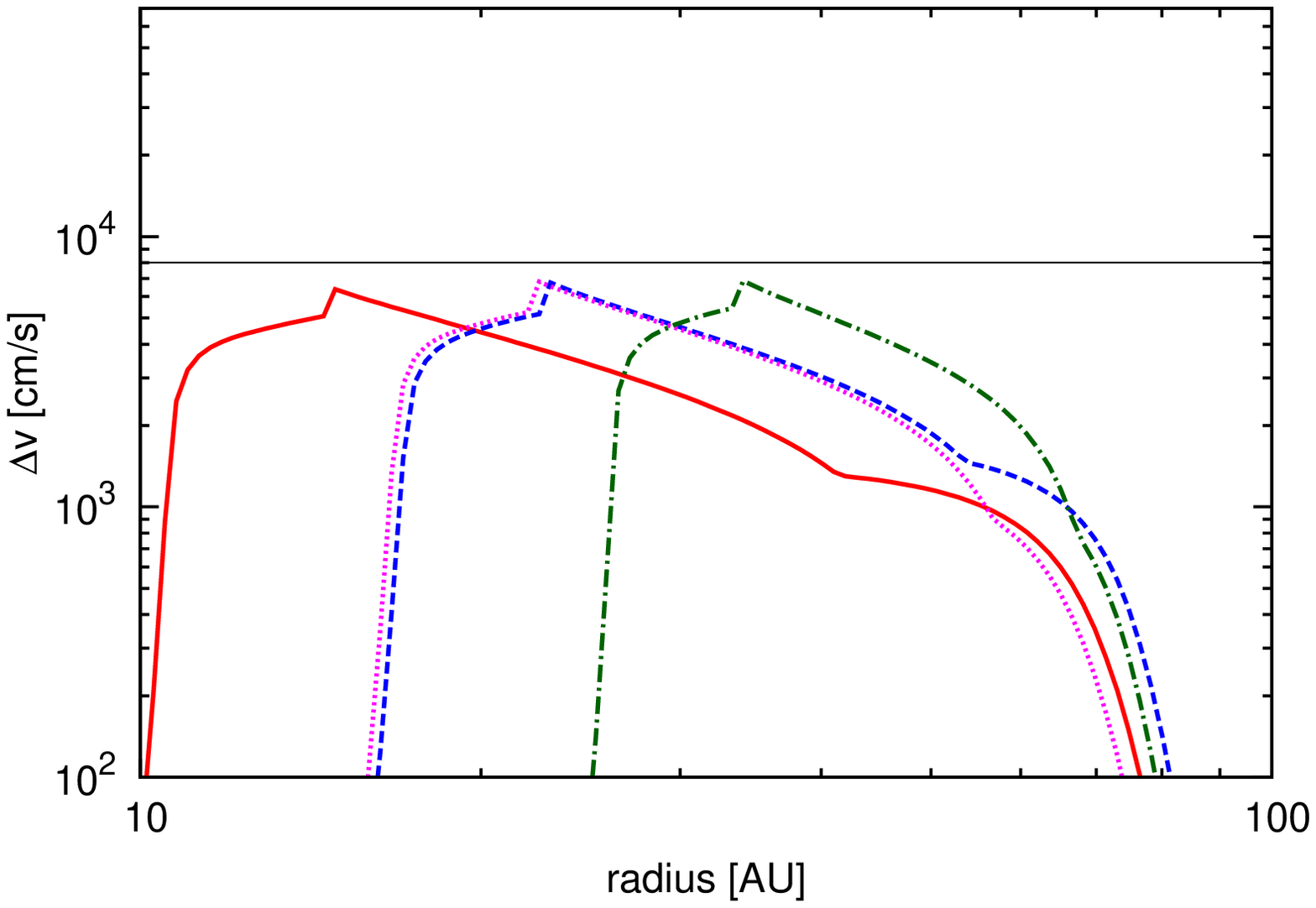}
\includegraphics[width=60mm,angle=-90]{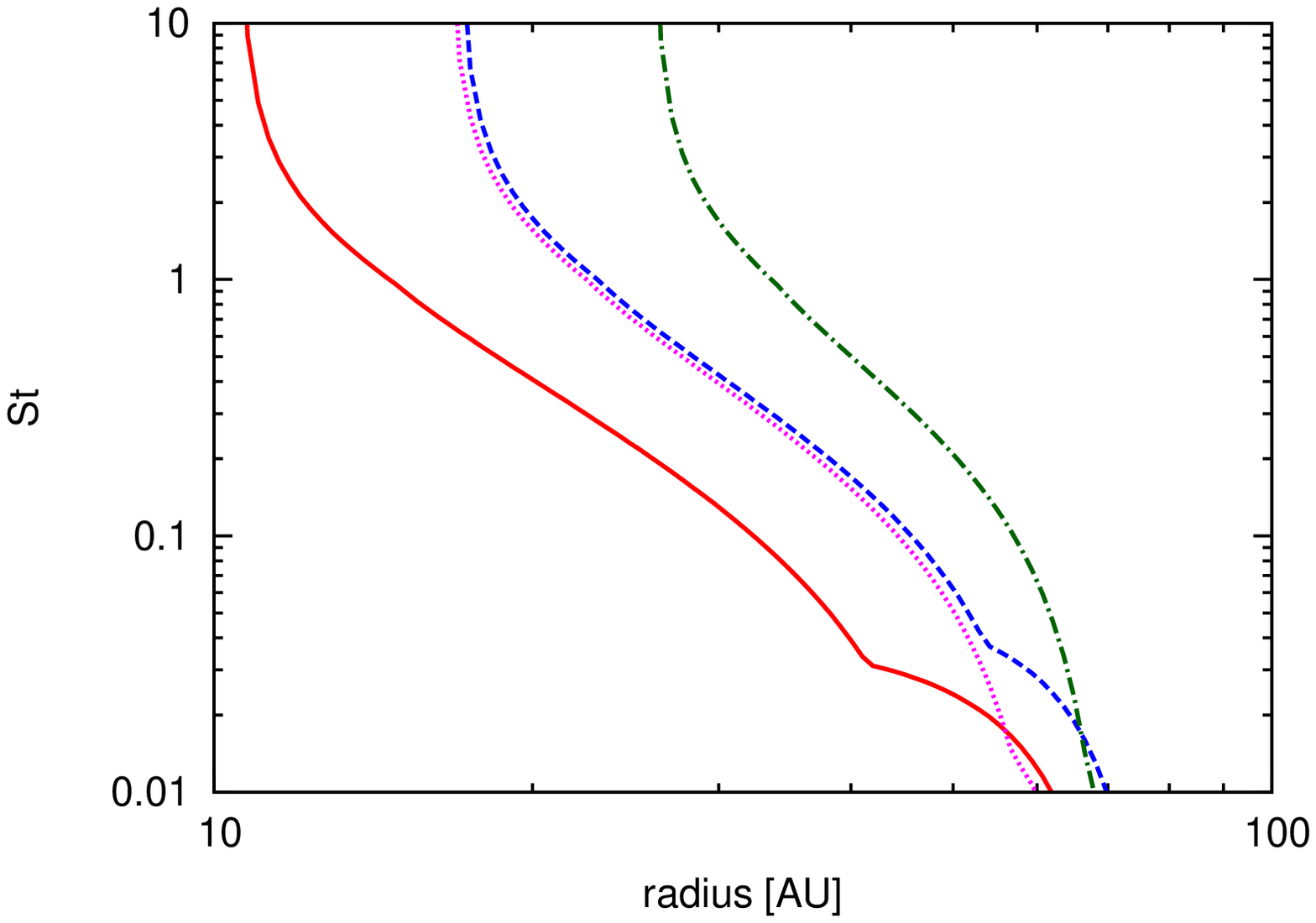}
\caption{
Radial profiles of dust surface density, Stokes number, 
and collision velocity in the steady state for 
models in which planetesimals form at $r>10 \AU$.
The red solid, blue dashed, and magenta dotted, green dashed-dotted lines
show the profiles 
of M1Mdot37r100f4, M2Mdot37r100f4, M1Mdot37r100f5,  
and M2Mdot37r100f5, respectively.
In all models shown in this figure, 
equation (\ref{vr_dust2}) is employed.
The black solid line in the right panel shows the threshold velocity. 
}
\label{profile_planetesimal}
\end{figure*}

In the left panel of figure \ref{profile_planetesimal}, 
we show the dust surface densities
of the models in which planetesimals form for $r>10 \AU$.
A sudden increase in the dust surface density indicates 
planetesimal formation.
$r_{\rm Stokes}$ and, therefore, $r_{\rm P}$ increase according to
the increase in $M_{\rm star}$ and $\dot{M}_{\rm gas}$, along
with the decrease in $Q_{\rm crit}$ and $f$.
Among these parameters, 
$M_{\rm star}$ and $f$ strongly affect $r_{\rm P}$.
The maximum $r_{\rm P}$ is realized in the model with 
$M_{\rm star}=2 \msun, ~\dot{M}_{\rm gas}=3 \times 10^{-7} \msunyear,~ f=10^{-5}$,
and has a value of  $r_{\rm P} \sim 30 \AU$.
On the other hand, if we consider 
the model with 
$M_{\rm star}=1 \msun, ~\dot{M}_{\rm gas}=3 \times 10^{-7} \msunyear,~ f=10^{-5}$,
the orbital radius within which planetesimals form is $r_{\rm P} \sim 20 \AU $.

The right panel of figure \ref{profile_planetesimal} 
shows the profile of collision velocity.
The collision velocity of all models considered in
this section is smaller than the threshold velocity
and neglecting the collisional fragmentation is still valid.
Although $r_{\rm P}$ can increase with larger $\dot{M}_{\rm gas}$,
the collision velocity becomes too large and exceeds the threshold 
velocity. Thus, the collisional fragmentation cannot be avoided, 
and larger $r_{\rm P}$ may not be achieved by increasing $\dot{M}_{\rm gas}$.

\begin{figure*} 
\includegraphics[width=60mm,angle=-90]{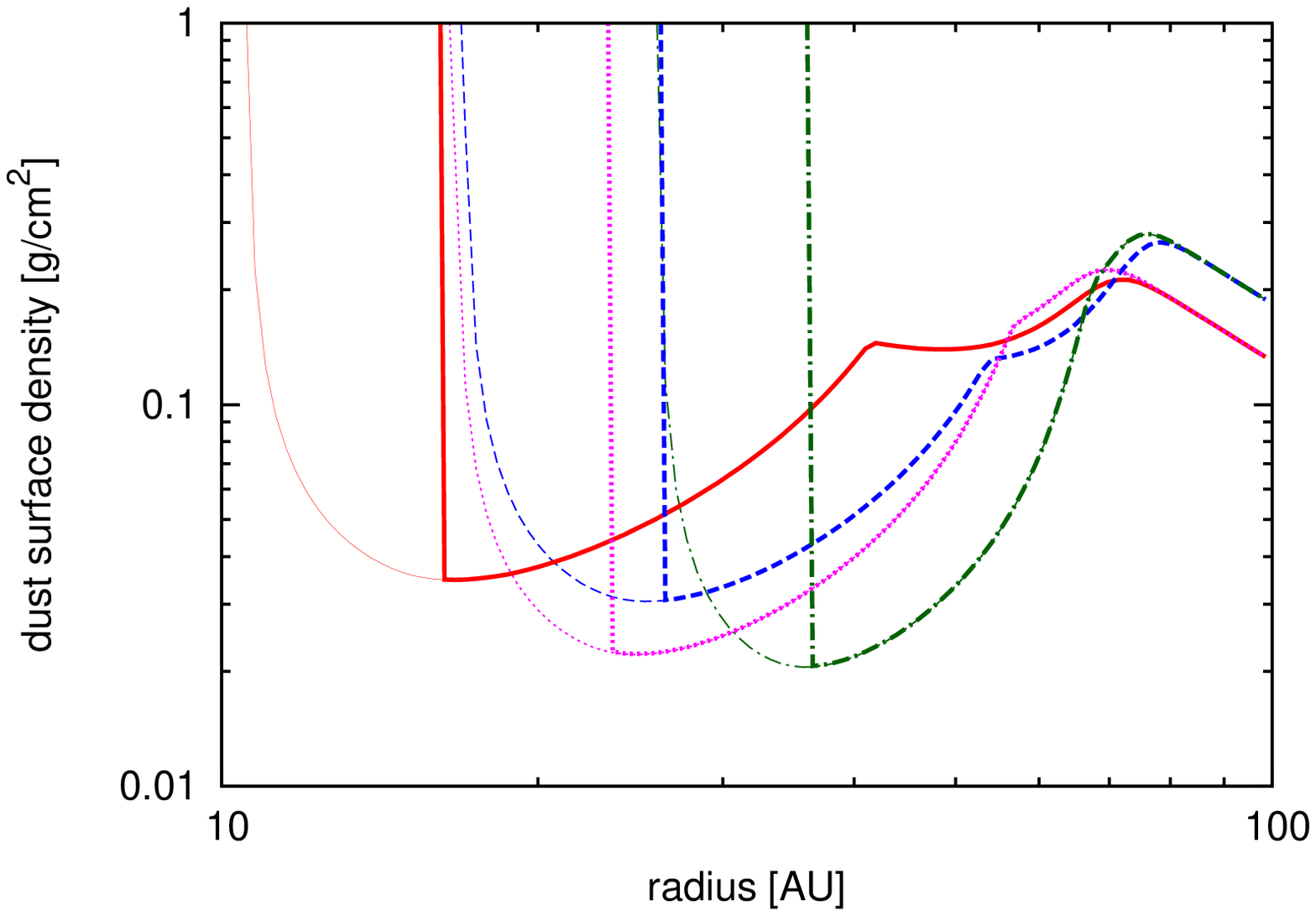}
\caption{
Radial profiles of dust surface density
in the steady state (at $t= 0.2$ Myr) of the
The red solid, blue dashed, and magenta dotted, green dashed-dotted lines
show the profiles 
of M1Mdot37r100f4, M2Mdot37r100f4, M1Mdot37r100f5,  
and M2Mdot37r100f5, respectively.
In all models shown with thick lines, 
equation (\ref{vr_dust}) is employed,
while equation (\ref{vr_dust2}) is employed
in the models shown with thin lines;
these models are identical to that in figure (\ref{profile_planetesimal}).
}
\label{profile_planetesimal2}
\end{figure*}

In figure \ref{profile_planetesimal2},
we compare  $r_{\rm P}$ in the simulations
with the equations (\ref{vr_dust}) and (\ref{vr_dust2}).
Thick lines indicate the results using the equation (\ref{vr_dust}).
As we discussed, $r_{\rm P}$ becomes large under equation (\ref{vr_dust}).
For example, in the model with
$M_{\rm star}= 1 \msun, ~\dot{M}_{\rm gas}=3 \times 10^{-7} \msunyear,~ f=10^{-5}$,
the orbital radius for planetesimal formation becomes $r_{\rm P} \gtrsim 20 \AU $,
which is approximately $10$ AU larger than that under equation (\ref{vr_dust2}).
This increase is caused by the large mass accretion 
rate and the large dust surface density at the beginning of the simulation.
In reality, planetesimals may form between the two radii obtained
using the equations (\ref{vr_dust}) and (\ref{vr_dust2}).
Summarizing the above discussion, we can conclude that
planetesimal formation is possible at $r \sim 20 \AU$ 
in gravitationally unstable disks around protostars
with $M_{\rm star}=1 \msun$.
This is the theoretical maximum value of $r_{\rm P}$
according to the planetesimal formation 
mechanism suggested by \citet{2012ApJ...752..106O} and \citet{2013A&A...557L...4K}.

\section{Summary and Discussion}
In this paper, we investigated the dust structure 
in a gravitationally unstable disk undergoing mass accretion
from an envelope, envisioning the application of our findings to Class 0/I YSOs.
We considered the growth and radial drift of dust particles 
with a single size approximation \citep{2001A&A...378..180K,2012A&A...539A.148B,
2016ApJ...821...82O,2016A&A...589A..15S}.
Comparing the dust thermal emission from 
a steady-state structure with that from a disk 
with ISM dust-to-gas mass ratio and micron-sized dust, 
we evaluated the apparent reduction rate of the gas disk mass estimated
from the dust thermal emission at $\lambda= 1.3 {\rm mm}$, and
showed that the disk mass is systematically underestimated. 

We also investigated the orbital radius within
which planetesimals form $r_{\rm P}$ in
a gravitationally unstable disk, using the
planetesimal formation mechanism suggested 
by \citet{2012ApJ...752..106O}. 
$r_{\rm P}$ of a gravitationally unstable disk is the theoretical maximum value 
because it increases with the gas surface density.
Thus, we derived the maximum $r_{\rm P}$ for arbitrary disks.

\subsection{Summary}
Our findings are summarized as follows.

\begin{enumerate}
\item
The dust disk quickly settles into the steady state and
the total dust mass in the steady state of gravitationally 
unstable disks in Class 0/I YSOs systematically decreases
from that of a disk with ISM dust-to-gas mass ratio.
The reduction rate $\mu_{\rm M}$ is 1/2 to 1/3
depending on the mass accretion rate,
disk radius, and mass of the central star.
The internal density of the dust aggregates 
only has a minimal effect on the reduction rate.
\item
The radiative flux of dust thermal emission 
from a steady state disk also systematically decreases 
by a factor of 1/3 to 1/5 from 
that of a disk with  ISM dust-to-gas mass ratio and micron-sized dust.
This reduction is caused both by dust depletion and 
a decrease in opacity owing to dust growth.
Because ISM dust-to-gas mass ratio (typically 1/100) and
micron-sized dust (typically $0.1 {\rm \mu m}$) are often assumed in the  estimate of the gas disk mass from
dust thermal emissions,
our results suggest that the gas mass of gravitationally unstable disk
is underestimated by a factor of 1/3 to 1/5.
\item
Even when a gravitationally unstable disk exists in Class 0/I YSOs,
the typical value of the apparent disk mass estimated from 
dust thermal emission is $\sim 0.03 \msun$
and has a  $0.01 \msun  \lesssim M_{\rm app} \lesssim 0.1 \msun$ range.
The apparent disk mass has positive dependence on the mass accretion
rate from the envelope and disk radius but is almost independent of 
the dust internal density and mass of the central star.
The apparent disk mass is consistent with the disk 
mass of Class 0/I YSOs estimated from observation, and
this apparent reduction suggests 
that a larger fraction 
Class 0/I YSO disks is gravitationally unstable
than previously believed.
\item
The asymptotic steady-state solutions for dust structures 
is derived (discussed in Appendix A).
The solutions are applicable to arbitrary disks
in which the radial drift determines the dust structure.
For example, in the gravitationally unstable gas disk considered in this paper, 
the dust surface density asymptotically 
obeys the power law of $\Sigma_{\rm dust}\propto r^{-31/28}$
in the Epstein regime, when 
$\Delta v=\sqrt{\alpha c_{\rm s}^2 \St}$ and 
$H_{\rm dust}=(\St/\alpha)^{-1/2} H_{\rm gas}$.
\item
In the gravitationally
unstable disk, planetesimals can form at $r\sim 20$ AU around a $1 \msun$ protostar
via the planetesimal formation 
mechanism suggested by \citet{2012ApJ...752..106O} and \citet{2013A&A...557L...4K}.
Because the gas surface density 
of the gravitationally unstable disk is the theoretical maximum and 
the orbital radius of planetesimal formation $r_{\rm P}$ 
is an increasing function of the gas surface density,
this value is the theoretical maximum. Thus,
planetesimal formation at $r\gtrsim 20$ AU 
for a $1 \msun$ protostar by the mechanism is highly unlikely
because of the radial drift barrier.
Some outer migration mechanisms, such as the outer migration of planets
\citep{1995AJ....110..420M, 2003Natur.426..419L,2011Icar..212..911T}
would be required to explain the existence of planetesimals
for $r\gtrsim 20$ AU in the solar system.
\end{enumerate}

\subsection{Discussion}
\subsubsection{Comparison to observations}
Our estimate of the apparent mass of a
gravitationally unstable disk is consistent
with the disk masses of Class 0/I YSOs estimated from 
dust thermal emission.
\citet{2007ApJ...671.1800A} observed YSOs in Ophiuchus and 
showed that the disk mass of Class I YSOs is typically $0.015 \msun$
and has a range of  $0.01 \msun<M_{\rm gas, obs}<0.1 \msun$.
\citet{2009A&A...507..861J} observed 20 Class 0/I YSOs and
showed that the typical disk mass of Class 0/I YSOs 
is approximately $0.05 \msun$.
\citet{2009A&A...507..861J} pointed out that their
theoretical model for disk evolution tends to produce a
larger mass than the observationally estimated value.
We propose that this inconsistency comes from the 
growth and depletion of the dust aggregate.
\citet{2011ApJS..195...21E} observed Class 0 YSOs
and suggested that the typical disk mass
is about $0.2 \msun$, which is larger than
the value suggested in other observational studies.
Because the apparent disk mass is an increasing 
function of mass accretion rate, as shown in figure \ref{disk_mass_mdot},
the large mass obtained in \citet{2011ApJS..195...21E} 
can be explained if the mass accretion from 
the envelope is large.
It is also possible that the dust disk has still not converged to
the steady-state structure in the Class 0 phase because of its
young age. In this case, the difference between 
$M_{\rm gas}$ and $M_{\rm app}$ becomes small.
Note also that contamination
of the envelope or temperature variation 
may also cause overestimation of the disk mass,
as was noted by these researchers.
Because the masses of Class 0/I YSOs are consistent with
or slightly larger than
our apparent disk mass, we conclude that a larger 
fraction of the Class 0/I disks
than previously considered are gravitationally unstable.

One may think that the decrease of dust-to-gas mass ratio
suggested in this paper is inconsistent with the recent observational
study that reports large dust-to-gas mass ratio $\sim 1/10$
of evolved disks in Lupus \citep{2016ApJ...828...46A}.
Note, however, that the gas mass is estimated from the CO line emissions
and, as authors noted, whether
the large dust-to-gas mass ratio indicates rapid gas loss
or depletion of CO by the chemical evolution is unclear (see also 
\citet{2016arXiv161201538M}). Note also that 
the age of the objects are $1-3$ Myr and they are not in Class 0/I phase.
Therefore, these observations do not directly contradict our results
because of uncertainty of chemical evolution and different evolutionary stages.

One may also think that the disk-mass independence of
the central-star mass suggested in this paper 
contradicts observational
results of positive correlation between stellar mass and
disk mass \citep{2011ARA&A..49...67W,2013ApJ...771..129A}.
However, the apparent disk-mass depends on the
mass accretion rate and the disk radius, and
it is possible that these parameters have correlations
to the central-star mass.
If this is the case, the disk-mass is indirectly correlated
to the central-star mass in Class 0/I YSOs.
Note also that the positive correlation
were reported based on the observations of Class II YSOs
and the results cannot be directly applied to Class 0/I YSOs.

\subsubsection{Empirical formula of dust-to-gas mass ratio for
a gravitationally unstable disk in Class 0/I YSOs}

Because $\mu_{\rm F}$ is the apparent reduction rate of the gas disk mass
and is useful for comparing theoretical and observational results, 
we derive the empirical formula for $\mu_{\rm F}$ using the conjugate gradient method as,
\begin{eqnarray}
\label{empirical_form_muf}
\mu_{\rm F} =0.17 \left(\frac{\dot{M}_{\rm gas}}{3 \times 10^{-7} \msunyear} \right)^{0.17} \nonumber \\
\left(\frac{M_{\rm star}}{1 \msun} \right)^{-0.53}\left(\frac{f}{10^{-1}} \right)^{0.0048}\left(\frac{r_{\rm disk}}{100\AU} \right)^{0.41}.
\end{eqnarray}
To derive this formula, we examined
the steady-state structures of 144 simulations
in total with different parameter sets.
We examined the following parameters:
$\dot{M}_{\rm gas} =\{1 \times 10^{-7},3 \times 10^{-7},1 \times 10^{-6},3 \times 10^{-6}\}\msunyear$,
$M_{\rm star} =\{0.5,1.0,2.0\}\msun$,
$f=\{10^{-4},10^{-3},10^{-2},10^{-1}\}$, and
$r =\{50, 100, 200\}\AU$.
We omitted the datasets with $f=1$ because 
they may be unlikely to occur in
realistic situations and the sudden jump 
at $f=1$ (see figure \ref{disk_mass_porus}) results in
an incorrect fit for $f\le10^{-1}$. Therefore, we cannot use 
the empirical formula for $f>10^{-1}$.

Because $\mu_{\rm F}$ indicates the extent to which
the disk mass is underestimated from that with
the ISM dust-to-gas mass ratio and the micron-sized dust,
we can calculate the "effective" 
dust-to-gas mass ratio $f_{\rm dg, eff}$ for a gravitationally unstable disk 
as
\begin{eqnarray}
\label{empirical_form_mueff}
f_{\rm dg, eff}\equiv \mu_{\rm F}f_{\rm dg, ISM} =1.7 \times 10^{-3}
\left(\frac{\dot{M}_{\rm gas}}{3 \times 10^{-7} \msunyear} \right)^{0.17} \nonumber \\
\left(\frac{M_{\rm star}}{1 \msun} \right)^{-0.53}\left(\frac{f}{10^{-1}} \right)^{0.0048}\left(\frac{r_{\rm disk}}{100\AU} \right)^{0.41} \left(\frac{f_{\rm dg, ISM}}{10^{-2}}  \right),
\end{eqnarray}
where $f_{\rm dg, ISM}$ is the dust-to-gas mass ratio of the ISM.

We can test whether
observed Class 0/I YSO disk is gravitationally unstable
using equation (\ref{empirical_form_mueff}) 
and the following procedure.
First, we calculate the trial gas disk mass using 
\begin{eqnarray}
M_{\rm trial}\equiv f_{\rm dg, eff}^{-1} M_{\rm dust},
\end{eqnarray}
where $M_{\rm dust}$ is the dust mass derived from the dust thermal emission
under the assumption of standard micron-sized dust opacity.
Then, we derive the mass of the gravitationally unstable disk as
\begin{eqnarray}
M_{\rm GI} = \int 2 \pi r \Sigma_{\rm GI} (r) dr 
\sim 8.6  \times 10^{-2} \left(\frac{M_{\rm star}}{1 \msun}\right)^{1/2}
\nonumber \\ 
\left(\frac{Q_{\rm crit}}{2}\right)^{-1}
(r_{\rm out}^{2/7}-r_{\rm in}^{2/7}) \msun,
\end{eqnarray}
Here,   $r_{\rm in}$ and $r_{\rm out}$ are the 
inner and outer radii of the disk in astronomical units, respectively.
$Q_{\rm crit}$ is the critical $Q$ value,
with the $ 1 \lesssim Q_{\rm crit} \lesssim 2$.
The temperature is assumed to be given by equation (\ref{sec2_irr}).
If $M_{\rm trial}$ is comparable to $M_{\rm GI}$,
we can conclude that the disk is gravitationally unstable.
On the other hand, when $M_{\rm trial}<M_{\rm GI}$, 
the disk is gravitationally stable.



\subsubsection{Appearance of gravitationally unstable disk: 
Do spiral arms always exist ?}
Although a gravitationally unstable disk is often considered to 
have spiral arms, this is not always true.
GI has an episodic nature and 
spiral arms emerge only when GI is triggered.
The viscous $\alpha$ could temporally increase to a value
as large as $\alpha_{\rm temp}\sim 1$
when the spiral arms are fully developed.
Once the surface density is redistributed by GI,
$\alpha_{\rm temp}$ decays to a small value until
the disk once again becomes sufficiently massive for gravitational 
instability to develop.
This episodic nature is neglected in the steady accretion
disk model using time-averaged value of $\alpha$.
When we consider a gravitationally unstable steady disk with 
$\alpha\sim 0.1$,
this means that the spiral arms appear only for a duration of 
$\sim (\alpha/\alpha_{\rm temp}) t_{\rm lifetime}\sim 0.05 $
Myr during the entire life time of a Class 0/I YSO where
$t_{\rm lifetime} \sim 0.5 $ Myr is the life time of a Class 0/I YSO.
Therefore, we expect that 
we have less chance to observe spiral arms in the Class 0/I phase,
even though the disk is marginally gravitationally unstable.

Note, however, that recent observations have found that 
some YSOs
have spiral structures possibly explainable by GI.
The grand designed spiral structures are found
in Elias 2-27, which is classified 
as Class II YSO \citep{2016Sci...353.1519P}.
Although \citet{2016Sci...353.1519P} 
suggested that the structures are spiral density waves, 
they can also be explained by gravitational instability 
as suggested by \citet[][]{2017ApJ...835L..11T}.
\citet{2016Natur.538..483T} found that the triple protostar 
system L1448 IRS 3B, which is classified as a Class 0 YSO,
has the spiral structures.
A compact multiple stellar system with 
spiral arms would be explained by fragmentation
of the gravitationally unstable disk \citep{
2009MNRAS.392..413S,2010ApJ...714L.133V,2010MNRAS.408L..36N,
2015MNRAS.446.1175T,2016MNRAS.458.3597T,2017MNRAS.465....2M}.
Although the degree of ubiquity of  spiral structures 
is still unclear, these observations suggest the
importance of investigating 
gravitationally unstable disks in Class 0/I YSOs 
in greater detail.

\subsubsection{Planetesimal formation in Class 0/I YSOs}

We suggest that Class 0/I YSOs are better sites for planetesimal
formation than Class II YSOs, which are commonly thought to be
the formation sites of planetesimals and planets.
One of the most important advantage 
for considering planetesimal formation in the Class 0/I phase
is that a large amount of solid material is available in this phase.
Because dust particles inevitably grow 
and radially migrate in a short timescale (see equation (\ref{time_grow})),
the total amount of dust that passes thorough the disk
is an important quantity for planetesimal formation.
Owing to envelope-to-disk accretion,
a significant amount of dust 
(approximately $10^{-2} \msun$ during the formation of
a $1 \msun$ protostar) is supplied to the disk during the  Class 0/I phase.
Therefore, the efficiency of planetesimal formation required
to produce planetesimals for a solar system-like planetary system
in which the  mass of solid material is about $10^{-4} \msun$ 
is only $\sim 1$ \%.
Thus, an inefficient planetesimal formation process is adequate.
On the other hand, 
the dust mass in Class II YSOs is typically 
$M_{\rm dust} \lesssim 10^{-4}\msun $ and there is no dust supply by
envelope-to-disk mass accretion.
Thus, the total amount of the dust passing through the disk is
approximately two orders of magnitude smaller than 
that in the Class 0/I phase
 and is comparable to that in the 
solar system. Therefore, almost 100 \% efficiency for 
planetesimal and subsequent planet formation is necessary to 
produce a solar system-like planetary system 
if planetesimal formation starts in the Class II phase.

Furthermore, it is also expected that the gas 
surface density in Class II phase is smaller than that in the Class 0/I phase,
which makes  planetesimals formation at larger orbital radius
difficult.
As shown in this paper, planetesimals can form
at $r \sim 20$ AU in a the gravitationally unstable disk.
On the other hand, 
in a disk with smaller gas surface density, overcoming the radial drift barrier becomes
more difficult because the mean free path increases as $\Sigma_{\rm gas}$ 
decreases and the dust aggregate enters the Stokes regime at a smaller radius.
As shown in \citet{2012ApJ...752..106O},
planetesimals can only form within 10 AU in the disk of minimum mass solar nebula model.
For these two reasons, 
we suggest that Class 0/I YSOs are preferable site
for planetesimal formation.

Once planetesimals form, they are decoupled
from the gas, and radial migration is negligible
until they grow to planets at which point Type I migration 
becomes important.
Therefore, 
solid material can be stored in the form of
planetesimals in the disks of Class 0/I YSOs.
Note that storing solid material in the form of small dust particles
is highly difficult because of its short growth timescale and
rapid radial migration \citep{2005A&A...434..971D,2007A&A...469.1169B,2009A&A...503L...5B}.

\subsubsection{Influence of outburst events}
Although there are several advantages for planetesimal formation
in Class 0/I YSOs, powerful outburst phenomena
expected in the Class 0/I phase can
possibly have negative impact on planetesimal formation by 
extending the snow line toward several tens of AU and 
reprocessing the icy dust particles.
Here, we focus on the
FU Ori outbursts because other outburst phenomena, such as
EX Lup outbursts, are relatively
weak \citep[see for example][]{2015ApJ...808...68H} and negligible.

During FU Ori outbursts,
the brightness increases by 4-6 magnitude on the timescale of months 
to years and 
gradually decays over a timescales of 100 years \citep{1989ESOC...33..233H}.
Once the stellar luminosity increases to several 
100 $L_\odot$, the H$_2$O snow line
moves to several tens of AU.
Recent observation of the FU Ori type star V883 Ori actually detected the
H$_2$O snow line at $\sim 40$ AU \citep{2016Natur.535..258C}.
Inside the snow line, H$_2$O is vaporized within a short time.
When an outburst ceases, the vaporized H$_2$O will
condense onto the dust particles.
Through this evaporation and condensation process, 
the properties of the dust particles change; in particular,
dust porosity would increase (dust particles are compacted),
which means that the outburst has a negative impact on
planetesimal formation via coagulation of porous dust aggregates.

Here, we roughly estimate the distance of snow line 
as a function of the luminosity.
By assuming the H$_2$O vaporization happens
at $T_{\rm vapor}=170$ K and disk temperature depends on the 
stellar luminosity as $T\propto L^{3/7}$ \citep{2007ApJ...654..606G},
the distance of the snow line from the central star
is calculated by solving $T_{\rm disk}=
150\times (L/L\odot)^{3/7}(r/1{\rm AU})^{-3/7}=T_{\rm vapor}$,
\begin{eqnarray}
\label{snow_line}
r_{\rm snow}\sim 41 \left(\frac{L}{400 L_\odot} \right)^{2/3} \AU.
\end{eqnarray}
This estimate is consistent with the snow line found in
V883 Ori, whose bolometric luminosity is estimated as $400 L_\odot$.


Whether burst events have a large impact on planetesimal formation
in the Class 0/I phase critically depends on the intervals 
between outburst events and on their magnitude.
Dust particles processed by an outburst
drift toward the central star and disappear, 
while unprocessed fresh dust particles 
are refilled, on the radial drift timescale.
Therefore, if the radial drift timescale or 
the refilling timescale is smaller than the outburst interval,
most of the dust does not undergo outburst processing.

Using equation (\ref{asymptotic_steady_solution_vr}) and (\ref{snow_line}), 
the radial drift 
(refiling) timescale in the Epstein regime is estimated as
\begin{eqnarray}
\label{refill_time_Epstein}
t_{\rm drift, Epstein}=\frac{r}{v_{\rm r,dust}}
\sim 2.6 \times10^4 \left(\frac{L}{400 L_\odot} \right)^{25/42}\nonumber \\
\left(\frac{\dot{M}}{10^{-6}\msunyear}\right)^{-1/2} {\rm years},
\end{eqnarray}
as a function of the stellar luminosity.

This is an upper limit on the drift timescale
because the porous dust enters the 
Stokes regime at several tens of AU and 
$v_{\rm r,dust}$ becomes much larger than the value of 
equation (\ref{asymptotic_steady_solution_vr}).
Using equation (\ref{Stokes_mu}) and (\ref{snow_line}), and assuming 
$\Delta v=\sqrt{\alpha c_{\rm s}^2 \St}~{\rm and}~ H_{\rm dust}=(\St/\alpha)^{-1/2} H_{\rm gas}$, the radial drift timescale in the Stokes regime is estimated as
\begin{eqnarray}
\label{drift_time_St}
t_{\rm drift,Stokes}\sim 2.7 \times 10^3
\left(\frac{L}{400 L_\odot}\right)^{103/63} \nonumber \\
\left(\frac{\dot{M}}{10^{-6}\msunyear}\right)^{-2/3}
\left(\frac{M}{1\msun}\right)^{-2/3}
\left(\frac{\rho_{\rm int}}{10^{-4}}\right)^{1/3} {\rm years}.
\end{eqnarray}

This is an lower limit on the drift timescale
because the dust radial profile does not quickly converge to
this asymptotic solution and the assumption of 
$\St \ll 1$ is no longer valid at the inner region of the disk.
In the realistic situation, it is expected that the refilling 
timescale is between $t_{\rm drift,Stokes}$ and $t_{\rm drift, Epstein}$.
These can be regarded as 
the timescales for which the system forgets an outburst with luminosity $L$.
If an outburst with a maximum luminosity of $L$
repeatedly occurs with an interval smaller than $t_{\rm drift}$ years,
it alters the planetesimal formation process suggested in this paper.

Although the event rate of outbursts is highly uncertain, 
it is estimated as $10^{-4}$ to $10^{-5}$ per year per protostar 
\citep{1996ARA&A..34..207H,2015ApJ...808...68H,2016ARA&A..54..135H},
suggesting a interval timescale $t_{\rm interval}\sim 10^4-10^5$ years.
$t_{\rm interval}$ seems to be comparable or slightly larger than $t_{\rm dirft}$.
Furthermore, $t_{\rm interval}$ is estimated using all FU Ori outbursts, 
but some of them are not strong enough to extend the snow line
to several tens of AU.
The bolometric luminosity of FU Ori objects 
ranges $20<L_{\rm bol}<550 L_\odot$  \citep{2001ApJS..134..115S} and 
a stellar luminosity of $L\sim 400 L_\odot$ seems to be closer 
to the maximum of the FU Ori outbursts.
Thus, it is expected that 
the event rate of strong outbursts that move the snow line
to several tens of AU is smaller than the
estimate of $10^{-4}$ to $10^{-5}$ per year per protostar.

Because the $t_{\rm drift}$ is $\sim 10^4$ years
and is expected to be smaller for porous dust aggregates, and
the interval timescale of the strong outburst 
is expected to be larger than $10^4$ years,
we conclude that the influence of outburst events 
is limited and that planetesimal formation in Class 0/I YSOs 
is still a promising pathway to producing planetesimals at the outer 
radius ($>10$ AU), although strong outburst events may 
decrease the efficiency of planetesimal formation in the Class 0/I phase. 

On the other hand, outbursts may have little influence on apparent disk mass reduction. 
As shown in equation (\ref{empirical_form_muf}), the reduction rate weakly
depends on the dust porosity. Furthermore, the outer region of the disk 
tends to have a larger contribution to the thermal radiation.
Thus, changes in the dust properties within several tens of AU
would not significantly affect the apparent disk-mass estimated from dust thermal
emission. 


\subsubsection{Importance of considering planet formation in Class 0/I YSOs}
Because planetesimals can form even in the Class 0/I phase,
investigating the planet formation process in this
phase is an interesting subject.
As discussed above, there are several advantages for 
planetesimal formation in the Class 0/I phase. 
In particular, planetesimal formation at $r>10$ AU
requires a large surface density corresponding to $Q\sim 1$.
There is also an advantage associated with considering 
planet formation in the  Class 0/I phase.
As shown in figure \ref{profile_mdot}, 
dust aggregates pass through the disk 
in the form of "pebbles".
Previous studies have shown that planetary seeds 
can grow quickly with pebble
accretion \citep{2010A&A...520A..43O,2012A&A...544A..32L,
2014A&A...572A.107L,2016A&A...591A..72I}.
In the Class 0/I phase, dust particles are supplied via envelope
accretion with a high mass accretion rate 
of $ \dot{M}_{\rm dust} \sim f_{\rm dg, ISM} \dot{M}_{\rm gas}
\sim 10^{-8}-10^{-9} \msunyear$. This period has a duration of $\sim 0.5$ Myr.
Under such a large accretion rate, protoplanet formation is
accelerated significantly, especially at 
$r>10 \AU$ \citep[][]{2016A&A...591A..72I}.
On the other hand, in the Class II phase, 
pebble accretion may stop within a short duration 
because of the depletion of dust at the outer edge of the disk.
As shown in \citet{2014A&A...572A.107L},
even starting from a disk with the ISM dust-to-gas mass 
ratio and micron-sized dust particles, 
dust particles in a disk of 100 AU in size deplete 
at $\sim 1$ Myr (this roughly corresponds
to the dust growth timescale at the outer edge).
As pointed out by \citet{2016A&A...591A..72I}, depletion of
dust at the outer disk edge 
is a serious problem for the pebble accretion scenario in an isolated disk.
Furthermore, at the beginning of the Class II phase, 
the dust-to-gas mass ratio may already be much smaller than $f_{\rm dg,ISM}$
and the dust particles have already grown to centimeter-size, even around the edge,
as shown in our simulations.
Therefore, dust depletion occurs over a
shorter period of time than that previously 
considered in the Class II phase.
Thus, planet formation via the
pebble accretion scenario may be preferred in Class 0/I YSOs.

Recent observations of HL Tau,
which is classified as being in the late Class I phase,
found the multiple ring structures 
in the dust disk \citep{2015ApJ...808L...3A} and 
possibly in the gas disk \citep{2016ApJ...820L..25Y}.
To explain the ring structures, 
several mechanisms, such as 
dust growth near the condensation front of volatiles \citep{
2015ApJ...806L...7Z,2016ApJ...818L..16Z},
sintering-induced ring formation
\citep{2016ApJ...821...82O},
secular gravitational instability 
\citep{2014ApJ...794...55T,2016AJ....152..184T},
or gap opening by planets \citep{2015MNRAS.453L..73D,
2015ApJ...806L..15K,2016PASJ...68...43K},
have been proposed.

If the gaps are induced by planets,
an obvious question is 
how these planets form in the very early phase of star formation (the
age of HL Tau is $\lesssim 1$ Myr).
In this paper, we proposed the possibility of
planetesimal formation in Class 0/I YSOs.
However, whether planetesimals can grow into the planets within 
the Class I phase is an open question.
The discovery of multiple rings also suggests the importance 
of investigating the possibility of planet 
formation in the Class 0/I phase.

\section *{Acknowledgments}
We thank Dr. H. Tanaka, Dr. H. Kobayashi, Dr. S. Takahashi, 
Dr. S. Inutsuka, Dr. Y. Imaeda, Dr. M. Kunitomo,
Dr. T. Muto, for their fruitful discussions.
We also thank the anonymous referee for his/her insightful comments.

\appendix
\section{A:Analytic solutions for steady-state structure of dust disk}
In this appendix, we derive the asymptotic steady-state 
solution for a dust disk in which dust radial drift dominates.
Although, we focus on the dust structure of a gravitationally unstable disk,
the solution is applicable to an arbitrary gas disk structure.

\subsection{governing equations}

Through omission of the time derivatives 
of equations (\ref{governing_eq1}) 
and (\ref{governing_eq2}), 
the governing equations for the steady-state solutions are given as 
\begin{eqnarray}
\label{governing_eq1_steady}
\frac{1}{r} \frac{\partial}{\partial r}(r v_{\rm r, dust}\Sigma_{\rm dust})=0,\\
\label{governing_eq2_steady}
v_{\rm r, dust} \frac{\partial m_{\rm dust}}{\partial r}=
\frac{m_{\rm dust}}{t_{\rm coll}}.
\end{eqnarray}

Equation (\ref{governing_eq1_steady}) is easily 
integrated and the mass conservation of dust in the steady state
is expressed as
\begin{equation}
\label{sigma_d1}
2 \pi r |v_{\rm r, dust}|\Sigma_{\rm dust}={\dot M}_{\rm dust}(={\rm const}).
\end{equation}

On the other hand, 
equation (\ref{governing_eq2_steady}) can be rewritten as
\begin{equation}
\label{g_eq_size}
\frac{\partial a_{\rm dust}(r)}{\partial r}=-
\left(\frac{1}{2 \pi} \right)^{3/2}
\left(\frac{m_{\rm gas}^2 v_{\rm K}^2 \dot{M}_{\rm dust}}{r k_B^2 T^2 (d \ln P/d \ln r)^2  \rho_{\rm int}} \right)\left(\frac{\Delta v}{H_{\rm dust} \St^2 } \right)
\end{equation}
where we use equation (\ref{sigma_d1}), $m_{\rm dust}=(4\pi/3)\rho_{\rm int} a^3_{\rm dust}$,
$n_{\rm dust}=\Sigma_{\rm dust}/(\sqrt{2 \pi} H_{\rm dust} m_{\rm dust})$, $v_{\rm r,dust}=- 2 \eta v_{\rm K} \St$, and assumed that $\St\ll 1$.

Note that the variables in the second bracket are constant or depend 
only on the gas disk structure;
on the other hand, those in the third bracket depend on $a_{\rm dust}$.
Thus, we can solve equation (\ref{g_eq_size}) by specifying
the relative velocity, dust scale height, and drag law.

\subsection{Steady-state solutions for dust disk}
First, we derive the steady-state solution for the case 
in which the drag law is given by the Epstein drag law and 
the relative velocity and dust scale height are given as
$\Delta v=\sqrt{\alpha c_{\rm s}^2 \St}$ and $H_{\rm dust}=(\St/\alpha)^{-1/2}H_{g}$,
respectively.
In this case, equation (\ref{g_eq_size}) becomes
\begin{equation}
\label{governing_eq_size2}
\frac{\partial a_{\rm dust}(r)}{\partial r}=-
\left(\frac{\dot{M}_{\rm dust}  m_{\rm gas}^2 v_{\rm K}^3 \Sigma_{\rm gas}}{(2\pi^5)^{1/2} k_B^2 T^2 (d \ln P/d \ln r)^2  \rho_{\rm int}^{2}} \right)r^{-2} a_{\rm dust}(r)^{-1}.
\end{equation}
By assuming that the gas disk profile can be expressed 
as the power law 
$\Sigma_{\rm gas}(r)=\Sigma_0 r^{-n_\Sigma},~
T(r)=T_0 r^{-n_T},~
\alpha(r)=\alpha_0 r^{n_\alpha},~
v_{\rm K}(r)=v_{\rm K,0} r^{-n_{K}}
$,
we can rewrite equation (\ref{governing_eq_size2}) as
\begin{eqnarray}
\label{governing_eq_size3}
\frac{\partial a_{\rm dust}(r)}{\partial r}=-
\left(\frac{\dot{M}_{\rm dust} m_{\rm gas}^2 v_{\rm K,0}^3 \Sigma_0}{(2 \pi^5)^{1/2} (d \ln P/d \ln r)^2 k_B^2 T_0^2   \rho_{\rm int}^{2}} \right)\times \nonumber \\
r^{-2+2 n_T-3n_{\rm K}-n_\Sigma} a_{\rm dust}(r)^{-1} \nonumber \\
\equiv A r^{-2+2 n_T-3n_{\rm K}-n_\Sigma} a_{\rm dust}(r)^{-1},
\end{eqnarray}
where $A$ is a negative constant.
Solution of (\ref{governing_eq_size3}) is given as
\begin{eqnarray}
\label{general_soluti
on}
a_{\rm dust}(r) =\left(\frac{2A}{-1+2 n_T-3n_{\rm K} -n_\Sigma} r^{-1+2 n_T-3n_{\rm K} -n_\Sigma} +C\right)^{\frac{1}{2}},
\end{eqnarray}
where $C$ is a constant.
Because the power law index of $r$, $(-1+2 n_T-3n_{\rm K} -n_\Sigma)$
is negative for the gas disk used in this paper, ($(-1+2 n_T-3n_{\rm K} -n_\Sigma)=-47/14 < 0$),
the dust size asymptotically converges to the power law 
\begin{eqnarray}
\label{asymptotic_solution}
a_{\rm dust}(r) \to \left(\frac{2A}{-1+2 n_T-3n_{\rm K} -n_\Sigma} r^{-1+2 n_T-3n_{\rm K} -n_\Sigma}\right)^{\frac{1}{2}} \\
\nonumber (r \to 0)
\end{eqnarray}
Using this asymptotic solution, we can show that the ratio of 
the collision timescale $t_{\rm coll}$ 
and the drift timescale $t_{\rm drift}\equiv r/v_{\rm r,dust}$
converges to a constant value as 
\begin{eqnarray}
\label{ratio_tcoll_tdrift_Ep}
\mu_{\rm dust} \equiv \frac{t_{\rm coll}}{t_{\rm drift}} \to \frac{2}{3(1-2 n_T+3n_{\rm K}+ n_\Sigma)}~(r \to 0).
\end{eqnarray}
The asymptotic value of $\mu_{\rm dust}$ is determined by the 
power indices of the gas disk.
For the gas disk model used in this paper, 
\begin{eqnarray}
\label{ratio_tcoll_tdrift_Ep_value}
\mu_{\rm dust} =\frac{2}{3(1-2 n_T+3n_{\rm K}+ n_\Sigma)}=\frac{28}{141},
\end{eqnarray}
where we use equations (\ref{GI_disk_sigma}) -- (\ref{GI_disk_alpha}) 
and $v_{\rm K}\propto r^{-1/2}$ .
The fact that $\mu_{\rm dust} \to {\rm const}~ (r\to 0)$  
has already been pointed out by \citet{2016ApJ...821...82O}.
The new finding in this section is  
that the value of $\mu_{\rm dust}$ depends on the gas disk structure
and there is no universal value for $\mu_{\rm dust}$.

Once the asymptotic solution for dust size 
(or equivalently, the value of $t_{\rm coll}/t_{\rm drfit}$) is determined, 
the dust surface density,
Stokes number,
collision velocity,
and radial drift velocity are
determined by 
equation (\ref{sigma_d1}),
$\St=(\pi \rho_{\rm int} a_{\rm dust})/(2 \Sigma_{\rm gas})$,
$\Delta v=\sqrt{\alpha c_{\rm s}^2 \St}$, and
$v_{\rm r,dust}=- 2 \eta v_{\rm K} \St$, respectively.
The asymptotic steady-state solution of the dust structure 
for our gas disk model
(equations (\ref{sec2_irr}), (\ref{GI_disk_sigma}), and (\ref{GI_disk_alpha}))
can be calculated as
\begin{align}
\label{asymptotic_steady_solution_sigma}
\Sigma_{\rm dust}&= 1.0 \left(\frac{r}{10 \AU}\right)^{-\frac{31}{28}}  \left(\frac{\dot{M}_{\rm gas}}{10^{-6} \msunyear}\right)^{\frac{1}{2}} \gscm& \\
\label{asymptotic_steady_solution_St}
\St &= 9.0 \times 10^{-2} \left(\frac{r}{10 \AU}\right)^{\frac{1}{28}}&  \nonumber \\
&\left(\frac{\dot{M}_{\rm gas}}{10^{-6} \msunyear}\right)^{\frac{1}{2}}
\left(\frac{M_{\rm star}}{1 \msun}\right)^{\frac{1}{2}}& \\
\label{asymptotic_steady_solution_size}
a_{\rm dust} &= 3.8 \times 10^2 \left(\frac{r}{10 \AU}\right)^{-\frac{47}{28}}  \left(\frac{\dot{M}_{\rm gas}}{10^{-6} \msunyear}\right)^{\frac{1}{2}}& \nonumber \\
&\left(\frac{M_{\rm star}}{1 \msun}\right) \left(\frac{\rho_{\rm int}}{0.1 \gcm}\right)^{-1} \cm &\\
\label{asymptotic_steady_solution_vr}
v_{\rm r,dust}&=-6.5 \times 10^2 \left(\frac{r}{10 \AU}\right)^{\frac{3}{28}}  \left(\frac{\dot{M}_{\rm gas}}{10^{-6} \msunyear}\right)^{\frac{1}{2}} \cms& \\
\label{asymptotic_steady_solution_dv}
\Delta v&= 2.8 \times 10^3 \left(\frac{r}{10 \AU}\right)^{\frac{1}{8}}& \nonumber \\
& \left(\frac{\dot{M}_{\rm gas}}{10^{-6} \msunyear}\right)^{\frac{3}{4}}\left(\frac{M_{\rm star}}{1 \msun}\right)^{\frac{1}{4}} \cms &
\end{align}
where we assume that the mass accretion rate of the dust is
given as $\dot{M}_{\rm dust}=f_{\rm dg,ISM} \dot{M}_{\rm gas}$.
We also assume that $\Delta v =\sqrt{2 \alpha c_{\rm s}^2 \St}$
instead of $\Delta v =\sqrt{\alpha c_{\rm s}^2 \St}$  for  
consistency with the numerical simulations conducted in this study.
The value of $\mu_{\rm dust}$ 
is not affected by the factor of difference of
$\Delta v$.
As already shown in figure \ref{profile3e-7}, 
this asymptotic solution describes the numerical results well.

The asymptotic value for 
$\mu_{\rm dust}$ in equation (\ref{ratio_tcoll_tdrift_Ep}) is obtained
by assuming $\Delta v=\sqrt{\alpha c_{\rm s}^2 \St}$, $H_{\rm dust}=(\St/\alpha)^{-1/2} H_{\rm gas}$
and the Epstein drag law; however, there are other possibilities. 
For example,
it is possible that the radial drift 
determines the collision velocity as $\Delta v=\eta v_{\rm K} \St$,
or that the dust scale height is equal to the gas scale height as 
$H_{\rm dust}= H_{\rm gas}$, or that the gas drag is determined by the Stokes drag law.
As expected from the term in the 
third bracket of equation (\ref{governing_eq_size3}),
the asymptotic value of $\mu_{\rm dust}$ depends on the forms
of $\Delta v$, $H_{\rm dust}$ and the drag law.
Following the same procedure as that described above, we can calculate the 
asymptotic value for $\mu_{\rm dust}$ with these different assumptions.
The asymptotic values for $\mu_{\rm dust}$ in the Epstein regime are given as 
\begin{equation}
\mu_{\rm dust}=
     \begin{cases}     
       \frac{2}{3(1-2 n_{\rm T}+3 n_{v_{\rm K}}+n_{\Sigma})}~ (\Delta v=\sqrt{\alpha c_{\rm s}^2 \St}~{\rm and}~ H_{\rm dust}=(\St/\alpha)^{-1/2} H_{\rm gas}) \\
       \frac{1}{2-3 n_{\rm T}+4 n_{v_{\rm K}}+n_{\alpha}+ n_{\Sigma}}~ (\Delta v=\eta v_{\rm K} \St~{\rm and}~ H_{\rm dust}=(\St/\alpha)^{-1/2} H_{\rm gas}) \\
       \frac{5}{3(2-4 n_{\rm T}+6 n_{v_{\rm K}}-n_{\alpha}+3 n_{\Sigma})}~ (\Delta v=\sqrt{\alpha c_{\rm s}^2 \St}~{\rm and}~ H_{\rm dust}=H_{\rm gas}) 
     \end{cases}
\end{equation}

In the case of the Stokes regime,
\begin{equation}
\label{Stokes_mu}
\mu_{\rm dust}=
    \begin{cases}     
      \frac{2}{-3 n_{\rm T}+4 n_{v_{\rm K}}}~ (\Delta v=\sqrt{\alpha c_{\rm s}^2 \St}~{\rm and}~ H_{\rm dust}=(\St/\alpha)^{-1/2} H_{\rm gas}) \\
      \frac{8}{3(2 - 5 n_{\rm T}+6 n_{v_{\rm K}}+2 n_{\alpha})}~ (\Delta v=\eta v_{\rm K} \St~{\rm and}~ H_{\rm dust}=(\St/\alpha)^{-1/2} H_{\rm gas}) \\
      \frac{16}{3(-2-5 n_{\rm T}+6 n_{v_{\rm K}}-2 n_{\alpha})}~ (\Delta v=\sqrt{\alpha c_{\rm s}^2 \St}~{\rm and}~ H_{\rm dust}=H_{\rm gas}). \\
     \end{cases}
\end{equation}
When $\mu_{\rm dust}$ becomes negative, 
the asymptotic solutions become complex 
and physical solutions under the given disk structures and conditions
do not exist.

Once $\mu_{\rm dust}$ is determined, 
the asymptotic solution for the dust size and the Stokes number 
are determined using
\begin{equation}
\label{St_mu}
\frac{t_{\rm coll}}{t_{\rm drift}}=\frac{3 \Delta v \dot{M}_{\rm  dust}}{2 (2 \pi)^{3/2} H_{\rm dust} \eta^2 \St^2 \rho_{\rm int}a_{\rm dust}}=\mu_{\rm dust},
\end{equation}
and equation (\ref{Stokes_num}).
Then, we can obtain the dust surface density and
the drift velocity
by equation (\ref{sigma_d1}) and
$v_{\rm r,dust}=- 2 \eta v_{\rm K} \St$, respectively.

\section{B:Comparison between viscous heating and irradiation heating}
In this paper, we assumed that the disk temperature is determined by 
stellar irradiation and viscous heating is negligible in outer region ($r>10$ AU).
To confirm that viscous heating is negligible 
in a gravitationally unstable disk in outer region,  
we estimate the temperature profile determined by the viscous heating only.
When viscous heating determines the disk temperature, 
the energy balance between 
the local viscous heating and the local radiation cooling,
\begin{equation}
\label{sec2_cooling0}
 \left| \frac{d \ln \Omega}{d \ln R} \right|^{2} \alpha \frac{c_{\rm s}^2}{\Omega}  \Sigma \Omega^2 = \frac{32 \sigma T_{\rm visc}^4}{3 (\tau_{\rm half}+\tau_{\rm half}^{-1})}.
\end{equation}
is realized in the disk,
where $T_{\rm visc}$ is the midplane temperature in the viscously heated disk,
$\tau_{\rm half}=(1/2) \kappa \Sigma_{\rm gas}$ is the vertical optical depth,
and $\kappa$ is the opacity.
We assume that the opacity obeys the power law 
\begin{equation}
\kappa(T)=\kappa_0 T^2~{\rm cm^2~g^{-1}}.
\end{equation}
This formula approximates the (gray) dust opacity 
in a low-temperature region, $T<100$ K \citep{1994ApJ...427..987B}.
$\kappa_0$ is typically $10^{-4} {\rm cm^2~g^{-1}}$
if we assume a dust-to-gas mass ratio of $1/100$. 
However, as shown in figure \ref{profile3e-7}, the dust-to-gas mass ratio 
becomes $\sim 1/1000$  and $\kappa_0$ becomes $\sim 10^{-5} {\rm cm^2~g^{-1}}$ 
in the steady state of our fiducial model at $r\sim 10$ AU.
By solving equations
(\ref{sec2_dynamics}), (\ref{sec2_qvalue}), (\ref{sec2_omega_kep}),
and (\ref{sec2_cooling0}) and assuming that 
the disk is vertically optically thick,
we obtain the temperature profile of a disk 
in which the local heating balances with the local radiative cooling
\begin{eqnarray}
T_{\rm visc, thick}=58 
 \left( \frac{r}{10 \AU} \right)^{-3} \left( \frac{\kappa_0}{10^{-5}  {\rm cm^2~g^{-1}}} \right)^{2/3} \nonumber  \\
\left( \frac{M_{\rm star}}{\msun }\right) \left( \frac{\dot{M}_{\rm gas}}{10^{-6} \msunyear }\right)^{2/3} ~{\rm K}.
\end{eqnarray}
Thus, $T_{\rm visc, thick}$ 
is nearly equal to the irradiated temperature $T_{\rm irr}$ 
at $10$ AU, $T=55$ K,
and  decreases more rapidly than
$T_{\rm irr}$ as $r$ increases.
Thus, $T_{\rm visc, thick}<T_{\rm irr}$
in almost the entire region of an optically thick disk.
On the other hand, the disk may become vertically 
optically thin in the outer region.
The temperature profile for an optically thin disk is given as
\begin{eqnarray}
T_{\rm visc, thin}= 16 
 \left( \frac{r}{100 \AU} \right)^{-3/13} \left( \frac{\kappa_0}{10^{-5}  {\rm cm^2~g^{-1}}} \right)^{-2/13} \nonumber  \\
 \left( \frac{\dot{M}_{\rm gas}}{10^{-6} \msunyear }\right)^{2/13} ~{\rm K}.
\end{eqnarray}
Although the radial dependence of $T_{\rm visc, thin}$ is shallower than
that of  $T_{\rm irr}$,
$T_{\rm visc, thin}$ is smaller than $T_{\rm irr}$ for 
$r\lesssim300 \AU$.
Therefore, at $r>10 $ AU, $T_{\rm visc}$ is smaller than 
$T_{\rm irr}$, in both the optically thick and thin cases
and we can adopt an approximation in which the stellar irradiation 
determines the temperature of the entire disk.

\bibliography{article}

\end{document}